\documentclass{jfm}

\usepackage{amsfonts}
\usepackage[fleqn,reqno]{amsmath}
\usepackage{amssymb}
\usepackage[titletoc]{appendix}
\usepackage{array}
\usepackage{enumitem}
\usepackage{filecontents}
\usepackage{graphics,graphicx}
\usepackage{lineno}
\usepackage{subcaption}
\usepackage{pgfplots}
\usepackage{tikz}
\usepackage{todonotes}
\usetikzlibrary{arrows}
\usepackage{comment}
\usepackage{float}
\usepackage{soul} 
\usepackage{epstopdf,epsfig}

\usepackage[pagebackref=false,bookmarks=false]{hyperref} 
\hypersetup{
  pdfcreator = {LaTeX with hyperref package}
}

\newcommand{\bd}{{\partial}}

\newcommand{\cc}{{\mathbf{c}}}
\newcommand{\CC}{{\mathbb{C}}}
\newcommand{\DD}{{\mathcal{D}}}
\newcommand{\DDD}{{\boldsymbol{\mathcal D}}}
\newcommand{\eeta}{{\boldsymbol\eta}}
\newcommand{\ff}{{\mathbf{f}}}
\newcommand{\grad}{{\nabla}}

\newcommand{\iin}{\mathrm{in}}
\newcommand{\llambda}{{\boldsymbol\lambda}}
\newcommand{\nn}{{\mathbf{n}}}
\newcommand{\NN}{{\mathcal{N}}}
\newcommand{\out}{\mathrm{out}}
\newcommand{\rr}{{\mathbf{r}}}

\newcommand{\RR}{{\mathbb{R}}}
\renewcommand{\ss}{{\mathbf{s}}}

\newcommand{\bary}{\mathrm{bary}}
\newcommand{\trap}{\mathrm{trap}}
\newcommand{\uu}{{\mathbf{u}}}
\newcommand{\UU}{{\mathbf{U}}}
\newcommand{\vv}{{\mathbf{v}}}
\newcommand{\xx}{{\mathbf{x}}}

\newcommand{\yy}{{\mathbf{y}}}

\def\gap{\hspace*{.2in}}

\newcommand{\pderiv}[2]{\frac{\partial #1}{\partial #2}}

\newcommand{\abs}[1]{\lvert #1 \rvert}

\newcommand{\thL}{$\theta$--$L$}
\newcommand{\eps}{\varepsilon}
\newcommand{\Vn}{V_\nn}

\newcommand{\CE}{C_E}

\shorttitle{Viscous Transport in Eroding Porous Media}
\shortauthor{Chiu, Moore and Quaife}

\title{Viscous Transport in Eroding Porous Media}

\author{Shang-Huan Chiu\aff{1}, M.~N.~J.~Moore\aff{2}, and Bryan
Quaife\aff{3}\corresp{\email{bquaife@fsu.edu}}}

\affiliation{
\aff{1}Department of Scientific Computing, Florida State University,
Florida State University, Tallahassee, FL 32306, USA
\aff{2}Department of Mathematics and Geophysical Fluid Dynamics
Institute, Florida State University, Tallahassee, FL 32306, USA
\aff{3}Department of Scientific Computing and Geophysical Fluid Dynamics
Institute, Florida State University, Tallahassee, FL 32306, USA
}

\begin{document}

\maketitle

\begin{abstract} 
  Transport of viscous fluid through porous media is a direct
  consequence of the pore structure. Here we investigate transport
  through a specific class of two-dimensional porous geometries, namely
  those formed by fluid-mechanical erosion.  We investigate the
  tortuosity and dispersion by analyzing the first two statistical
  moments of tracer trajectories. For most initial configurations,
  tortuosity decreases in time as a result of erosion increasing the
  porosity.  However, we find that tortuosity can also increase
  transiently in certain cases.  The porosity-tortuosity relationships
  that result from our simulations are compared with models available in
  the literature.  Asymptotic dispersion rates are also strongly
  affected by the erosion process, as well as by the number and
  distribution of the eroding bodies. Finally, we analyze the pore size
  distribution of an eroding geometry. The simulations are performed by
  combining a high-fidelity boundary integral equation solver for the
  fluid equations, a second-order stable time stepping method to
  simulate erosion, and new numerical methods to stably and accurately
  resolve nearly-touching eroded bodies and particle trajectories near
  the eroding bodies.
\end{abstract}

\section{Introduction}
\label{sec:intro}
Porous media flow plays an important role in many environmental and
industrial applications.  Depending on the application, length scales
can vary from $10^{-6}\mathrm{m}$ to
$10^{-1}\mathrm{m}$~\citep{mil-chr-imh-mcb-ped1998} and velocity scales
can be as small as $10^{-1}\mathrm{m/day}$~\citep{kut-scr-dav-ham1995}.
Moreover, for a single porous geometry, the pore sizes and velocities
can range over several orders of magnitude.  Numerical methods that
resolve this range of scales offer the ability to: (i) characterize
dispersion~\citep{saf1959}, (ii) quantify
mixing~\citep{leb-den-dav-bol-car-dec-bou2011, den-leb-eng-bij2011}, and
(iii) develop meaningful constitutive relationships that link the
microscopic and macroscopic realms~\citep{mil-chr-imh-mcb-ped1998}.
Examples of coarse-grained models for porous media flow include
permeability-porosity relationships~\citep{dar-mcc1998, car1937},
tortuosities~\citep{mat-kha-koz2008, dud-koz-mat2011, kop-kat-tim1996},
geometry connectivity~\citep{knu-car2005}, anomalous
dispersion~\citep{den-cor-sch-ber2004}, and more.  

Flow in porous media is further complicated when boundaries evolve 
dynamically in response to the fluid flow. This coupling between
geometry and flow occurs, for example, in applications involving
melting~\citep{bec-vis1998, rycroft2016asymmetric, jambon2018singular,
favier2019rayleigh, morrow2019moving},
dissolution~\citep{kan-zha-che-he2002, mac2015shape, moo2017,
wykes2018self}, deposition~\citep{joh-eli1995, hewett2018modelling},
biofilm growth~\citep{tan-val-wer2015}, and crack
formation~\citep{cho2019crack}. We focus on erosion, a fluid-mechanical
process that is prevalent in many geophysical, hydrological, and
industrial applications~\citep{ris-moo-chi-she-zha2012,
berhanu2012shape, hewett2017evolution, lachaussee2018competitive,
lopez2018cfd, allen2019sde, amin2019role}.

When a porous medium erodes, certain qualitative characteristics are
unveiled that affect transport through the geometry.  For example, an
eroded geometry may contain channels of high porosity, which, though few
in number and modest in volume fraction, transmit a large portion of the
flux~\citep{qua-moo2018}. This arrangement results in velocities that
vary over several orders of magnitude~\citep{all-hea-lab-rei2002}.
Moreover, channelization creates heterogeneous and anisotropic medium
properties, which affect the transport of tracers such as
contaminants~\citep{cve-che-wen1996, dag1987, kon-bre1978} and
heat~\citep{nil-sto1990, ree-sto1995}.

\begin{figure}
\begin{center}
\includegraphics[height = 0.3 \textwidth]{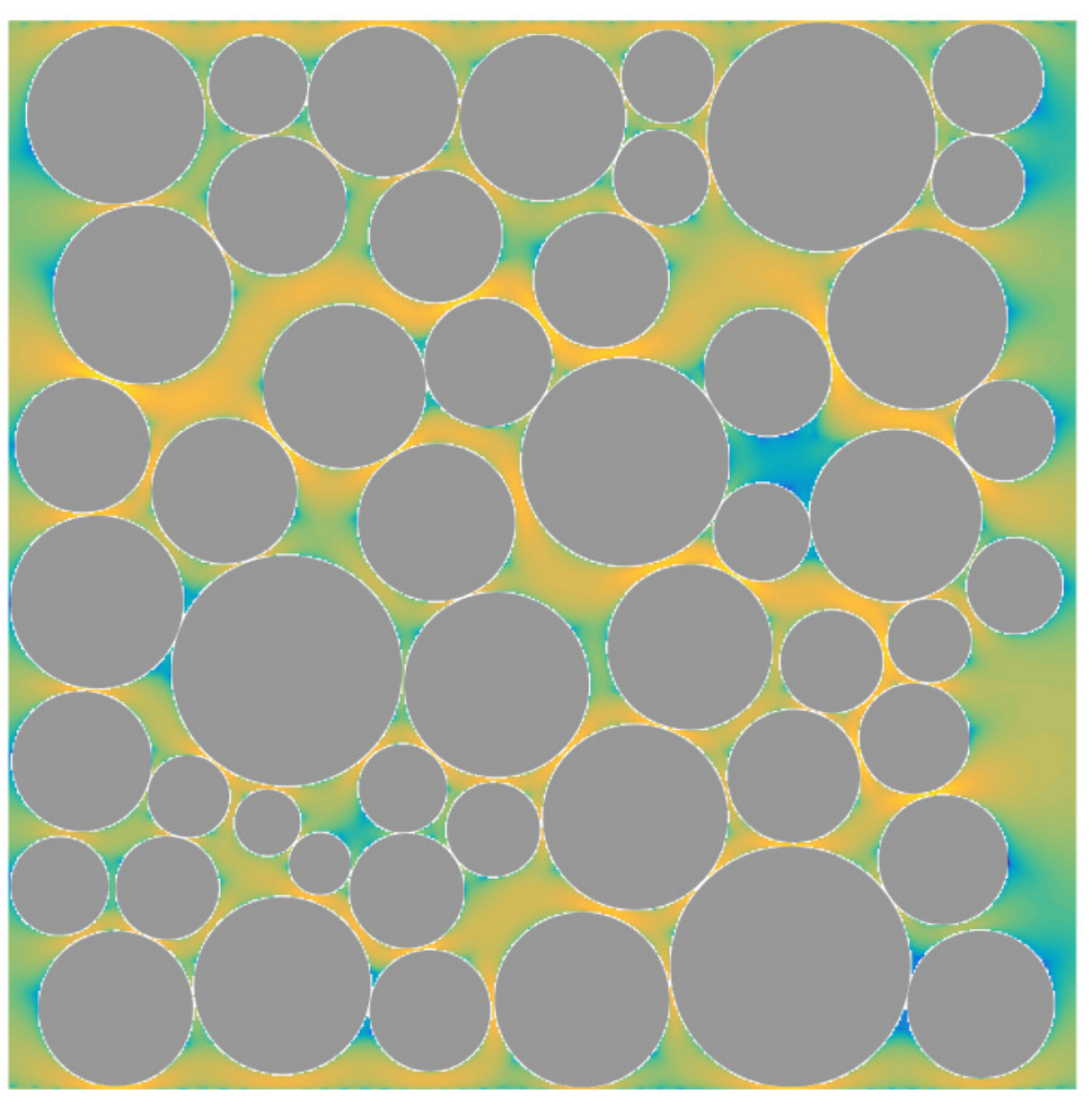}
\includegraphics[height = 0.3 \textwidth]{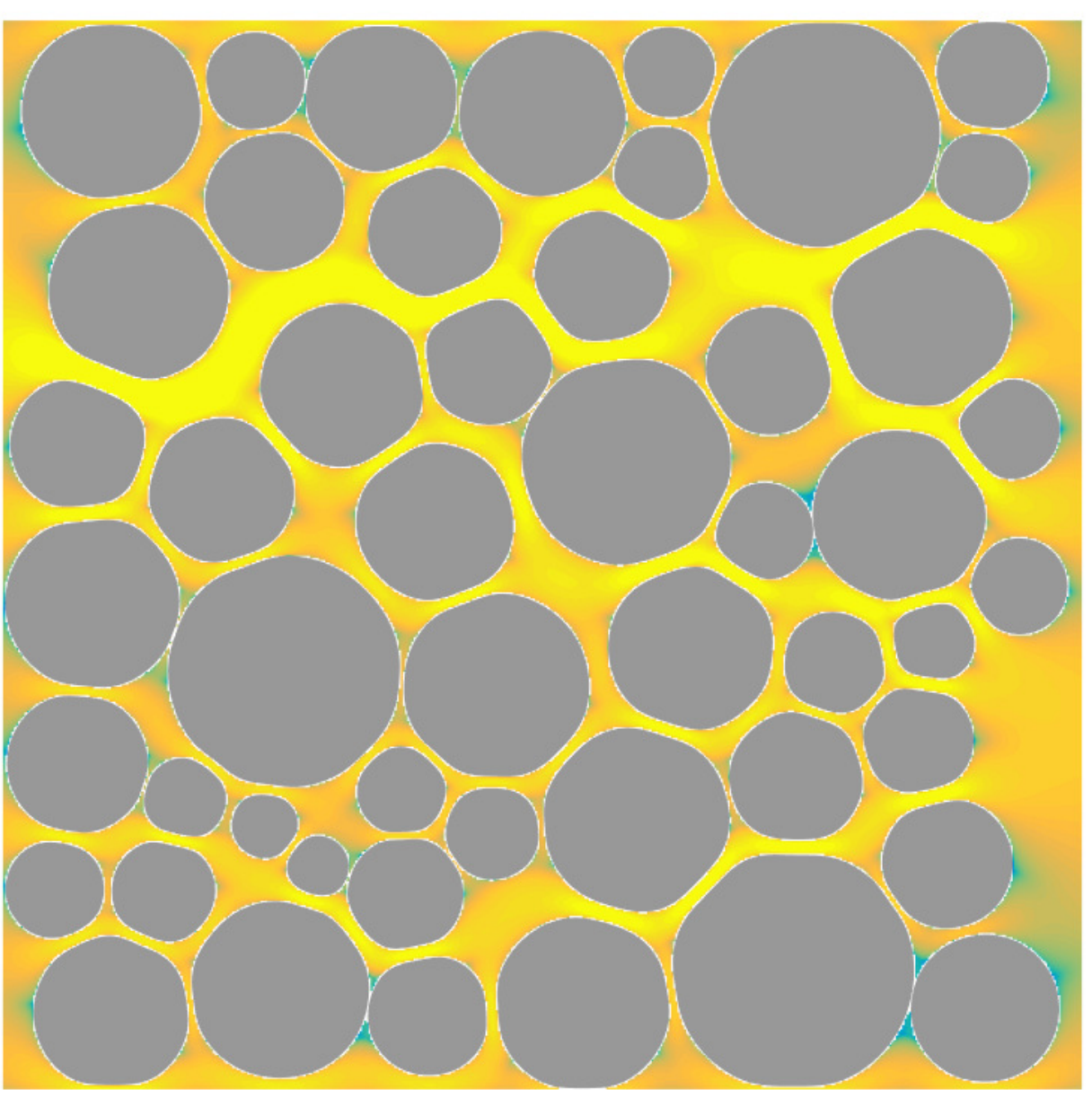}
\includegraphics[height = 0.3 \textwidth]{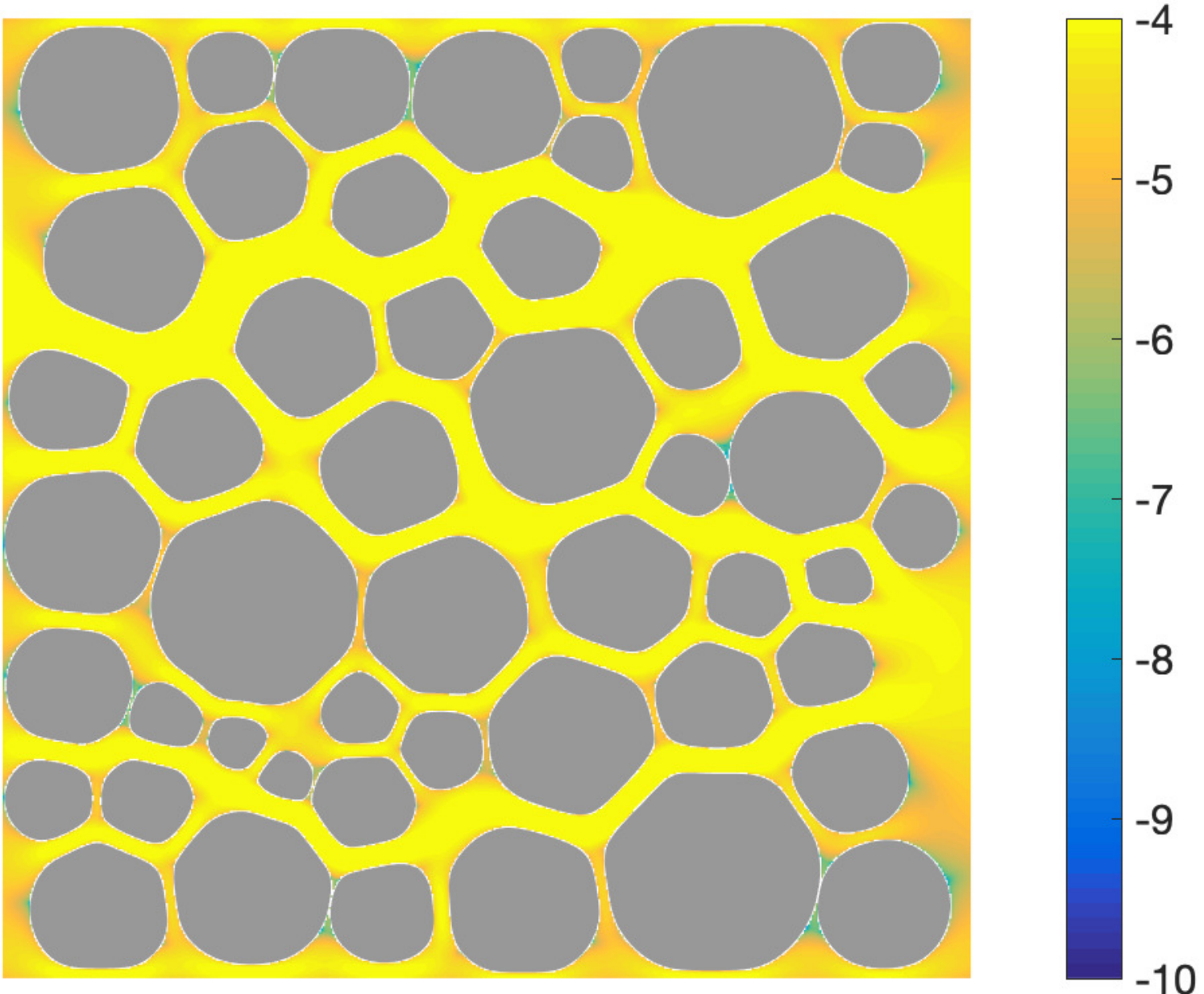}\\
\includegraphics[height = 0.3 \textwidth]{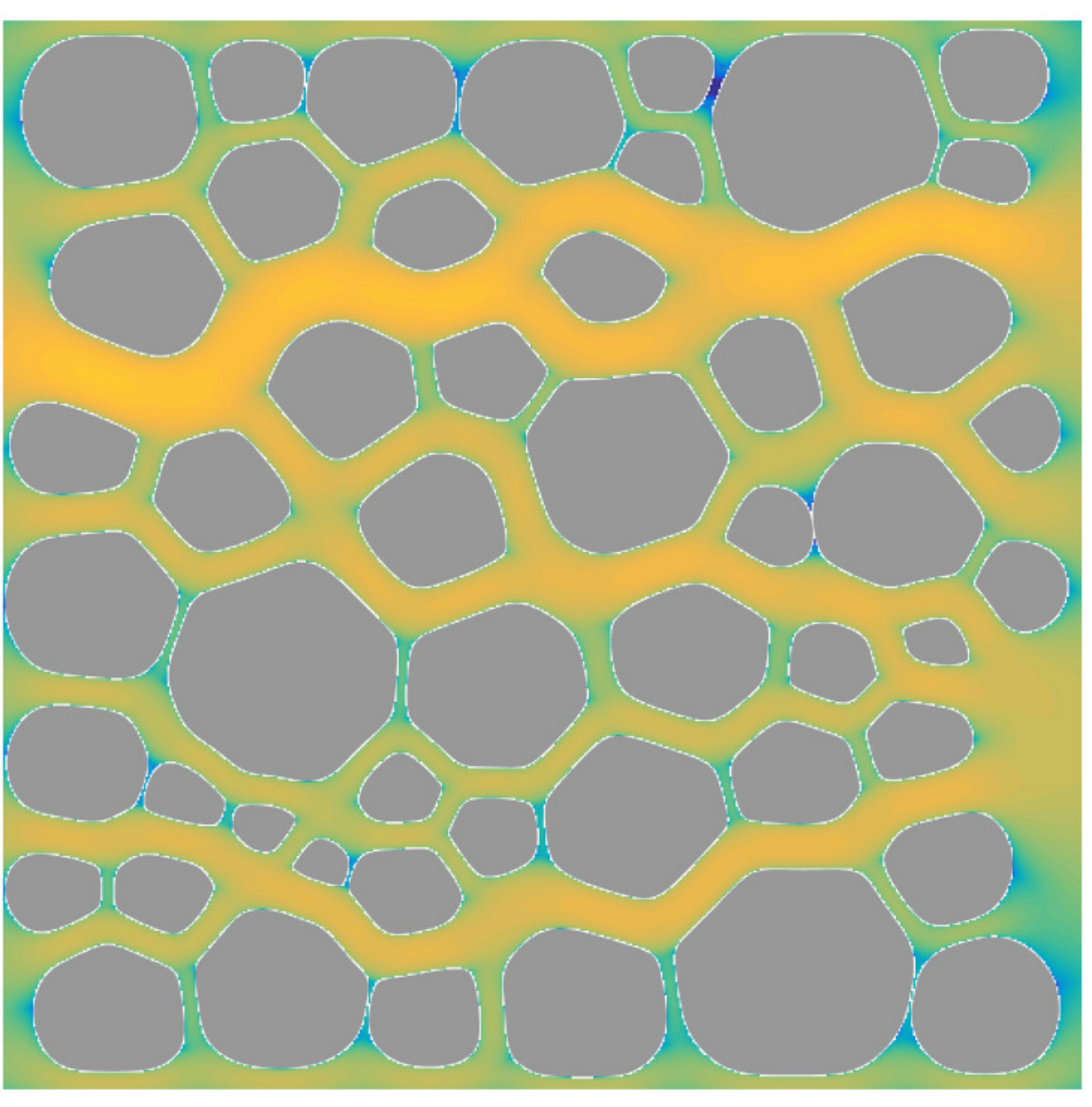}
\includegraphics[height = 0.3 \textwidth]{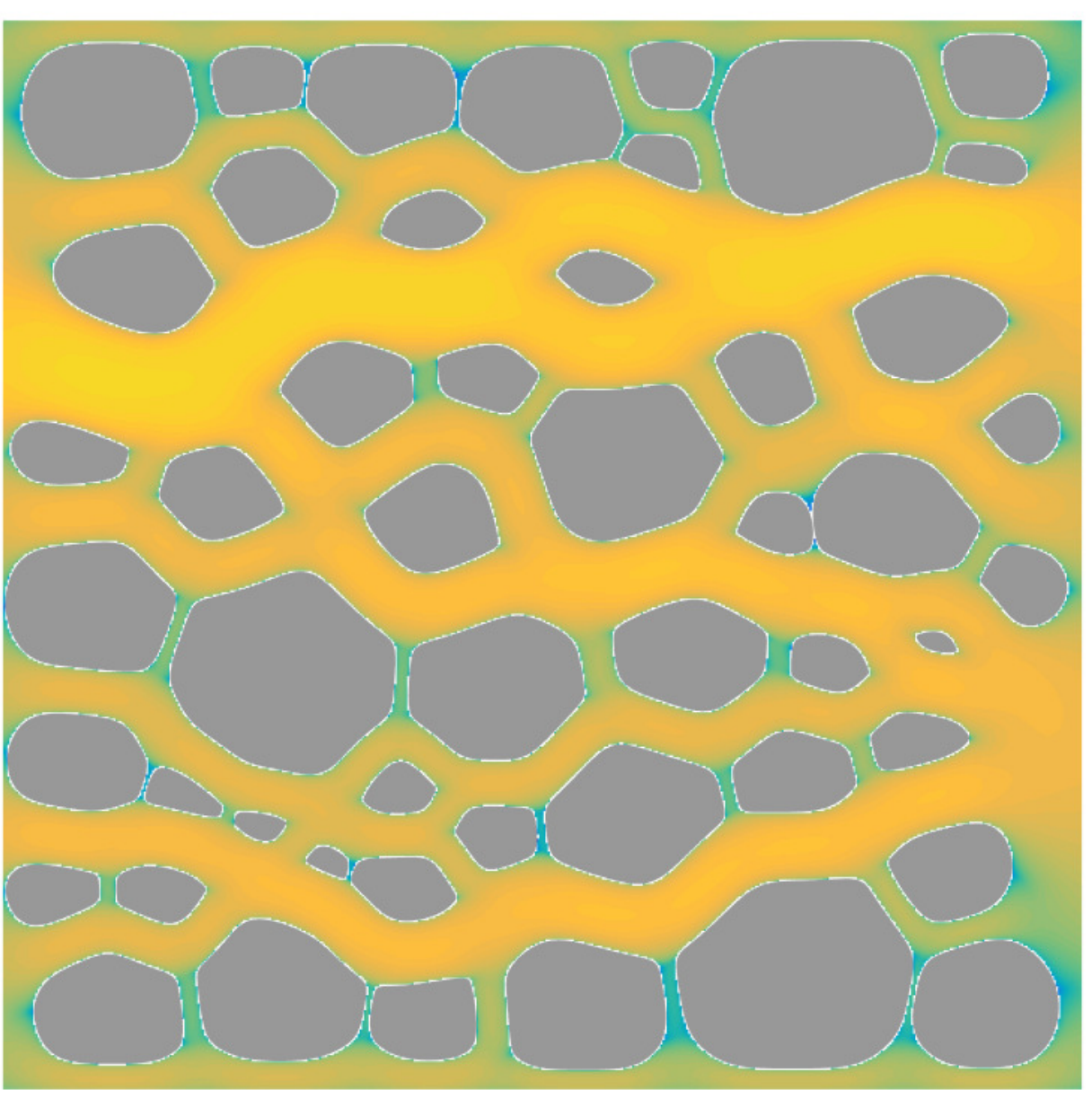}
\includegraphics[height = 0.3 \textwidth]{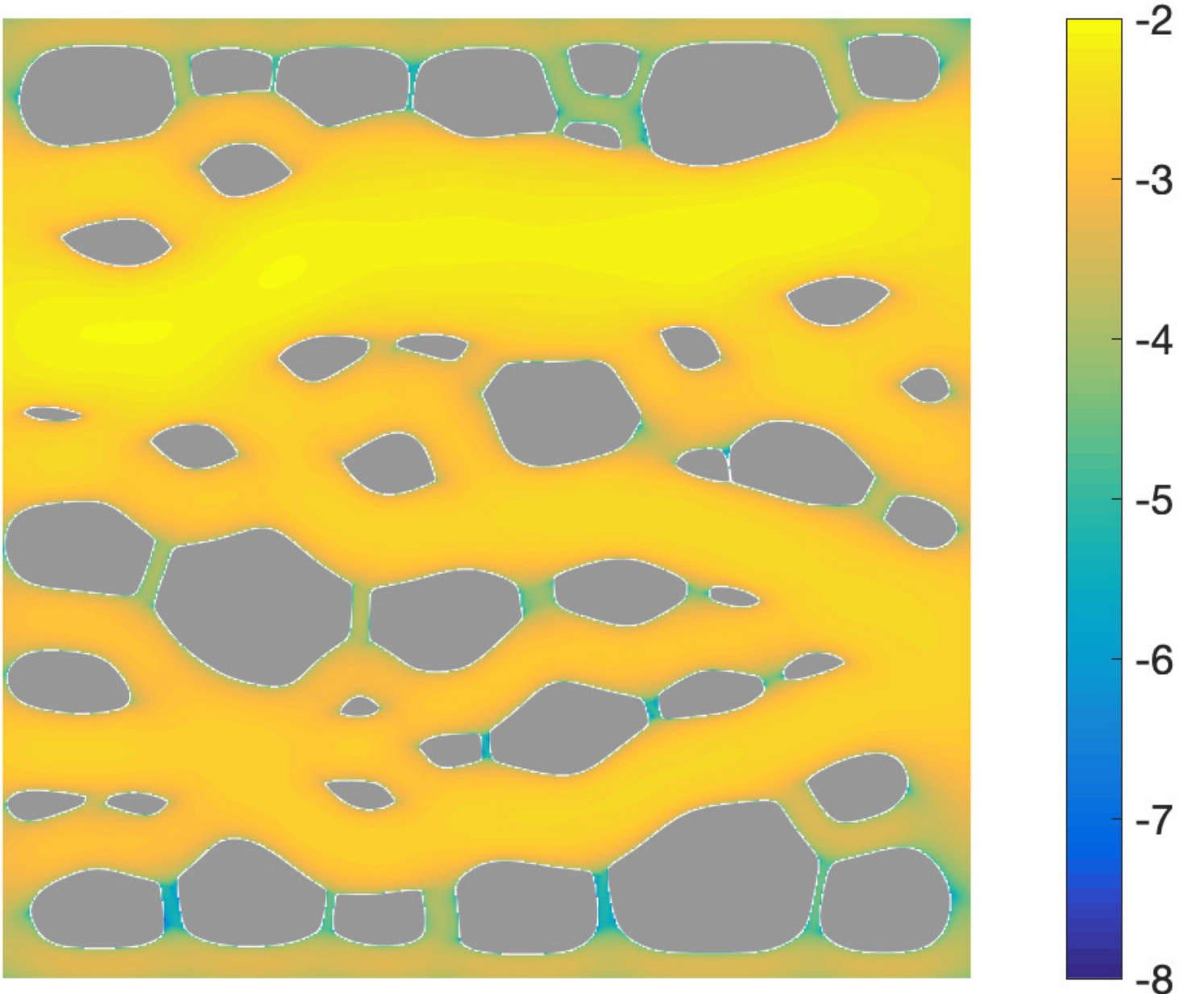}
\caption{\label{fig:Eroding50vel} 50 bodies eroding in a
Hagen-Poiseuille flow.  The six snapshots are equispaced in time, and
the color is the magnitude of the fluid velocity in a logarithmic scale.
The flow velocity varies over several orders of magnitude, the grains
form irregular shapes with large aspect ratios, and the geometry becomes
channelized and anisotropic.}
\end{center}
\end{figure}

\begin{figure}
\begin{center}
\includegraphics[height = 0.3 \textwidth]{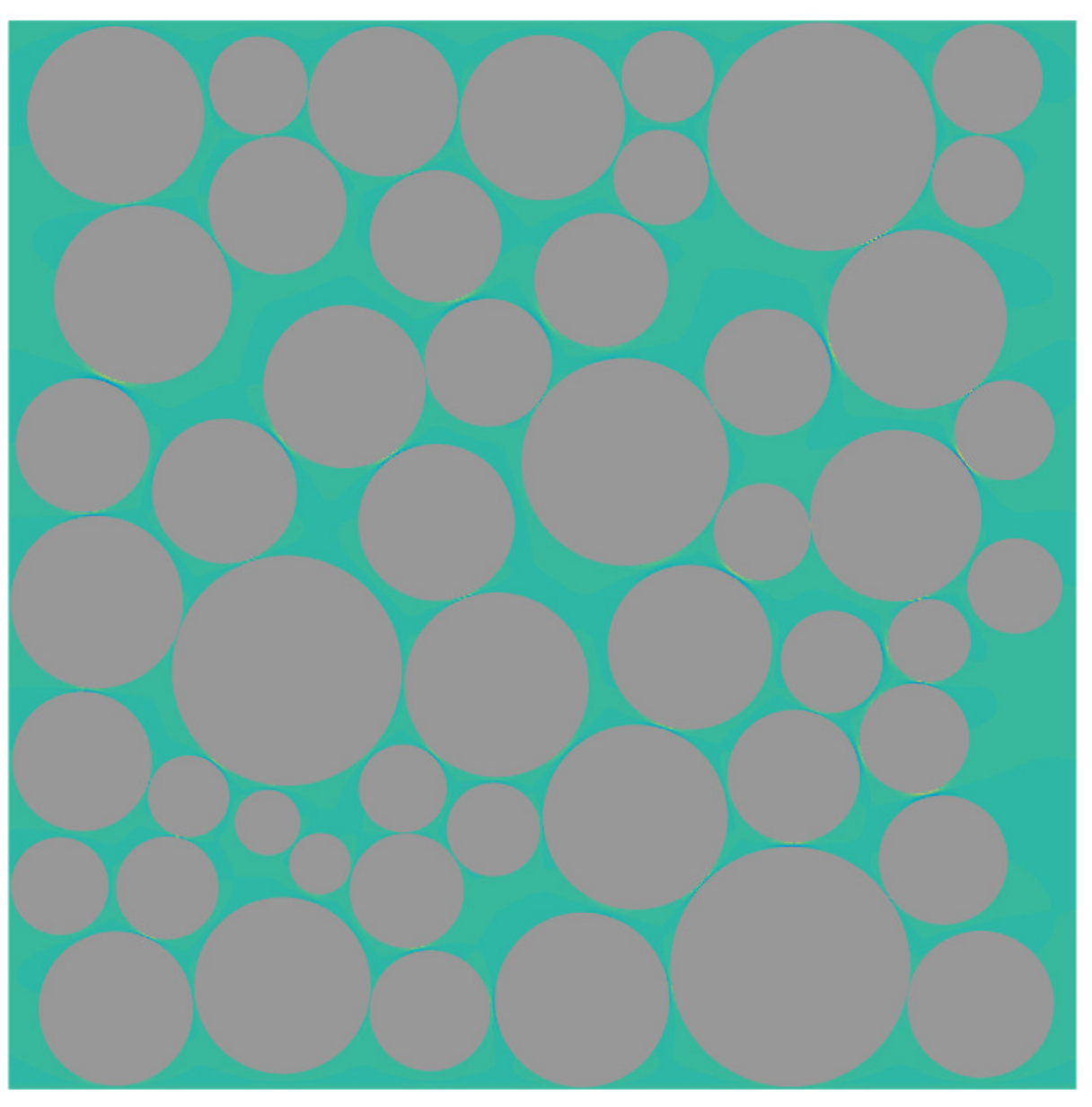}
\includegraphics[height = 0.3 \textwidth]{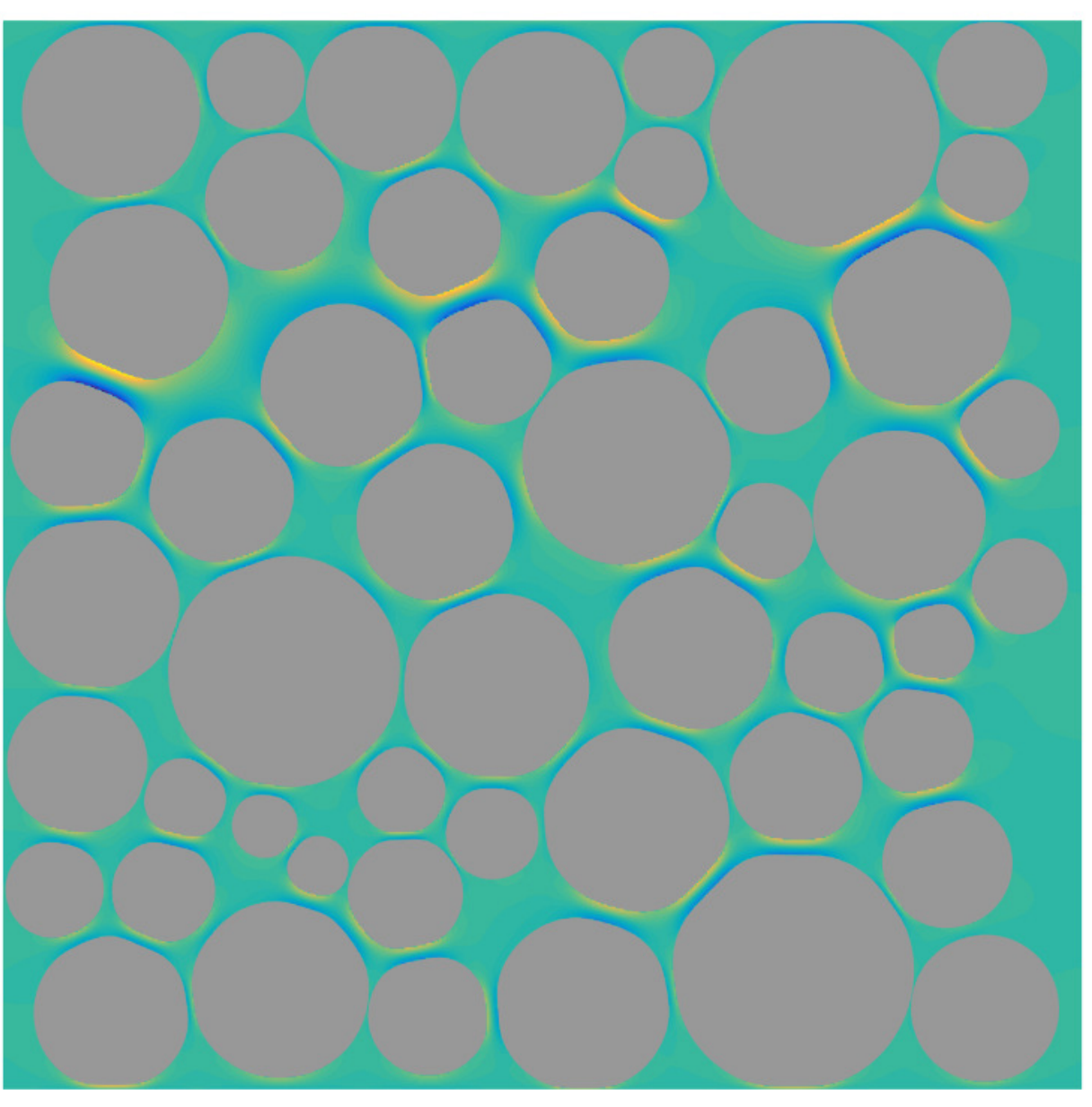}
\includegraphics[height = 0.3 \textwidth]{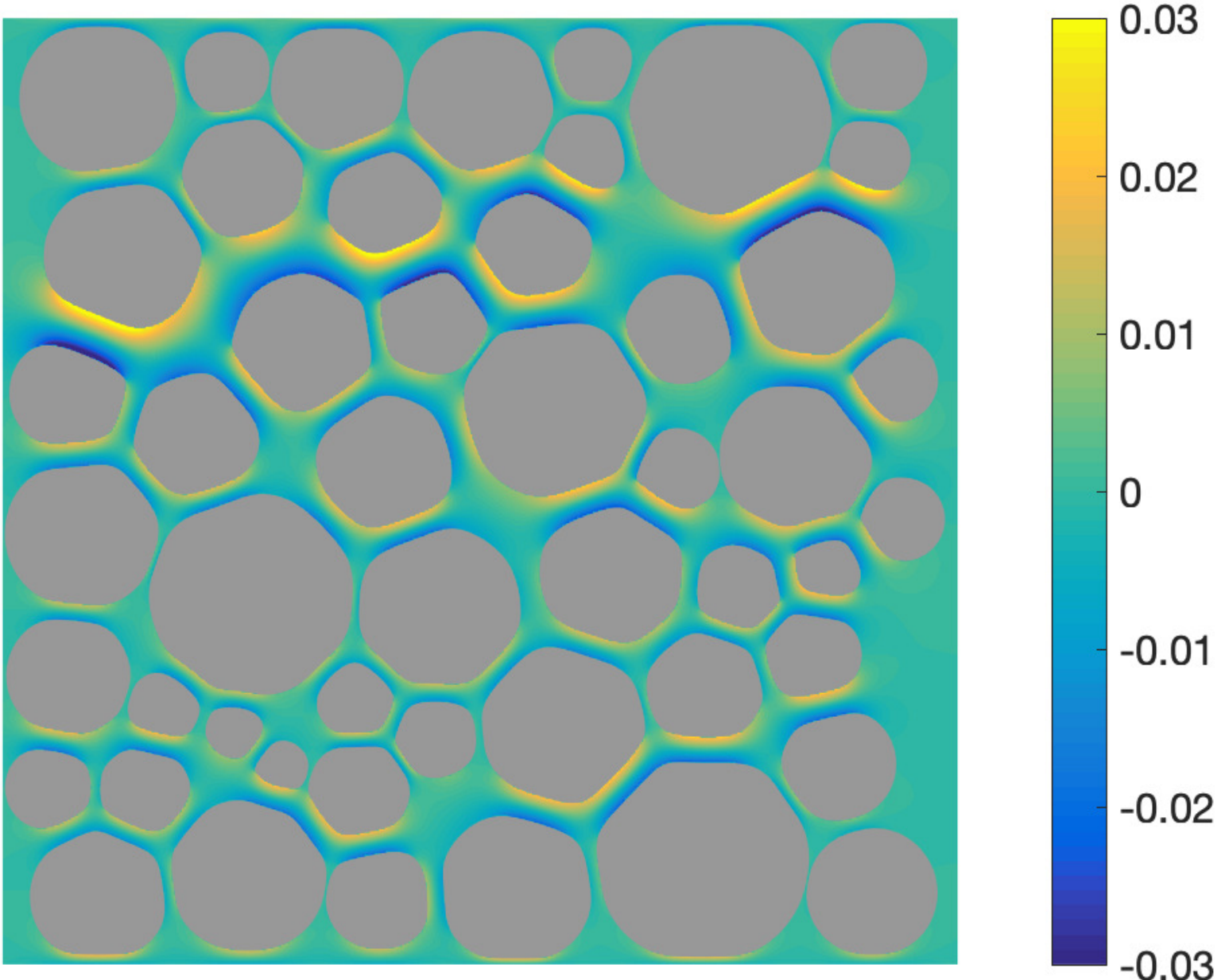}\\
\includegraphics[height = 0.3 \textwidth]{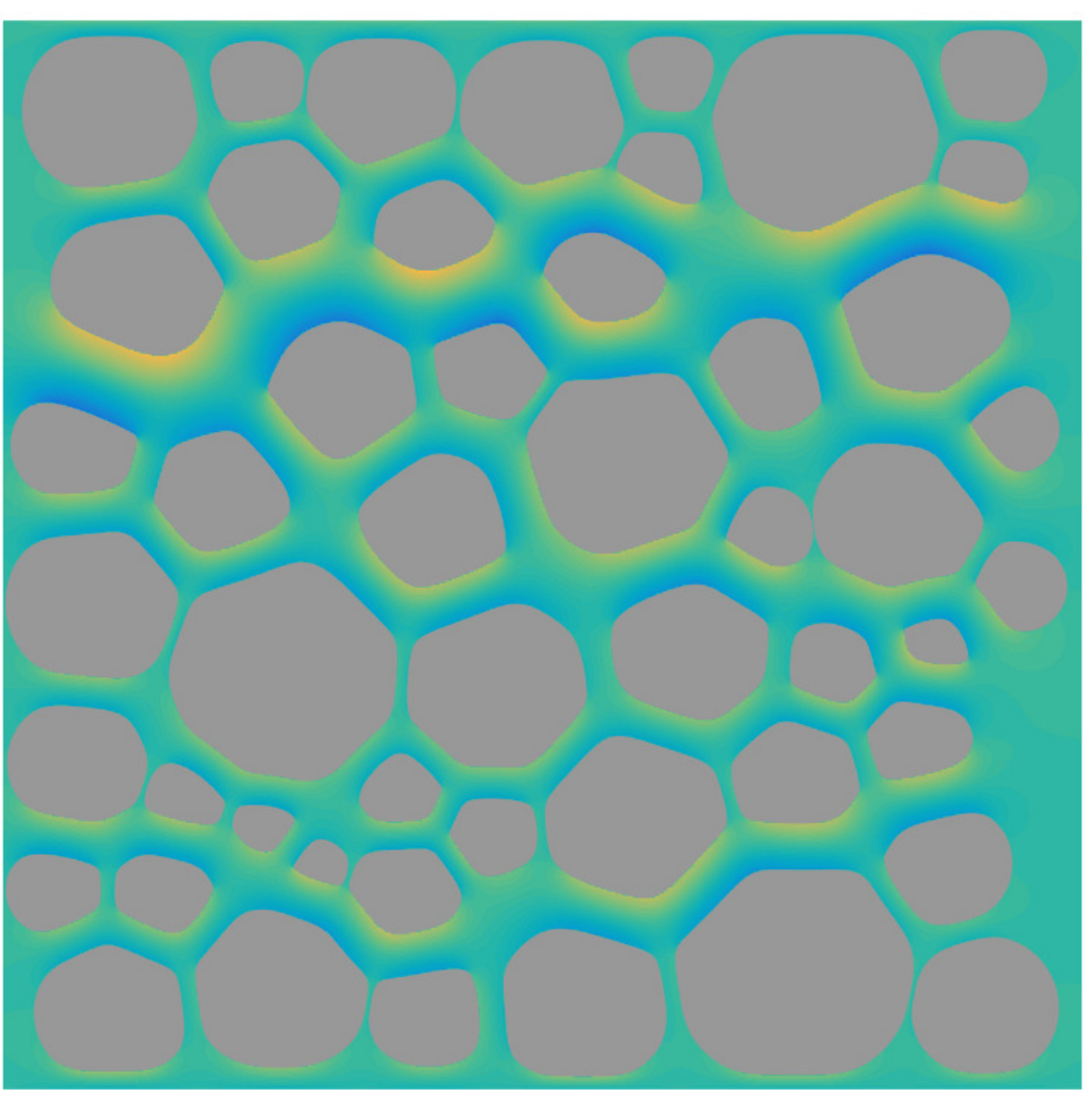}
\includegraphics[height = 0.3 \textwidth]{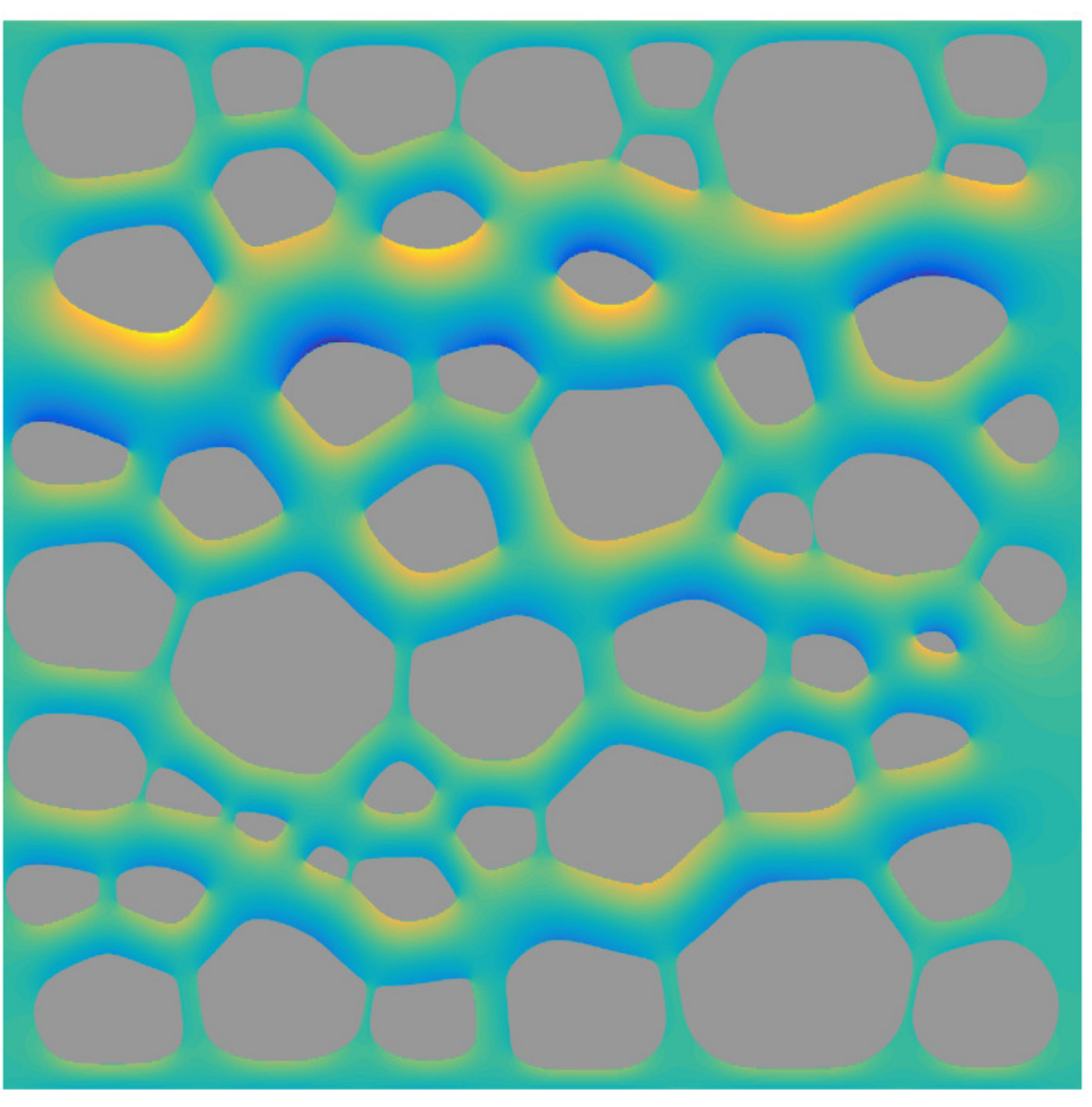}
\includegraphics[height = 0.3 \textwidth]{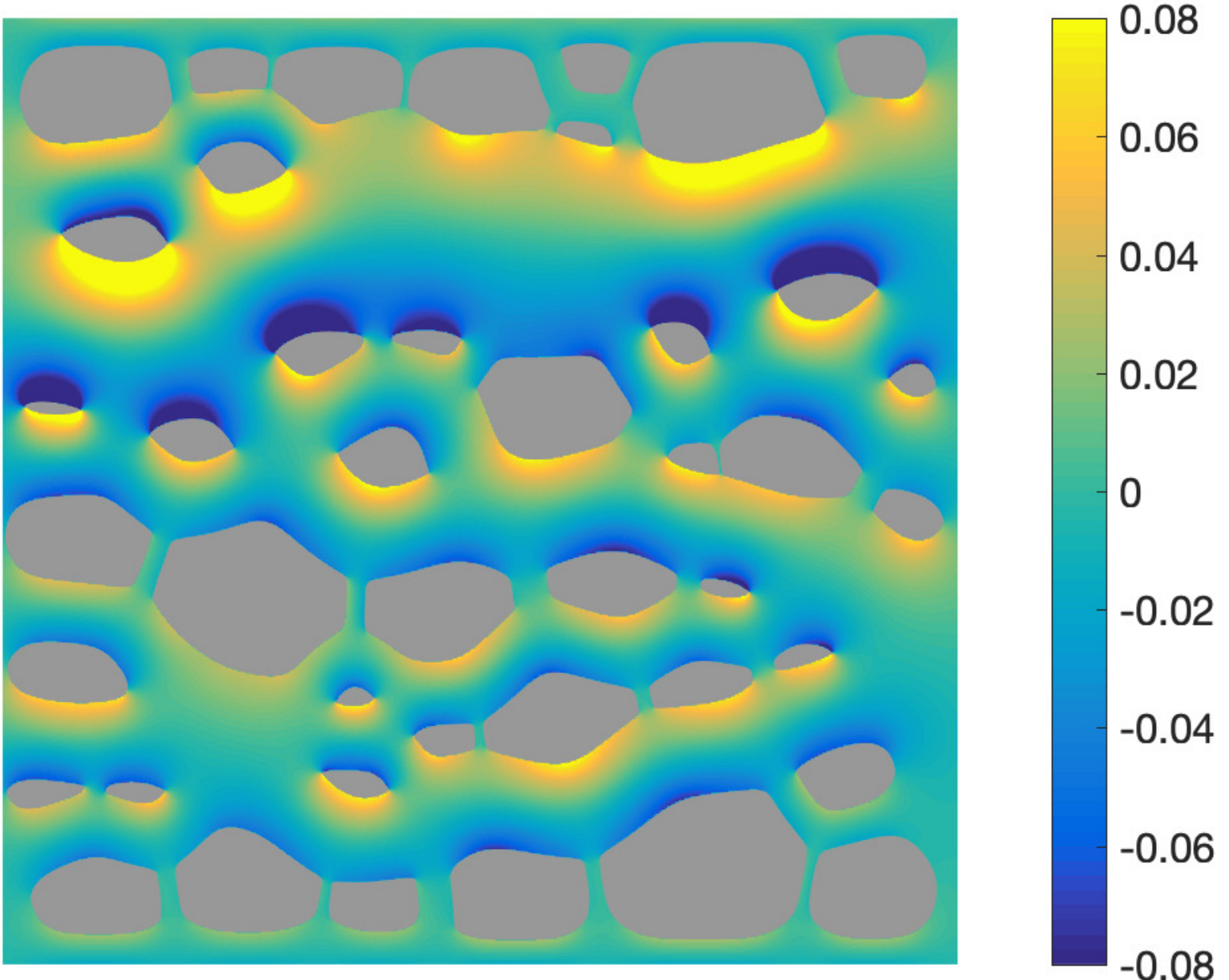}
\caption{\label{fig:Eroding50vort} The same six snapshots as
figure~\ref{fig:Eroding50vel}.  The color is the vorticity of the fluid.
Since the rate of erosion is equivalent to the magnitude of the
vorticity, erosion is fastest in the yellow and blue regions and slowest
in the green regions.}
\end{center}
\end{figure}

This study consists of two main undertakings: first, high-fidelity
simulations of eroding porous media, and, second, characterization of
tracer transport through the resulting eroded geometries. Our modeling
efforts build on previous work~\citep{ris-moo-chi-she-zha2012,
moore2013self, moo2017}, in particular recent numerical methods
developed to simulate erosion in the Stokes-flow
regime~\citep{qua-moo2018}. We, however, make key improvements to the
numerical methods to enable simulations of more realistic, dense
suspensions of eroding grains (figures~\ref{fig:Eroding50vel}
and~\ref{fig:Eroding50vort}).  Then, to characterize transport through
these configurations, we examine coarse-grained variables through
statistical analysis of tracer trajectories.

Owing to the scales present in groundwater flow~\citep{bea1972}, we
model the hydrodynamics with the two-dimensional incompressible Stokes
equations.  Meanwhile, individual grains erode at a rate proportional to
the hydrodynamic shear stress~\citep{wan-fel2004,
ris-moo-chi-she-zha2012, moore2013self, par-izu2000}.  Since the fluid
equations are linear and homogeneous, they are converted to a boundary
integral equation (BIE), and this allows us to naturally resolve the
non-negligible interactions between bodies.  We also compute the
vorticity in the fluid bulk since, on solid boundaries, vorticity
reduces to shear and thus provides a convenient way to simultaneously
visualize local erosion rates and changes in the surrounding flow
(figure~\ref{fig:Eroding50vort}).

To compute stable simulations of erosion, we use methods of high-order
in both space and time. The time integration is unchanged from previous
work~\citep{qua-moo2018}.  We apply a mild regularization and a
smoothing term to eliminate numerical instabilities that can be
triggered by changes in sign of the shear stress, and we use a stable
second-order Runge-Kutta method applied to the {\thL}
coordinates~\citep{hou-low-she1994} of the eroding grains.  In this
work, we introduce a new quadrature method to resolve dense suspensions.
The accuracy of the trapezoid rule, which was used in previous work, is
adequate for bodies that are sufficiently separated~\citep{tre-wei2014},
but not for grains in close contact.  One of the earliest quadrature
methods for nearly-singular integrands was developed
by~\citet{bak-she1986}, and in recent years, many other schemes have
followed~\citep{kli-tor2018, hel-oja2008a, bea-yin-wil2016, bea-lai2001,
klo-bar-gre-one2013}.  We use a Barycentric quadrature
method~\citep{bar2014, bar-wu-vee2015} since it is a non-intrusive
modification of the trapezoid rule, and the error is guaranteed to be
uniformly bounded.  We extend the original quadrature method to compute
the velocity gradient, which is needed to evaluate the shear stress and
the fluid vorticity.

To characterize transport through the resulting configurations, particle
trajectories must be computed. Depending on the application, microscale
transport can be modelled as pure advection~\citep{dea-qua-bir-jua2018,
cve-che-wen1996, puy-gou-den2019},
advection-diffusion~\citep{cus-hu-den1995, dag1987, den-ica-hid2018}, or
with a random walk~\citep{saf1959, bij-blu2006, ber-sch-sil2000}.  In
this work, we assume trajectories to be governed by pure advection
(ie.~no diffusion), so the particle trajectories are identical to the
streamlines. We compute trajectories $\ss(t)$ that are initialized at
$\ss_0$ by solving the advection equation
\begin{align}
  \frac{d\ss}{dt} = \uu(\ss,t), \quad \ss(0) = \ss_0,
  \label{eqn:tracers}
\end{align}
where $\uu$ is the fluid velocity.  Since there is no stiff diffusive
term, we solve~\eqref{eqn:tracers} with a fourth-order explicit
Runge-Kutta time stepping method. The Barycentric quadrature rule is
used to accurately compute trajectories that are close to an eroding
grain. 

Once the streamlines are computed, we characterize transport by
analyzing three different metrics: the tortuosity, the anomalous
dispersion, and the pore size distribution.  The local tortuosity of a
streamline that connects the inlet to the outlet is defined as the
streamline's length normalized by the linear inlet-to-outlet distance.
In porous media, the local tortuosity can be greater than
1.5~\citep{kop-kat-tim1996, mat-kha-koz2008} or even
2~\citep{dud-koz-mat2011}, depending on several factors such as the
porosity. The tortuosity of a geometry is defined by averaging the local
tortuosity over all streamlines initialized at the inlet, and the
geometry's tortuosity characterizes average particle
motions~\citep{hak-com-den2019}.  To characterize spreading, the fluid
dispersion is defined as the variance of the streamline lengths. In
porous media, this spreading is often
super-dispersive~\citep{kan-dea-nun-bij-blu-jua2014, cus-hu-den1995,
dea-leb-den-tar-bol-dav2013}. Since anomalous dispersion results from
streamlines spending time in both the high and low velocity
regimes~\citep{ber-sch2001}, it is crucial to accurately resolve
streamlines near grain boundaries, as achieved in this work.  Finally,
we construct the pore-size distribution throughout the erosion process.
These distributions are required to quantify velocity
distributions~\citep{ali-par-wei-bre2017, dea-qua-bir-jua2018},
channelization~\citep{sie-ili-pri-riv-gua2019},
connectivity~\citep{knu-car2005, wes-blo-gra2001}, and to develop
network models~\citep{bry-kin-mel1993, bry-mel-cad1993, bij-blu2006}. 

This paper is organized as follows. In section~\ref{sec:formulation}, we
summarize the erosion model that is described in more detail in previous
work~\citep{qua-moo2018}.  In section~\ref{sec:DLP}, we recast all the
governing equations as layer potentials defined in both $\RR^2$ and in
$\CC$.  Section~\ref{sec:transport} describes measures for
characterizing the geometry and transport. Section~\ref{sec:method}
describes the numerical methods, with special attention paid to the new
quadrature method for computing the shear stress and the vorticity.
Section~\ref{sec:results} presents numerical examples for a variety of
dense packings of bodies.  Finally, concluding remarks are made in
section~\ref{sec:conclusions}.

\section{Governing Equations}
\label{sec:formulation}
We start by defining the main variables used to model erosion.  We only
briefly summarize the model, and a more detailed description can be
found in previous work~\citep{qua-moo2018}.  We consider flows inside a
confined geometry $\Omega$ that contains $M$ eroding bodies with
boundaries $\gamma_\ell$, $\ell = 1,\ldots,M$.  The boundary of the
fluid domain is $\bd \Omega = \Gamma \cup \gamma_1 \cup \cdots \cup
\gamma_M$, where $\Gamma$ is the outer boundary, taken to be a slightly
smoothed version of the boundary of $[-3,3] \times [-1,1]$.  All eroding
bodies are placed in $[-1,1] \times [-1,1]$ to create a buffer region
that allows the flow profile imposed at the inlet to transition to the
more complex flow intervening between the bodies. Neglecting inertial
forces, the governing equations are
\begin{equation}
\label{eqn:erosionModel}
  \begin{split}
    \mu \Delta \uu = \grad p, &\hspace{20pt} \xx \in \Omega, \gap 
      &&\mbox{\em conservation of momentum}, \\
    \grad \cdot \uu = 0, &\hspace{20pt} \xx \in \Omega, \gap 
      &&\mbox{\em conservation of mass}, \\
    \uu = \mathbf{0}, &\hspace{20pt} \xx \in \gamma, \gap 
      &&\mbox{\em no slip on the eroding bodies}, \\
    \uu = \UU, &\hspace{20pt} \xx \in \Gamma, \gap 
      &&\mbox{\em outer wall velocity}, \\
    \Vn = \CE \, \abs{\tau}, &\hspace{20pt} \xx \in \gamma,
      &&\mbox{\em erosion model}.
  \end{split}
\end{equation}
Here $\uu$ is the fluid velocity, $p$ is the pressure, $\UU$ is a
prescribed Hagen-Poiseuille velocity field, and $\Vn$ is the normal
velocity of $\gamma$. The shear stress on $\gamma$ is
\begin{align}
  \tau = -(\nabla \uu + \nabla \uu^T) \nn \cdot \ss
  \label{eqn:shearStress}
\end{align}
where, $\nn$ is the normal vector pointing into the body, and $\ss$ is
the unit tangent vector pointing in the counterclockwise direction. We
simulate erosion by alternating between solving the fluid equations and
advancing the eroding grains.  The strength of $\UU$ is adjusted at each
time step to achieve a constant pressure drop across the channel,
motivated by the geological situation of a porous medium connecting two
regions of  fixed hydraulic heads.

\section{Boundary Integral Equation Formulation}
\label{sec:DLP}
To accurately solve the governing equations~\eqref{eqn:erosionModel} in
complex two-dimensional geometries, we reformulate the equations as a
BIE.  This has the advantage that only the one-dimensional boundary of
the domain must be discretized, and, with appropriate quadrature
formulas and fast summation methods, the result is a high-fidelity
numerical simulation with near-optimal computational complexity.

\subsection{Double-Layer Potential Formulation in $\RR^2$}
Applying the same approach as our previous work~\citep{qua-moo2018}, we
start with the double-layer potential 
\begin{align}
  \DDD[\eeta](\xx) = \int_{\bd\Omega} D(\xx,\yy) \eeta(\yy)\, ds_\yy = 
  \frac{1}{\upi}\int_{\bd\Omega} 
    \frac{\rr \cdot \nn}{\rho^2} \frac{\rr \otimes \rr}{\rho^2}
    \eeta(\yy) \, ds_\yy, \quad \xx \in \Omega,
  \label{eqn:velocityDLP}
\end{align}
where $D$ is the kernel of the integral operator, $\rr = \xx - \yy$,
$\rho = \|\rr\|$, $\nn$ is the unit outward normal at $\yy$, and $\eeta$
is an unknown density function.  We complete the integral equation
formulation by adding the $M$ Stokeslets,
$S[\llambda_\ell,\cc_\ell](\xx)$, and $M$ rotlets,
$R[\xi_\ell,\cc_\ell](\xx)$, where $\cc_\ell$ is a point inside the
$\ell^{th}$ body~\citep{pow-mir1987}.  Here $\llambda_\ell$ and
$\xi_\ell$ are the Stokeslet and rotlet strengths, respectively,
corresponding to the $\ell^{th}$ body.  Then, for any sufficiently
smooth geometry $\Omega$, the solution of the incompressible Stokes
equation with a Dirichlet boundary condition $\ff$ is
\begin{align}
  \uu(\xx) = \DDD[\eeta](\xx) + 
    \sum_{\ell=1}^M S[\llambda_\ell,\cc_\ell](\xx) + 
    \sum_{\ell=1}^M R[\xi_\ell,\cc_\ell](\xx), \quad \xx \in \Omega,
\end{align}
where the density function, Stokeslets, and rotlets satisfy
\begin{subequations}
\label{eqn:BIE}
\begin{alignat}{3}
  \ff(\xx) &= -\frac{1}{2}\eeta(\xx) + \DDD[\eeta](\xx) + 
    \NN_0[\eeta](\xx) \nonumber \\
    &\quad + \sum_{\ell=1}^M S[\llambda_\ell,\cc_\ell](\xx) + 
    \sum_{\ell=1}^M R[\xi_\ell,\cc_\ell](\xx), 
    \quad &&\qquad\xx \in \bd\Omega, \\
  \llambda_\ell &= \frac{1}{2\upi} \int_{\gamma_\ell} 
    \eeta(\yy)\, ds_\yy, &&\qquad \ell = 1,\ldots,M, \\
  \xi_\ell &= \frac{1}{2\upi} \int_{\gamma_\ell}
    (\yy - \cc_\ell)^\perp \cdot \eeta(\yy)\, ds_\yy, 
    &&\qquad \ell = 1,\ldots,M.
\end{alignat}
\end{subequations}
Here, the null space associated with the flux-free condition of $\ff$ is
addressed with  $\NN_0$ which is the integral operator with kernel
$N_0(\xx,\yy) = \nn(\xx) \otimes \nn(\yy)$, $\xx,\yy \in \Gamma$.  In
this work, $\ff$ is the prescribed velocity, which is equal to $\UU$ on
the outer wall, $\Gamma$, and equal to zero on the eroding bodies,
$\gamma_\ell$, $\ell=1,\ldots,M$.

Once~\eqref{eqn:BIE} is solved for the density function $\eeta$, the
corresponding deformation tensor, pressure, and vorticity at $\xx \in
\Omega$ are written in terms of layer potentials~\citep{qua-moo2018}.
To compute the deformation tensor for $\xx \in \gamma$, we include the
jump term
\begin{align}
  \frac{1}{2} \left(\pderiv{\eeta}{\ss} \cdot \ss \right) \left[
    \begin{array}{cc}
      s_x^2 - s_y^2 & 2s_x s_y \\ 2s_x s_y & s_y^2 - s_x^2
    \end{array}
  \right].
  \label{eqn:deformationJump}
\end{align}
Finally, the deformation tensor, pressure, and vorticity due to the
Stokeslets and rotlets are readily available~\citep{poz1992}. Having
computed the deformation tensor on $\gamma$, the shear stress is
computed using equation~\eqref{eqn:shearStress}. 

\subsection{Cauchy Integral Representation of the Double-Layer
Potential}
\label{sec:DLPcomplex}
The velocity double-layer potential~\eqref{eqn:velocityDLP}, and its
corresponding deformation tensor, pressure, and vorticity are all
written as layer potentials in $\RR^2$.  However, the quadrature method
we introduce in section~\ref{sec:method} requires complex-valued
representations. The first step to form a complex representation is to
write the Laplace double-layer potential as the complex integral
\begin{align}
  \DD[\eeta](\xx) = \frac{1}{2\upi} \int_{\bd\Omega} 
    \frac{\rr \cdot \nn}{\rho^2}\eeta(\yy)\, ds_\yy = \Real (v(x)),
\end{align}
where
\begin{align}
  v(x) = \frac{1}{2\upi i} \int_{\bd\Omega}
    \frac{\eta(y)}{x - y} \, dy, \quad x \in \Omega.
  \label{eqn:laplaceComplex}
\end{align}
Here $x = x_1 + i x_2,y = y_1 + i y_2 \in \CC$ are the complex
counterparts of $\xx = (x_1,x_2),\yy = (y_1,y_2) \in \RR^2$, and $\eta =
\eta_1 + i \eta_2$ is the complex counterpart of $\eeta =
(\eta_1,\eta_2)$. Therefore, depending on the formulation of the layer
potential, $\Omega$ is interpreted as a subset of $\RR^2$ or $\CC$.
Equation~\eqref{eqn:laplaceComplex} is converted to a Cauchy integral by
first finding the boundary data of $v$. If $\Omega$ is a
simply-connected interior domain, then the boundary data of $v$
satisfies the Sokhotski-Plemelj jump relation
\begin{align}
  \label{eqn:SPrelation}
  v(x) = - \frac{1}{2} \eta(x) + \frac{1}{2\upi i} \int_{\bd\Omega}
    \frac{\eta(y)}{x-y}\, dy, \quad x \in \bd\Omega.
\end{align}
For exterior domains, the jump term changes from $-1/2$ to $1/2$, and
for multiply-connected domains, such as a porous media, $\bd\Omega$ is
decomposed into its different connected components and the appropriate
jump relation is applied.  Having computed the boundary data of the
holomorphic function $v$, by the Cauchy integral theorem we have
\begin{subequations}
  \label{eqn:cauchy}
  \begin{alignat}{3}
  v(x) &= \frac{1}{2\upi i}\int_{\bd\Omega} 
    \frac{v(y)}{y-x} \,dy, \\
  v'(x) &= \frac{1}{2\upi i} \int_{\bd\Omega}
    \frac{v(y)}{(y-x)^2} \, dy, \\
  v''(x) &= \frac{1}{\upi i} \int_{\bd\Omega}
    \frac{v(y)}{(y-x)^3} \, dy,
  \end{alignat}
\end{subequations}
for $x \in \Omega$.  Since $v(x)$ depends on the complex-valued density
function $\eta$, we use the notation $v[\eta](x)$ for the holomorphic
function defined in equation~\eqref{eqn:laplaceComplex}, and its first
two derivatives are written as $v'[\eta](x)$ and $v''[\eta](x)$.  
  
Finally, the Stokes double-layer potential~\eqref{eqn:velocityDLP} can
be written using a Laplace double-layer
potential~\eqref{eqn:laplaceComplex} and its gradients 
\begin{equation}
  \label{eqn:Stokes2Laplace}
  \begin{aligned}
    \DDD[\eeta](\xx) &= 
      \frac{1}{2\upi} \int_{\bd\Omega} 
        \frac{\nn}{\rho^2} (\rr \cdot \eeta) \, ds_\yy + 
      \frac{1}{2\upi} \nabla \int_{\bd\Omega}
        \frac{\rr \cdot \nn}{\rho^2} (\yy \cdot \eeta) \, ds_\yy \\
      &- \frac{1}{2\upi} x_1 \nabla \int_{\bd\Omega}
        \frac{\rr \cdot \nn}{\rho^2}\eta_1(\yy) \, ds_\yy -
      \frac{1}{2\upi} x_2 \nabla \int_{\bd\Omega}
        \frac{\rr \cdot \nn}{\rho^2}\eta_2(\yy) \, ds_\yy.
  \end{aligned}
\end{equation}
Therefore, the Stokes double-layer potential can be written as a sum of
Cauchy integrals and its first derivative~\citep{bar-wu-vee2015}
\begin{equation}
  \begin{aligned}
    u_1(x) &= \Real (v[\psi_1](x)) + \Real (v'[y\cdot\eta](x)) 
             -x_1\Real (v'[\eta_1](x)) - x_2\Real (v'[\eta_2](x)), \\
    u_2(x) &= \Real (v[\psi_2](x)) - \Imag (v'[y\cdot\eta](x)) 
         +x_1\Imag (v'[\eta_1](x)) + x_2\Imag (v'[\eta_2](x)),
  \end{aligned}
  \label{eqn:cauchyVelocity}
\end{equation}
where $y \cdot \eta = y_1 \eta_1 + y_2 \eta_2$, 
\begin{align} 
  \psi_1=(\eta_1+i\eta_2)\frac{\Real(n)}{n}, \quad
  \psi_2=(\eta_1+i\eta_2)\frac{\Imag(n)}{n},
\end{align}
and $n \in \CC$ is the complex counterpart of the outward unit normal
$\nn \in \RR^2$.

\subsection{Cauchy Integral Representation for the Gradient of the
Double-Layer Potential}
\label{sec:gradDLPcomplex}
Computing the shear stress and vorticity requires a complex-valued layer
potential representation of the velocity gradient.  The deformation
tensor at $x \in \Omega$ is found by computing the derivatives of the
expressions for $u_1$ and $u_2$ in equation~\eqref{eqn:cauchyVelocity}  
\begin{equation}
\label{eqn:cauchyGradient}
  \begin{aligned}
    \pderiv{u_1}{x_1} &= +\Real (v'[\psi_1](x)) + 
    \Real (v''[y\cdot\eta](x)) - \Real (v'[\eta_1](x)) \\
    &- x_1\Real (v''[\eta_1](x)) - x_2\Real (v''[\eta_2](x)), \\
    \pderiv{u_1}{x_2} &= - \Imag (v'[\psi_1](x)) - 
    \Imag (v''[y\cdot\eta](x)) + x_1\Imag (v''[\eta_1](x)) \\
    &- \Real (v'[\eta_2](x)) + x_2\Imag (v''[\eta_2](x)), \\
    \pderiv{u_2}{x_1} &= +\Real (v'[\psi_2](x)) - 
    \Imag (v''[y\cdot\eta](x)) + \Imag (v'[\eta_1](x))  \\
    &+ x_1\Imag (v''[\eta_1](x)) + x_2\Imag (v''[\eta_2](x)), \\
    \pderiv{u_2}{x_2} &= -\Imag (v'[\psi_2](x)) - 
    \Real (v''[y\cdot\eta](x)) + x_1\Real (v''[\eta_1](x)) \\
    &+ \Imag (v'[\eta_2](x)) + x_2\Real (v''[\eta_2](x)).
  \end{aligned}
\end{equation}
The same expressions are used to compute the deformation tensor for $x
\in \bd\Omega$, except that the jump
condition~\eqref{eqn:deformationJump} is included.  Finally, to compute
the shear stress, the deformation tensor on $\bd\Omega$ is applied to
the normal and tangent vectors as in equation~\eqref{eqn:shearStress}.
The velocity gradient is also used to compute the vorticity in the fluid
bulk.  For $x \in \Omega$, the Cauchy integral representation of the
vorticity at $x \in \Omega$ is
\begin{align}
  \omega(x) = 
    \Real (v'[\psi_2](x)) + \Imag (v'[\psi_1](x))+ 
    \Real (v'[\eta_2](x))+ \Imag (v'[\eta_1](x)).
\end{align}

\section{Transport, Tracers, and Tortuosity}
\label{sec:transport}
Erosion in porous media leads to phenomena such as
channelization~\citep{berhanu2012shape}, and we are interested in
characterizing transport in such geometries.  In our previous
work~\citep{qua-moo2018}, we examined the effect of erosion on the area
fraction, flow rate, and the total drag.  However, to characterize
macroscopic signatures of the transport, other quantities must be
examined.  Here, we compute the anomalous dispersion rate, the
tortuosity, and the distribution of the pore sizes.  The first two
metrics are defined in terms of streamlines governed by the autonomous
advection equation~\eqref{eqn:tracers}.

\subsection{Anomalous Dispersion}
\label{sec:dispersion}
The spreading of fluid in a porous media is often characterized in terms
of anomalous dispersion~\citep{kla-rad-sok2008, den-cor-sch-ber2004}.
The anomalous dispersion rate depends on the porosity and
permeability~\citep{koc-bra1988}, but is also affected by the
distribution and shape of the grains. We calculate the anomalous
dispersion rates by analyzing the streamlines governed by
equation~\eqref{eqn:tracers} in eroded geometries. Given a set of $N_p$
trajectories, we define $\lambda_j(t)$ to be the arclength of the
trajectory
\begin{align}
  \lambda_j(t) = \int_{0}^t \|\ss'_j(\tilde{t})\|\, d\tilde{t}, 
    \quad j=1,\ldots,N_p.
\end{align}
Then, the first and second ensemble moments are
\begin{align}
  \label{eqn:moments}
  \langle \lambda \rangle (t) = 
    \frac{1}{N_p} \sum_{j=1}^{N_p} \lambda_j(t), \quad 
    \sigma_\lambda^2(t) = \frac{1}{N_p} \sum_{j=1}^{N_p}
    \left[\lambda_j(t) - \langle \lambda \rangle(t) \right]^2,
\end{align}
and $\sigma_\lambda$ characterizes the dispersion.  At early times, the
particles have not explored much of the geometry, and we expect a
ballistic motion $\sigma_\lambda \sim t$.  However, as the particles
pass the grains, their trajectories are altered, and we expect that
$\sigma_\lambda \sim t^\alpha$, with $\alpha \in (0.5,1)$, indicating
that the flow is super-dispersive.

To establish an asymptotic anomalous dispersion rate, the trajectories
must pass several grains. The geometries that we consider are too short
to observe asymptotic dispersion, so we use a reinsertion method to form
longer trajectories. Similar to the work of
others~\citep{dea-qua-bir-jua2018, puy-gou-den2019}, once a particle
reaches the outlet of the porous region, it is reinserted at the inlet.
To minimize errors caused by reinsertion, the particle is initialized at
the discretization point that has the closest velocity to the particle's
velocity at the outlet.  After a single trajectory is formed, it has
undergone a collection of reinsertions.  Then, as a post-processing
step, the trajectory is made continuous by attaching the tail of each
segment to the origin of the next segment.

\subsection{Tortuosity}
The tortuosity is a dimensionless number that quantifies the amount of
twisting of streamlines. Unlike the dispersion calculations, we do not
use reinsertion to form long trajectories.  The local tortuosity is
\begin{align}
  \tau(y_0) = \frac{\lambda(y_0)}{d}.
  \label{eqn:localTort}
\end{align}
Here the streamline originates on the inlet cross-section $x=x_0$ at
$(x_0,y_0)$, and its arclength, $\lambda(y_0)$, is calculated until the
streamline passes the parallel outlet cross-section $x = x_0 + d$.  In
this work, we consider streamlines originating at $x=-1$ and terminating
at $x=1$, so $d=2$.  However, other choices for the terminal point when
computing the tortuosity are sometime used~\citep{dud-koz-mat2011}.  The
hydraulic tortuosity is defined by taking the average over all points on
the inlet cross-section
\begin{align}
  T = \frac{1}{d}\left(\int_{S}u_1(x_0,y_0)\lambda(y_0)\,dy_0 \right)
  \Bigg/
  \left(\int_{S}u_1(x_0,y_0)\,dy_0 \right),
  \label{eqn:tortuosity1}
\end{align} 
where $S$ is the inlet cross-section $x = -1$, $u_1(y_0)$ is the
$x$-component of the velocity at the initial point of the streamline,
and $d=2$ is the distance between the inlet and outlet.  Note that $T
\geq 1$, and $T=1$ only if no grains are present.

The tortuosity can also be computed with an area integral. Assuming that
the flow is incompressible and not re-entrant, meaning that all
streamlines connect the two cross-sections, the tortuosity in
equation~\eqref{eqn:tortuosity1} is equivalent
to~\citep{dud-koz-mat2011}
\begin{align}
  T = \left(\int_\Omega \|\uu(\yy)\|\, d\yy \right)
    \Bigg/
      \left(\int_\Omega u_1(\yy)\, d\yy \right),
  \label{eqn:tortuosity2}
\end{align}
where $\Omega$ is the fluid region between the inlet and outlet
cross-sections.  Recirculation zones are possible in viscous
fluids~\citep{hig1985}, but they are very small in the examples we
consider and have a negligible effect on the tortuosity.  Since
equation~\eqref{eqn:tortuosity2} does not require the additional work of
computing particle trajectories at every time step, we use this
definition for the majority of the tortuosity calculations.  However, we
do compare the two definitions for the tortuosity at several porosities
in section~\ref{sec:results}.

There have been efforts to relate the tortuosity, $T$, to the porosity,
$\phi$.  For example,~\citet{mat-kha-koz2008} propose the
models
\begin{subequations}
  \label{eqn:tortuosityModels}
  \begin{align}
    \widehat{T}(\phi) &= \phi^{-p}, \\
    \widehat{T}(\phi) &= 1-p \log \phi, \\
    \widehat{T}(\phi) &= 1+p (1-\phi), \\
    \widehat{T}(\phi) &= (1+p (1-\phi))^2, 
  \end{align}
\end{subequations}
where $p>0$ is a fitting parameter.  In section~\ref{sec:results}, we
compare these four models with the tortuosity of eroding porous
geometries.

\subsection{Pore Throat Size}
\label{sec:throats}
\begin{figure}
\begin{subfigure}[b]{0.5\textwidth}
\includegraphics*[height =0.8\linewidth]{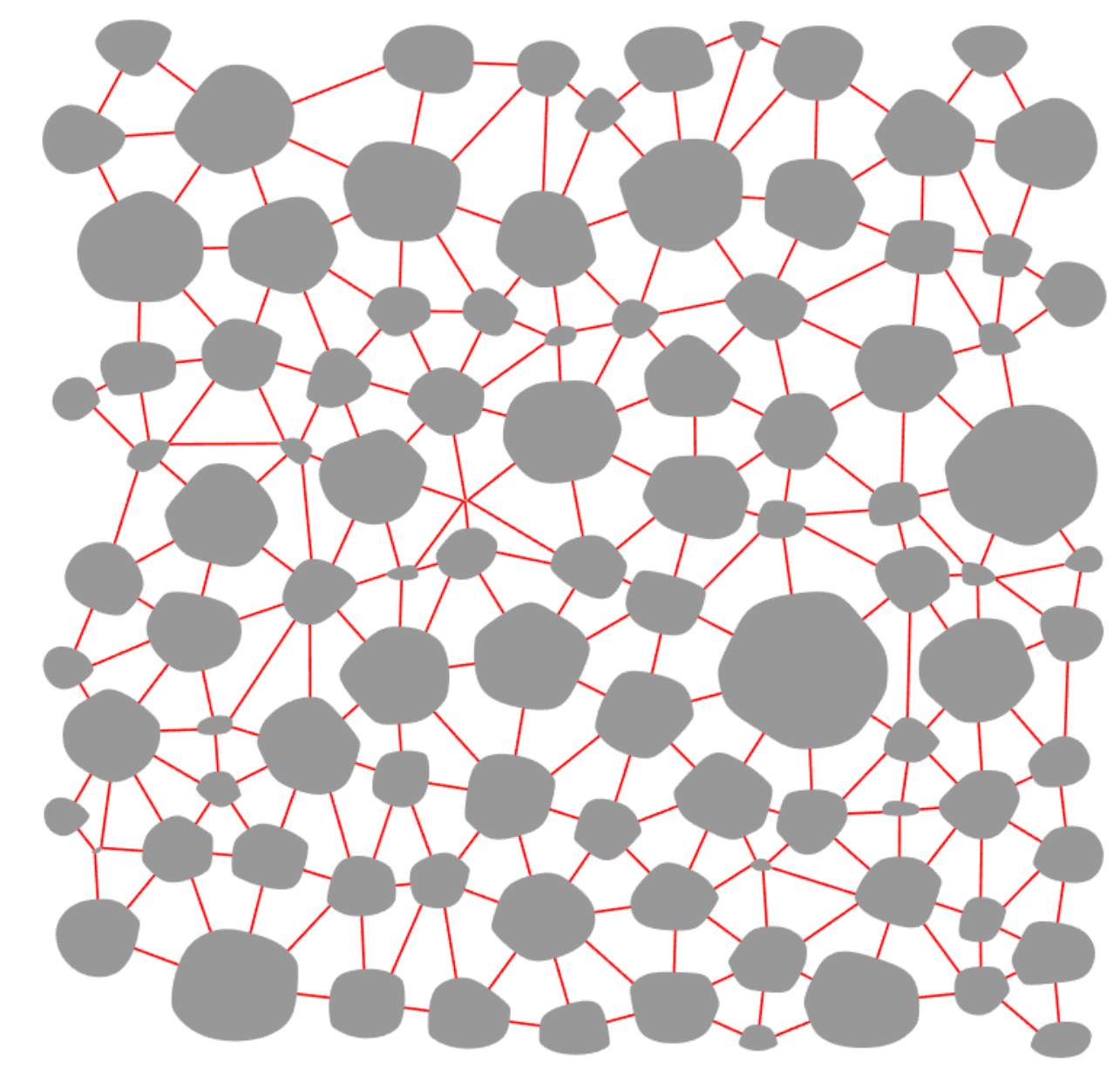}
\caption{}
\end{subfigure}
\begin{subfigure}[b]{0.5\textwidth}
\includegraphics*[height = 0.8\linewidth]{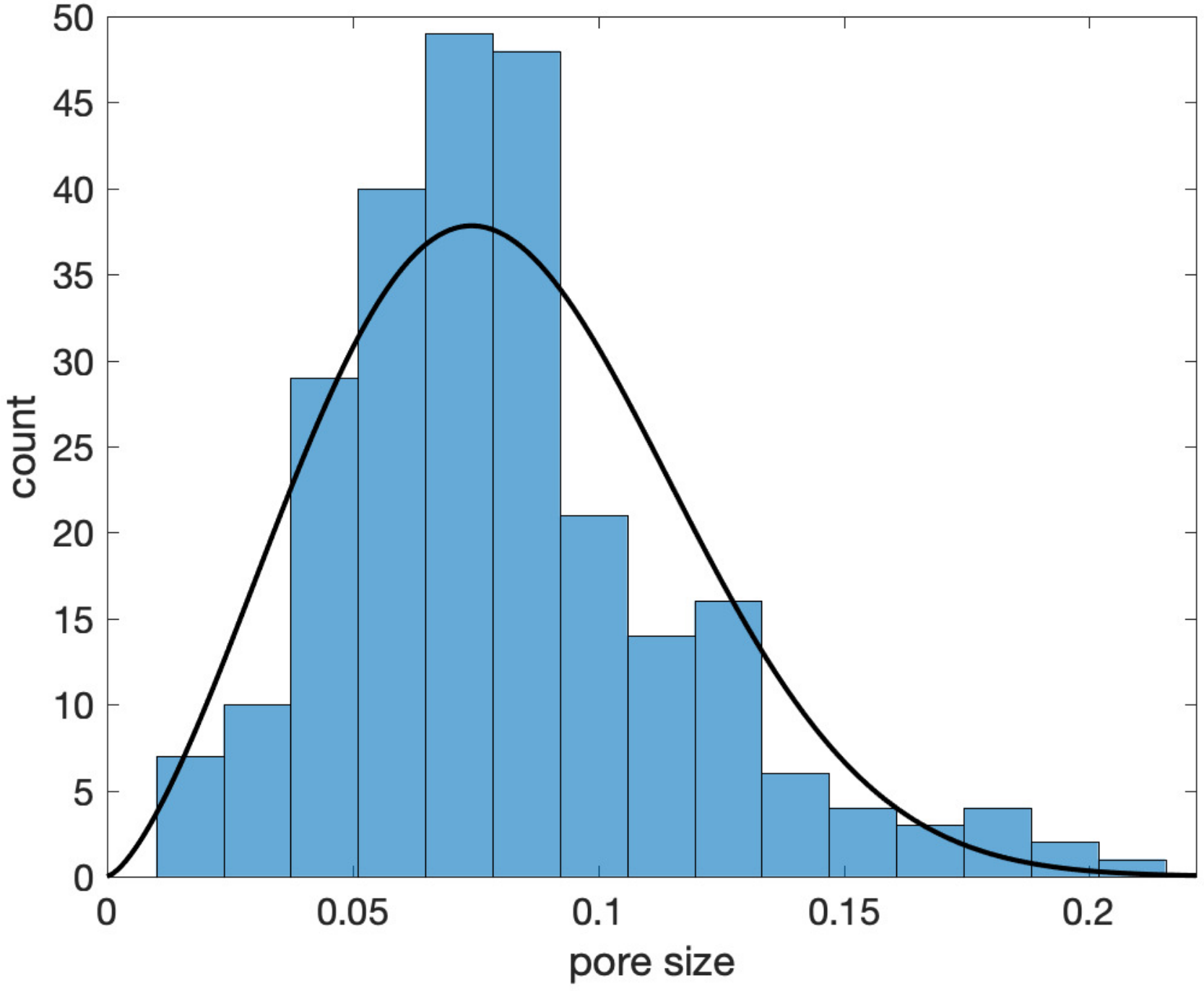}
\caption{}
\end{subfigure}
\caption{\label{fig:Eroding100gap} (a) The pore sizes between
neighboring grains. Two grains are neighbors if they share an edge of
the Delanuay triangulation with nodes at the center of each eroding
grain. (b) The pore sizes of an eroded geometry. The black curve is the
Weibull distribution with the same first two moments as the data.}
\end{figure}
While transport in porous media depends on the porosity, it also depends
on the placement of the grains.  In particular, grain placement affects
velocity scales~\citep{ali-par-wei-bre2017}, correlation
structures~\citep{leb-ded-dav-bou2007}, contaminant
transport~\citep{knu-car2005},
channelization~\citep{sie-ili-pri-riv-gua2019,berhanu2012shape}, and
pore network models~\citep{bry-kin-mel1993, bry-mel-cad1993,
bij-blu2006}. To characterize the grain placement, we compute
distributions of pore sizes between neighboring grains.  To define
neighboring grains, we form the Delaunay triangulation using nodes
placed at the center of each eroding grain and at a collection of points
around the boundary of the porous media.  Then, we say that two grains
are neighbors if their centers share an edge of the
triangulation~\citep{dea-qua-bir-jua2018}. The pores of an eroded
geometry are illustrated in figure~\ref{fig:Eroding100gap}(a). We do not
consider pores between eroding bodies and the solid wall $\Gamma$, so
some of the grains near the porous region boundary only have two
neighbors. Having defined the pore sizes, we plot its distribution in
figure~\ref{fig:Eroding100gap}(b) and compare it with the Weibull
distribution, a distribution used by others to characterize pore
sizes~\citep{ioa-cha1993}.  In section~\ref{sec:Eroding100}, we
investigate the effect of erosion on the pore size distribution.

\section{Numerical Methods}
\label{sec:method}
In line with our previous work~\citep{qua-moo2018}, we use two meshes to
simulate erosion. The integral equation is solved by discretizing the
boundary of the geometry at a set of collocation points distributed
equally in arclength (section~\ref{sec:spatialDiscretization}).
Depending on the proximity of the target point to the source points, the
quadrature rule is either the trapezoid rule or the Barycentric
quadrature rule (section~\ref{sec:bary}).  The criteria that determines
which quadrature method is applied is described in
section~\ref{sec:fmm}.  Once the shear stress is computed, the bodies
are eroded a single time step by using a {\thL}
discretization~\citep{hou-low-she1994, moore2013self}.  The time
stepping methods for erosion and passive particles are described in
section~\ref{sec:time}.

\subsection{Spatial Discretization}
\label{sec:spatialDiscretization}
Since we use a BIE formulation, we only need to discretize the
one-dimensional boundary of the domain.  We discretize each eroding
grain $\gamma_k$ with $N_\iin$ points and discretize the outer wall
$\Gamma$ with $N_\out$ points.  The $j^{th}$ discretization point on
$\Gamma$ and $\gamma_k$ are denoted by $\yy_j^0$ and $\yy^k_j$,
respectively.  The discretization points are initially distributed
evenly in arclength, and this equispacing is maintained throughout the
entire simulation by using the {\thL} formulation.  In addition, we
apply regularization~\citep{qua-moo2018} to slightly smooth the corners
that inevitably develop during erosion.

Given the discretization points of $\bd\Omega$, the trapezoid rule
results in the collocation method for~\eqref{eqn:BIE}
\begin{subequations}
\label{eqn:trapLinearSystem}
  \begin{alignat}{3}
  \UU_\ell &= \sum_{j=1}^{N_\out} 
    w^0_j D(\yy^0_\ell,\yy^0_j) \eeta_j^0 +
  \sum_{k=1}^M \sum_{j=1}^{N_\iin}
    w^k_j D(\yy^0_\ell,\yy^k_j) \eeta^k_j +
  \sum_{j=1}^{N_\out} w^0_j N_0(\yy^0_\ell,\yy^0_j)\eeta_j^0 
    \nonumber \\
  &+\sum_{k=1}^M S[\llambda_k,\cc_k](\yy^0_\ell) + 
  \sum_{k=1}^M R[\xi_k,\cc_k](\yy^0_\ell), 
  &&\hspace{-120pt}\ell = 1,\ldots,N_\out, \\
  \mathbf{0} &= \sum_{j=1}^{N_\out} 
    w^0_j D(\yy^m_\ell,\yy^0_j) \eeta_j^0 +
  \sum_{k=1}^M \sum_{j=1}^{N_\iin}
    w^k_j D(\yy^m_\ell,\yy^k_j) \eeta^k_j \nonumber \\
  &+\sum_{k=1}^M S[\llambda_k,\cc_k](\yy^m_\ell) + 
  \sum_{k=1}^M R[\xi_k,\cc_k](\yy^m_\ell),
    &&\hspace{-120pt}m=1,\ldots,M, \: \ell = 1,\ldots,N_\iin,  \\
  \llambda_m &= \frac{1}{2\upi} \sum_{j=1}^{N_\iin} 
    w_j^m \eeta^m_j, 
  &&\hspace{-120pt}m = 1,\ldots,M \\ 
  \xi_m &= \frac{1}{2\upi} \sum_{j=1}^{N_\iin} 
    w_j^m (\yy^m_j - \cc_m)^\perp \cdot \eeta^m_j,
  &&\hspace{-120pt}m=1,\ldots,M,
\end{alignat}
\end{subequations}
where $w^k_j$ are quadrature weights that depend on $N_\iin$, $N_\out$,
and the lengths of $\gamma^k$ and $\Gamma$, and $D$ is the kernel of the
Stokes double-layer potential defined in
equation~\eqref{eqn:velocityDLP}.  Since the kernel is smooth, the
diagonal terms $D(\yy_j^m,\yy_j^m)$ are replaced with the appropriate
curvature-dependent limiting term.  The linear
system~\eqref{eqn:trapLinearSystem} is a well-conditioned second-kind
integral equation and is solved iteratively with GMRES.  If the number
of discretization points is sufficiently large, then the solution
of~\eqref{eqn:trapLinearSystem} is an accurate approximation of the
density function, Stokeslets, and rotlets. Then, for $\xx \in \Omega$,
the double-layer potential is approximated as
\begin{align}
  \label{eqn:trap}
  \uu(\xx) &= \sum_{j=1}^{N_\out} w^0_j D(\xx,\yy^0_j) \eeta_j +
  \sum_{k=1}^M \sum_{j=1}^{N_\iin} w^k_j D(\xx,\yy^k_j) \eeta^k_j. 
\end{align}
Similarly, the corresponding layer potentials for the deformation tensor
and vorticity are approximated with the trapezoid rule. The
contributions due to the Stokeslets and rotlets require no quadrature
and are easily included in the velocity, deformation tensor, and
vorticity.  Finally, Fourier differentiation is used to compute the jump
term~\eqref{eqn:deformationJump} of the shear stress, and then the
tensor is applied to the normal and tangent vectors as defined in
equation~\eqref{eqn:shearStress}.

Once the velocity is computed in $\Omega$, the tortuosity can be
computed with the Eulerian velocity field~\eqref{eqn:tortuosity2}.  We
compute the velocity at $\xx_{ij} = (-1 + i\Delta x, -1 + j\Delta y)$,
$i,j=1,\ldots,N$, where $\Delta x = \Delta y = 2/N$, and the velocity at
points inside an eroding body are assigned a value of 0.  Then, the
tortuosity is approximated as
\begin{align}
  T = \left(\sum_{i,j=1}^N \|\uu(\xx_{ij})\|\Delta x \Delta y \right) 
      \Bigg/ 
      \left(\sum_{i,j=1}^N u_1(\xx_{ij}) \Delta x \Delta y \right)
\end{align}

\subsection{Barycentric Quadrature Formulas}
\label{sec:bary}
While the trapezoid rule is spectrally accurate for smooth, periodic
functions~\citep{tre-wei2014}, the derivative of the integrand grows as
the target point $\xx$ approaches $\bd\Omega$.  Therefore, if the
trapezoid rule is applied when bodies are in near-contact, or if a layer
potential is evaluated at a point close to $\bd\Omega$, then the result
become unreliable and the simulation ultimately becomes unstable.  We
thus desire a quadrature method whose error bound does not depend on the
target location.

We showed in sections~\ref{sec:DLPcomplex} and~\ref{sec:gradDLPcomplex}
that the velocity, shear stress, and vorticity of the double-layer
representation can all be written as the sum of Cauchy integrals and its
first two derivatives.  Therefore, we require quadrature rules with a
uniform error bound for Cauchy integrals and its
derivatives~\eqref{eqn:cauchy}. \citet{ioa-pap-per1991} developed
quadrature rules, that we call {\em Barycentric quadrature rules}, to
compute Cauchy integrals and their derivatives with an error bound that
is independent of $x \in \CC$.  Then,~\citet{bar-wu-vee2015} used these
quadrature rules to compute the Stokes double-layer potential
representation of the velocity~\eqref{eqn:velocityDLP}. After briefly
summarizing this method, we extend the work to compute the second
derivative so that the shear stress and vorticity can be computed with a
uniform error bound. 

We present the quadrature rules for a simply-connected interior domain
$\Omega \subset \CC$, with any point $a \in \Omega$, and we consider
target points $x \in \Omega$ and $x \in \Omega^c$. Then, the quadrature
rules can be applied to individual components of a multiply-connected
domain to compute the velocity, vorticity, and deformation tensor in an
eroding porous media.  The method starts with an underlying quadrature
rule, and we use the spectrally accurate $N$-point trapezoid rule.
Since the quadrature points are uniformly distributed, the quadrature
weights are $w_j = L/N$, $j=1,\ldots,N$, where $L$ is the length of
$\bd\Omega$.

Again, given a complex-valued density function $\eta$, the boundary data
of $v[\eta](x)$, as defined in equation~\eqref{eqn:laplaceComplex},
satisfies the Sokhotski-Plemelj jump relation~\eqref{eqn:SPrelation}.
Since the limiting boundary data of $v$ differs when considering $x \in
\Omega$ and $x \in \Omega^c$, we denote the boundary data as $v^-$ for
$x \in \Omega$, and as $v^+$ for $x \in \Omega^c$.  Rather than directly
applying the trapezoid rule to approximate $v(x)$ in
equation~\eqref{eqn:cauchy}, we start with the identity
\begin{align}
  \int_{\bd\Omega} \frac{v^{-}(y) - v(x)}{y - x} \,dy = 0,
    \quad x \in \Omega.
\end{align}
Since the integrand is bounded for all $x \in \Omega$, we can apply the
trapezoid rule
\begin{align}
  \sum_{j=1}^{N} \frac{v^{-}(y_j) - v(x)}{y_j - x} w_j \approx 0,
\end{align}
and the error is independent of $x$.  Rearranging for $v(x)$, we have
the interior Barycentric quadrature rule
\begin{align}
  v(x) = \left(\sum_{j=1}^N \frac{v^{-}(y_j)}{y_j - x} w_j \right)
  \Bigg/
  \left(\sum_{j=1}^N \frac{1}{y_j - x} w_j \right), 
  \quad x \in \Omega.
  \label{eqn:BaryvInterior}
\end{align}
Using a similar construction and letting $a$ be any point inside
$\Omega$, the exterior Barycentric quadrature rule is
\begin{align}
  v(x) = \frac{1}{x-a} 
    \left(\sum_{j=1}^N \frac{v^+(y_j)}{y_j - x}w_j \right)
    \Bigg/
    \left(\sum_{j=1}^N \frac{(y_j - a)^{-1}}{y_j - x}w_j \right),
    \quad x \in \Omega^c.
  \label{eqn:BaryvExterior}
\end{align}

Similar constructions can be used to derive quadrature rules
for $v'(x)$.  For $x \in \Omega$, we have the identity
\begin{align}
  \int_{\bd\Omega} \frac{v^-(y) - v(x) + (y-x)v'(x)}{(y - x)^2} = 0,
\end{align}
and the integrand is bounded for all $x \in \Omega$.  Therefore, after
applying the trapezoid rule and rearranging for $v'(x)$, we have the
interior Barycentric quadrature rule
\begin{align}
  v'(x) = \left(\sum_{j=1}^{N}
    \frac{v^{-}(y_j) - v(x)}{(y_j-x)^2} w_j \right)
  \Bigg/
  \left(\sum_{j=1}^{N} \frac{1}{y_j-x} w_j\right), 
  \quad x \in \Omega.
  \label{eqn:BaryvprimeInterior}
\end{align}
Using a similar construction, the exterior Barycentric quadrature rule
for the first derivative is
\begin{align}
  v'(x) = \frac{1}{x-a} \left(\sum_{j=1}^N
    \frac{v^+_j - v(x)}{(y_j - x)^2} w_j \right)
    \Bigg/
    \left(\sum_{j=1}^N \frac{(y_j-a)^{-1}}{y_j - x} w_j\right),
    \quad x \in \Omega^c.
  \label{eqn:BaryvprimeExterior}
\end{align}
Note that $v(x)$ is required to compute $v'(x)$ for both the interior
and exterior case, and this is available using the Barycentric
quadrature rules~\eqref{eqn:BaryvInterior}
and~\eqref{eqn:BaryvExterior}.

To compute the shear stress and vorticity, we require a Barycentric
quadrature rule for $v''(x)$.  The derivation is largely based on the
work of~\citet[see equation (2.12)]{ioa-pap-per1991}.  We start with
the second derivative of the Cauchy integral theorem
\begin{align}
  0 = \frac{1}{2\upi i} \int_{\bd\Omega} 
      \frac{2v^{-}(y)}{(y-x)^3}\,dy - v''(x).
\end{align}
For the interior case, $x \in \Omega$, we use the identity
\begin{align}
  \frac{1}{2\upi i}\int_{\bd\Omega} \frac{1}{(y-x)^n}\, dy = 
  \left\{
    \begin{array}{ll}
      1, & n = 1, \\
      0, & n = 2,3,\ldots. \\
    \end{array}
  \right.
\end{align}
Combining this identity with the Cauchy integral representation of
$v''(x)$, we have
\begin{align}
  0 = \frac{1}{2\upi i} \int_{\bd\Omega} 
      \frac{2v^{-}(y) - v''(x)(y-x)^2 - 2v(x) - 2(y-x)v'(x)}
      {(y-x)^3}\, dy.
\end{align}
This integrand is constructed so that it is bounded for all $x \in
\Omega$, and applying the trapezoid rule, we have
\begin{align}
  0 \approx  \sum_{j=1}^{N} 
      \frac{2v^{-}(y_j) - v''(x)(y_j-x)^2 - 2v(x) - 2(y_j-x)v'(x)}
      {(y_j-x)^3} w_j,
\end{align}
where the accuracy is independent of $x$.  Solving for $v''(x)$, the
Barycentric quadrature rule for the interior second derivative at $x \in
\Omega$ is
\begin{align}
  v''(x) \approx \left(2\sum_{j=1}^N 
    \frac{v^{-}_{j} - v(x) - (y_j-x)v'(x)}{(y_j-x)^3}w_j \right)
    \Bigg/
    \left(\sum_{j=1}^N \frac{1}{y_j-x}w_j\right).
\end{align}
For the exterior case, $x \in \Omega^c$, we start with the identity
\begin{align}
\frac{1}{x-a} &= -\frac{1}{2\upi i}\int_{\bd\Omega} 
    \frac{(y-a)^{-1}}{y-x}\, dy. 
\end{align}
Combining this identity with the Cauchy integral representation of
$v''(x)$, we have
\begin{align}
  0 = \frac{1}{2\upi i} \int_{\bd\Omega} 
    \frac{2v^+(y) - 2v(x) - 2(y-x)v'(x) + (y-a)^{-1} (x-a) (y-x)^2 v''(x)}
    {(y-x)^3}.
\end{align}
As in the interior case, the integrand is chosen so that it is bounded
for all $x \in \Omega^c$.  Therefore, after applying the trapezoid rule
and solving for $v''(x)$, we have the Barycentric quadrature rule for
the exterior second derivative at $x \in \Omega^c$
\begin{align}
  v''(x) \approx \frac{1}{x-a}\left(2\sum\limits_{j=1}^N
    \frac{v^{+}_{j} - v(x) - (y_j-x)v'(x)}{(y_j-x)^3}w_j\right)
    \Bigg/
    \left(\sum_{j=1}^N \frac{(y_j-a)^{-1}}{y_j-x}w_j\right).
\end{align}
The quadrature rule for $v''(x)$ requires $v(x)$ and $v'(x)$, and theses
are computed using the quadrature rules in
equations~\eqref{eqn:BaryvInterior},~\eqref{eqn:BaryvExterior},~\eqref{eqn:BaryvprimeInterior},
and~\eqref{eqn:BaryvprimeExterior}. In section~\ref{sec:results}, these
quadrature rules are used to form simulations of nearly-touching eroding
grains, and to study dynamics of the flow in regions arbitrarily close
to eroding grains.

\subsection{Efficiently Applying the Quadrature}
\label{sec:fmm}
By using the Barycentric quadrature rule, the velocity, shear stress,
and vorticity are computed with an error that is bounded independent of
the target location. However, applied directly, it requires
$O(N^2)$ operations, where $N$ is the total number of source
and target points.  By using a fast summation method, such as the fast
multipole method (FMM)~\citep{gre-rok1987}, the cost can be reduced to
$O(N)$ operations.  However, each application of the Barycentric
quadrature rules involve several $N$-body calculations, rendering the
computational cost prohibitive, so we introduce a hybrid method that
combines the Barycentric quadrature rule and an accelerated trapezoid
rule.  Note that the source points of the layer potential is always one
of the eroding bodies or the outer wall, but the target point can either
be on another component of $\bd\Omega$ or it can be in the fluid bulk
$\Omega$.

To compute the velocity double-layer potential~\eqref{eqn:velocityDLP},
we start by applying the trapezoid rule~\eqref{eqn:trap} accelerated
with the FMM.  This calculation requires $O(N)$ operations, and we call
the resulting velocity $\vv_\trap(\xx)$.  Since the trapezoid rule is
spectrally accurate, the error of $\vv_\trap(\xx)$ is small for points
sufficiently far from $\bd\Omega$, and this region depends on the number
of discretization points $N_\iin$ and $N_\out$.  However, the trapezoid
rule needs to be replaced with a more accurate quadrature rule for
points that are too close to $\bd\Omega$.  Note that since a point is
typically only close to one or two components of $\bd\Omega$, only the
contribution of these nearby bodies needs to be replaced.  Assuming that
$\xx$ is too close to $\gamma_k$, we first subtract the inaccurate
trapezoid rule approximation of the double-layer potential due to
$\gamma_k$.  Then, the Barycentric quadrature rule is used to compute
the velocity due to $\gamma_k$ with more accuracy.  Finally, the
velocity at $\xx$ is
\begin{align}
  \vv(\xx) = \vv_\trap(\xx) - \sum_{j=1}^{N_\iin} w_j^k
    D(\xx,\yy^k_j) \eeta(\yy_j) + \vv^k_\bary(\xx),
  \label{eqn:velocityDecomp}
\end{align}
where $\vv^k_\bary(\xx)$ is the velocity at $\xx$ resulting from
applying the Barycentric quadrature rule to the double-layer potential
due to $\gamma_k$.  This strategy naturally extends to points that are
close to $\Gamma$, and to points that are simultaneously close to
multiple components of $\bd\Omega$.  While the term $\vv_\trap(\xx)$ in
equation~\eqref{eqn:velocityDecomp} is computed for all target points
using the FMM, the other two terms are computed with a direct summation.
However, these terms are only required for target points near an eroding
body or the outer wall, and these points make up only a small fraction
of the total number of points.  

An identical strategy is used to compute the vorticity and the
deformation tensor.  That is, the trapezoid rule is used as a first pass
to form the vorticity and deformation tensor, and then local corrections
are made to amend the inaccuracies of the trapezoid rule.  However,
since the shear stress and vorticity are only computed once per time
step, the trapezoid rule is applied with a direct summation rather than
the FMM.  Relative to the cost of computing the velocity at each GMRES
iteration with the FMM, the additional once-per-time-step costs to
compute the vorticity and deformation tensor are minimal.

Per grid point, applying the Barycentric quadrature rules dominate the
computational cost, so it is imperative that it is only applied when
necessary. As a rule of thumb, the trapezoid rule due to $\gamma_k$
achieves machine epsilon accuracy if~\citep{bar2014}
\begin{align}
  \label{eqn:TrapCutoff}
  d(\xx,\gamma_k) = \inf_{\yy \in \gamma_k} \|\xx - \yy\| > 
    5 \frac{L_k}{N_\iin}.
\end{align}
Instead of checking if all target points $\xx$
satisfy~\eqref{eqn:TrapCutoff}, we first check, for all pairs of eroding
grains, if
\begin{align}
  \label{eqn:BodyCutoff}
  \|\cc_i - \cc_j\| < \frac{L_i}{2\upi} + \frac{L_j}{2\upi} + 
    \alpha_\iin \left(\frac{L_i}{N_\iin} + \frac{L_j}{N_\iin} \right),
\end{align}
where $\cc_i$ is the center of grain $i$, $L_i$ is the length of its
boundary, and $\alpha_\iin \geq 1$ is a parameter that needs to be
determined.  In this manner, rather than using an expensive all-to-all
algorithm to compute the distance between pairs of discretization
points, we compute the distance between pairs of circle centers.  This
criteria allows us to quickly determine bodies that contain
discretization points where the Barycentric quadrature rule might need
to be applied, and the parameter $\alpha_\iin$ accounts for the
approximation that the grains are circular.  Assuming that the two
bodies $\gamma_i$ and $\gamma_j$ satisfy
condition~\eqref{eqn:BodyCutoff}, for each point $\xx \in \gamma_j$, we
check if
\begin{align}
  \label{eqn:PointsBodyCutoff}
  \|\xx - \cc_i\| < \frac{L_i}{2\upi}
+ \alpha_\iin \left(\frac{L_i}{N_\iin} \right).
\end{align}
To determine if points on $\gamma_i$ are too close to the outer wall, we
recall that the eroding bodies are all contained in $[-1,1] \times
[-1,1]$, so a target point can only be close to the lines $y = \pm 1$.
Therefore, we first check if
\begin{align}
  \left\|\cc_i - \left[
    \begin{array}{c}
      x_i \\ \pm 1
    \end{array}
    \right]
  \right\| < \frac{L_i}{2 \upi} + \alpha_\out \frac{L_\out}{N_\out},
\end{align}
where $x_i$ is the $x$-coordinate of $\cc_i$. If body $\gamma_i$
satisfies this condition, for each point $\xx=(x,y) \in \gamma_i$, we
apply the Barycentric rule to points that satisfy
\begin{align}
  \label{eqn:PointsWallCutoff}
  |y \pm 1| < \frac{L_i}{2 \upi} +\alpha_\out \frac{L_\out}{N_\out}.
\end{align}
Finally, to determine if a target point $\xx$ in the fluid bulk requires
the Barycentric quadrature rule, we only check
conditions~\eqref{eqn:PointsBodyCutoff}
and~\eqref{eqn:PointsWallCutoff}.  

To determine appropriate values for $\alpha_\iin$ and $\alpha_\out$,
we fixed an eroded geometry and computed an accurate solution by using
the Barycentric quadrature rule for all discretization points.  Then,
for multiple values of $\alpha_\iin$ and $\alpha_\out$, we computed the
velocity field with the trapezoid rule for all points that do not
satisfy conditions~\eqref{eqn:PointsBodyCutoff}
and~\eqref{eqn:PointsWallCutoff}.  By comparing these two velocities, we
find that $\alpha_\iin = 4$ and $\alpha_\out = 4$ give sufficient
accuracy to maintain stability while keeping the number of points that
require the expensive Barycentric quadrature rule to a minimum.  We use
these values for all of our numerical simulations.

\subsection{Time Integration}
\label{sec:time}
We use the time stepping method outlined in our previous
work~\citep[see][section 3.3]{qua-moo2018} which we briefly summarize
here.  The erosion rate loses differentiability if the shear stress
changes sign, and this leads to corners developing on $\gamma$ and
numerical instabilities.  Therefore, we modify the erosion rate, $V_\nn$
in equation~\eqref{eqn:erosionModel}, with
\begin{align}
  \Vn = \CE \, \abs{\tau} + \epsilon \langle\abs{\tau}\rangle \left(
    \frac{L}{2\upi} \kappa - 1 \right),
\end{align}
where $\epsilon \ll 1,$ $\langle \cdot \rangle$ is the spatial average,
$L$ is the length of $\gamma$, and $\kappa$ is the curvature of
$\gamma$.  The new erosion model penalizes regions of high curvature,
but does not change the total length of each body.  Moreover, to
increase the overall stability of the method, a narrow Gaussian filter
is applied to the erosion rate at each time step.

Rather than tracking the $(x,y)$ coordinates, the {\thL} coordinates are
tracked. In addition, tangential velocity fields are used to maintain
an equispaced discretization. Time stepping is performed with a
second-order Implicit-Explicit Runge-Kutta method. In particular, the
diffusive term corresponding to the curvature penalization term is
discretized implicitly, and all other terms, which are non-stiff, are
treated explicitly.  By using this time stepping method in conjunction
with the Barycentric rule, we stably evolve the eroding bodies.

To examine the tortuosity and the anomalous dispersion rates
(section~\ref{sec:transport}), we require accurate 
streamlines governed by equation~\eqref{eqn:tracers}. If a low-order
time stepping method is used, or if $\uu(\ss(t))$ is inaccurate, then a
trajectory $\ss(t)$ can unphysically enter a grain, rendering the
trajectory meaningless. However, simply ignoring trajectories that pass
close to a grain could significantly bias the characterization of
transport.  Therefore, we use high-order quadrature and time stepping
methods.  In particular, depending on the proximity of $\ss(t)$ to
$\bd\Omega$ (section~\ref{sec:fmm}), we apply the trapezoid rule or the
Barycentric quadrature rule.  For time stepping, we use a fourth-order
explicit Runge-Kutta method.  By using these high-order methods, we are
able to simulate dynamics very close to the eroding bodies (see
figures~\ref{fig:Eroding20tracer} and~\ref{fig:Eroding100tort}).

Once a collection of trajectories are formed, they are used to quantify
the dispersion and the tortuosity.  We use $N_p = 1000$ streamlines so
that the statistics have converged~\citep{bel-sal-rin1992}. As described
in section~\ref{sec:dispersion}, a reinsertion method is used to compute
trajectories that are sufficiently long to observe an asymptotic
anomalous dispersion rate.  To compute the tortuosity using
equation~\eqref{eqn:tortuosity1}, we consider trajectories crossing
between the two cross-sections $x=-1$ and $x=1$, and approximate the
tortuosity with
\begin{align}
  T = \frac{1}{d}\left(\sum_{i=1}^{N_p} 
    u_1(y_i) \lambda(y_i) \Delta y \right)
  \Bigg/
  \left(\sum_{i=1}^{N_p} u_1(y_i) \Delta y\right), 
\end{align}
where $\Delta y = 2/(N_p + 1)$ and $y_i = -1 + i \Delta y$,
$i=1,\ldots,N_p$.  

\section{Numerical Results}
\label{sec:results}
We now present numerical results of dense grain packings eroding in
Stokes flow and analyze transport through the evolving geometries.  Each
body is initialized as a circle of center $\cc_i$, radius $r_i$, and
length $L_i = 2\upi r_i$.  The center and radius are chosen at random,
and the body is accepted if it is contained in $[-1,1] \times [-1,1]$
and is sufficiently separated from all other bodies.  Owing to our
adaptive quadrature rule, we can consider bodies that are separated by
less than 10\% of an arclength spacing. The randomized method is
repeated until the initial geometry reaches a desired initial porosity.

For all simulations, we discretize each eroding grain with $N_\iin =
256$ points and the outer wall $\Gamma$ with $N_\out=1024$ points.  A
no-slip boundary condition is imposed on each eroding body $\gamma_i$,
and a Dirichlet boundary condition on $\Gamma$ is used to approximate a
far-field boundary condition. For all but the first example, the
Dirichlet boundary condition is a Hagen-Poiseuille flow, and the flow
rate is adjusted at each time step to maintain a constant pressure drop.
Since the fluid equations are linear, this is achieved by computing the
pressure near the inlet and outlet at each time step, and then scaling
the flow rate appropriately~\citep{qua-moo2018}. We also compute the
vorticity in the fluid bulk to help visualize the erosion rate.

The erosion rate loses regularity at stagnation points, which inevitably
leads to corner formation on the bodies.  As described in
section~\ref{sec:time} and our previous work~\citep{qua-moo2018}, we
ameliorate corner formation by introducing a curvature penalization term
with parameter $\eps$ and a Gaussian smoothing step with parameter
$\sigma$.  For all examples, we use the smoothing parameters
$\eps=15/256$ and $\sigma=10/256$, and the time step size is $\Delta t =
10^{-4}$.

The common characteristic of each of the experiments is near-contact
between the eroding bodies, outer walls, and streamlines. We use our
numerical methods to simulate, analyze, and visualize the following
examples:
\begin{itemize}
  \item{\bf Single Body Close to a Wall}: We consider a single eroding
  body close to the outer wall at $y=-1$.  We impose a shear flow
  centered at $y=-1$ and compare the eroding body's shape to a similar
  experiment of~\citet{mit-spa2017}.

  \item{\bf 20 Bodies at a Medium Porosity}: We consider 20 eroding
  bodies with a medium initial porosity.  After computing accurate
  streamlines, the tortuosity and anomalous dispersion rates are
  computed and compared to those of an open channel.

  \item{\bf 20 Bodies at a Low Porosity}: We consider 20 eroding bodies
  with a low initial porosity.  We examine the effect of the lower
  porosity on the tortuosity and anomalous dispersion rates.

  \item{\bf 100 Bodies at a Medium Porosity}: We consider 100 eroding
  bodies with a medium porosity.  We compute the tortuosity, anomalous
  dispersion rates, and the pore throat size distributions.
\end{itemize}

\subsection{A Single Body Close to a Wall}
Consider a single eroding body close to a solid wall with the shear flow
$\UU(\xx) = (y+1,0)$ imposed on $\Gamma$. \citet{mit-spa2017} performed
a similar three-dimensional experiment using a second-order quadrature
method.  Their initial body is a sphere with its center located $1.5$
radii above the solid wall.  We initialize the two-dimensional eroding
body with radius $r = 0.4$, and we conduct numerical experiments where
the initial distance between the grain and the solid wall is $h$, $h/2$,
and $h/10$, where $h = 2\upi r/N_\iin$.  If we used the trapezoid rule
and required an error that is comparable to the Barycentric quadrature
rule, the body with an initial distance of $h/10$ from the solid wall
would require $6,400$ discretization points, and the outer wall would
require $50,000$ discretization points.

\begin{figure}
\begin{center}
\begin{subfigure}[b]{0.32\textwidth}
\includegraphics[height = 0.74\textwidth]{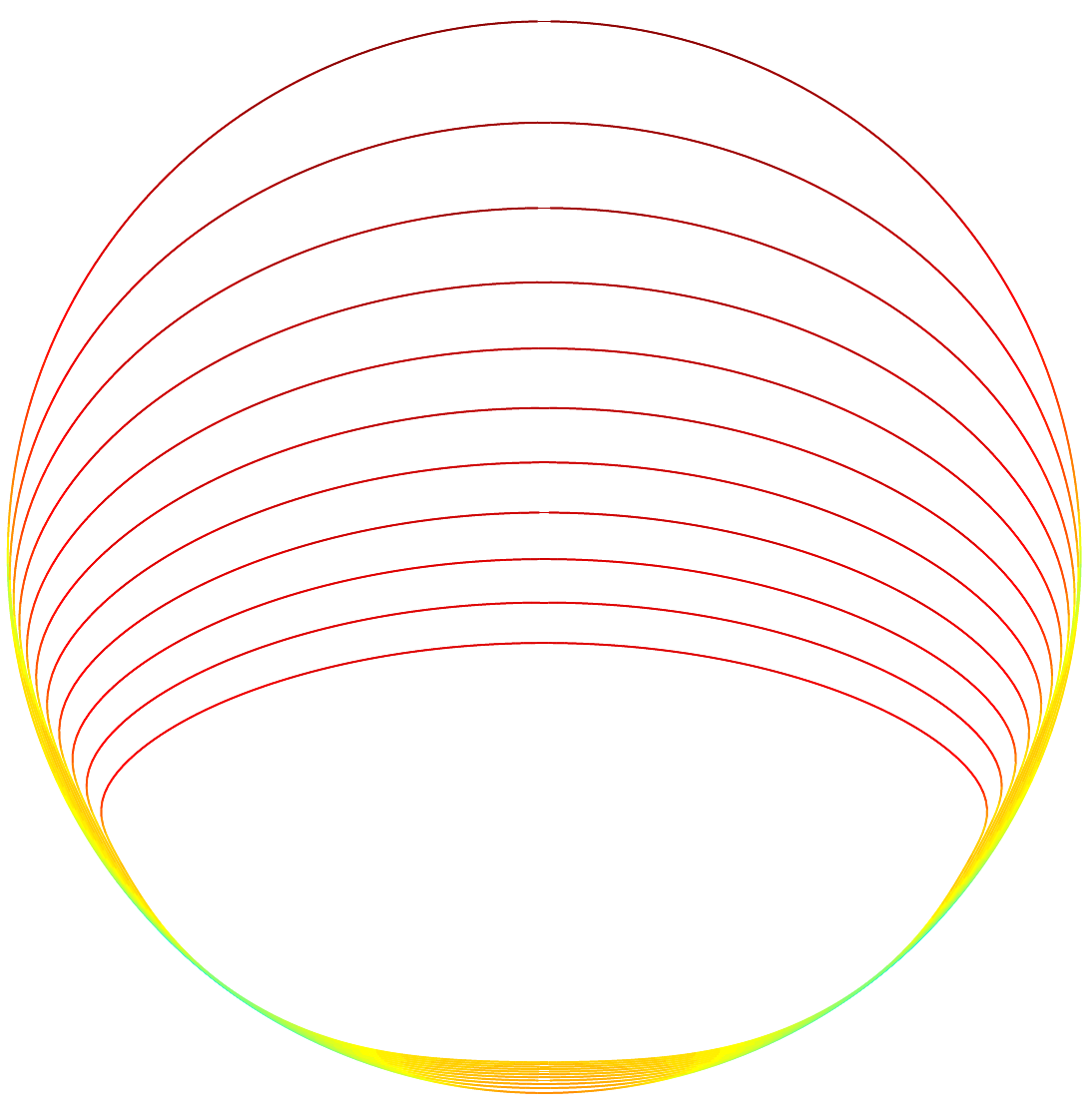}
\caption{}
\end{subfigure}
\begin{subfigure}[b]{0.32\textwidth}
\includegraphics[height = 0.74\textwidth]{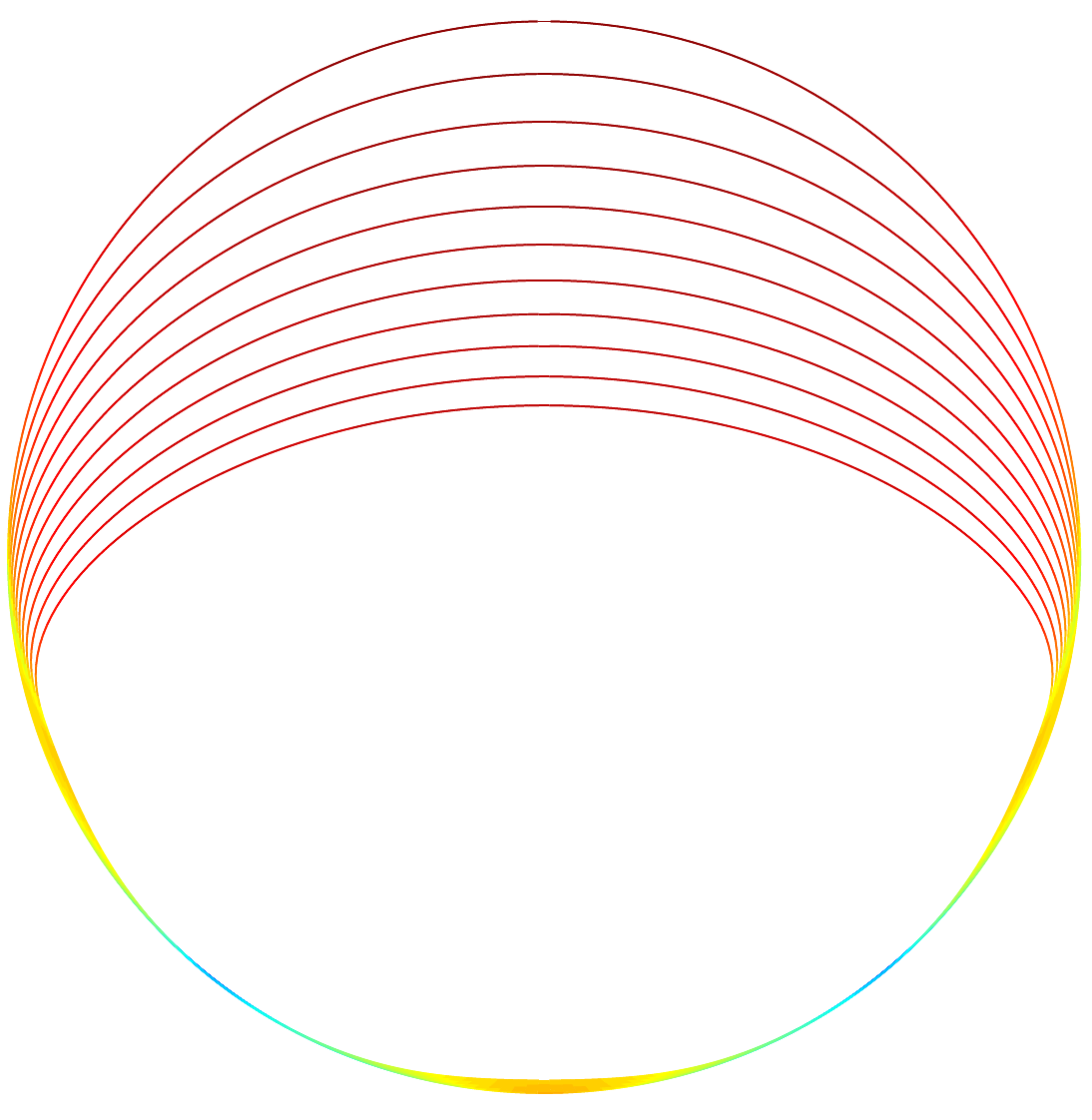}
\caption{}
\end{subfigure}
\begin{subfigure}[b]{0.32\textwidth}
\includegraphics[height = 0.74\textwidth]{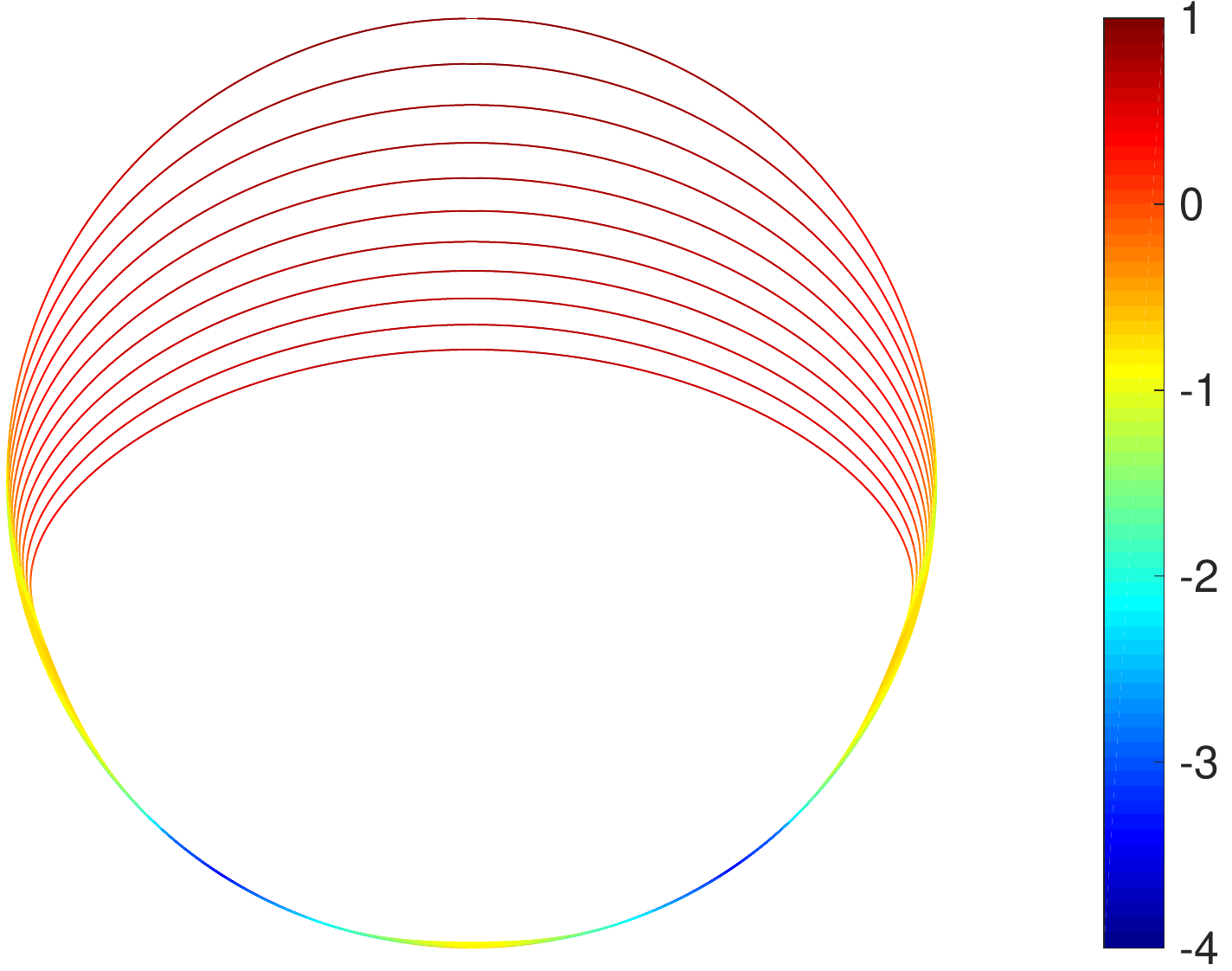}
\caption{}
\end{subfigure}
\end{center}
\caption{\label{fig:NearWall} A single body eroding in a shearing Stokes
flow.  The color is the logarithm of the shear stress. Therefore,
erosion is fastest in the red regions (upper half) and slowest in the
blue regions (lower half).  The body is initialized at three different
distances from the lower wall: (a) $h$, (b) $h/2$, and (c) $h/10$.}
\end{figure}

\begin{figure}
\begin{center}
\begin{subfigure}[b]{0.32\textwidth}
\includegraphics[height = 0.53\textwidth]{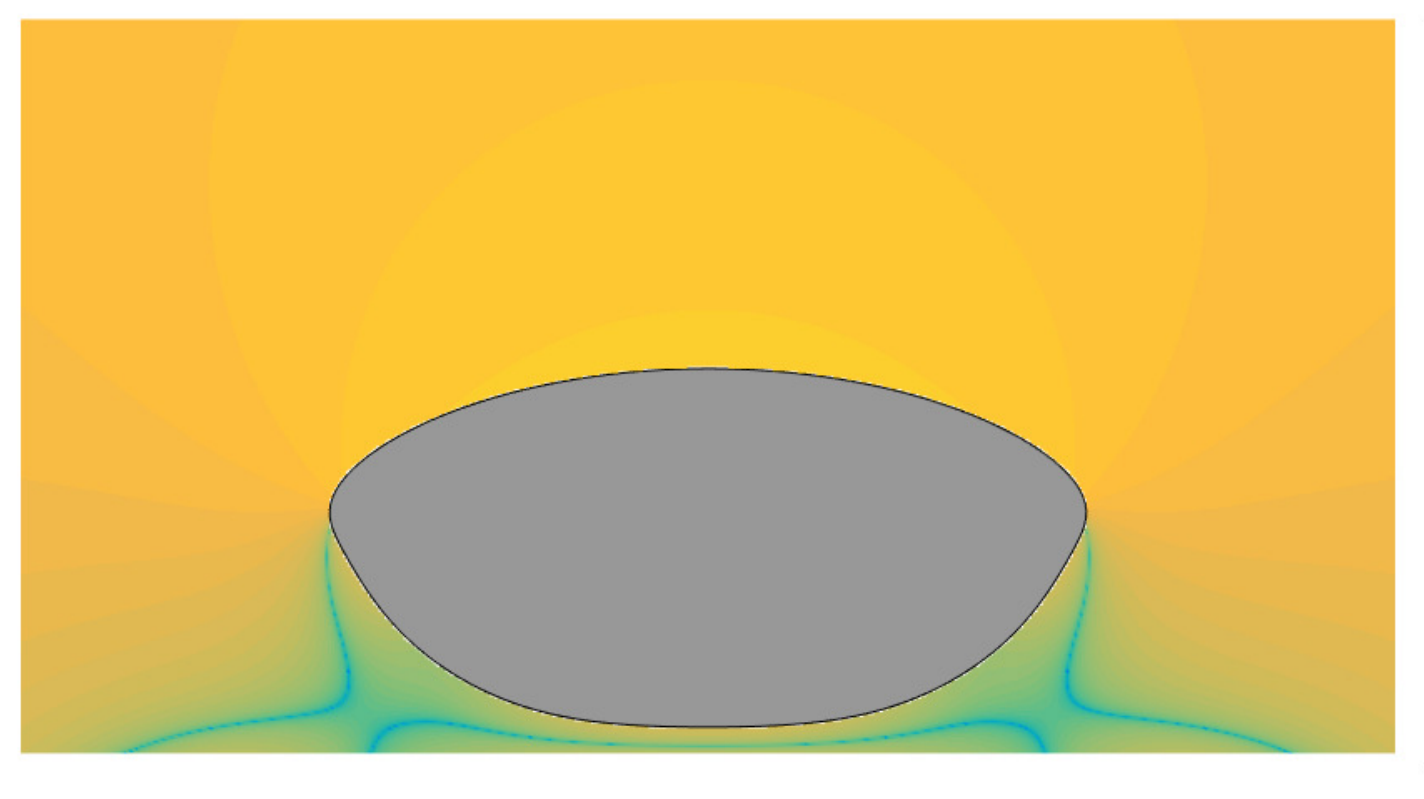}
\caption{}
\end{subfigure}
\begin{subfigure}[b]{0.32\textwidth}
\includegraphics[height = 0.53\textwidth]{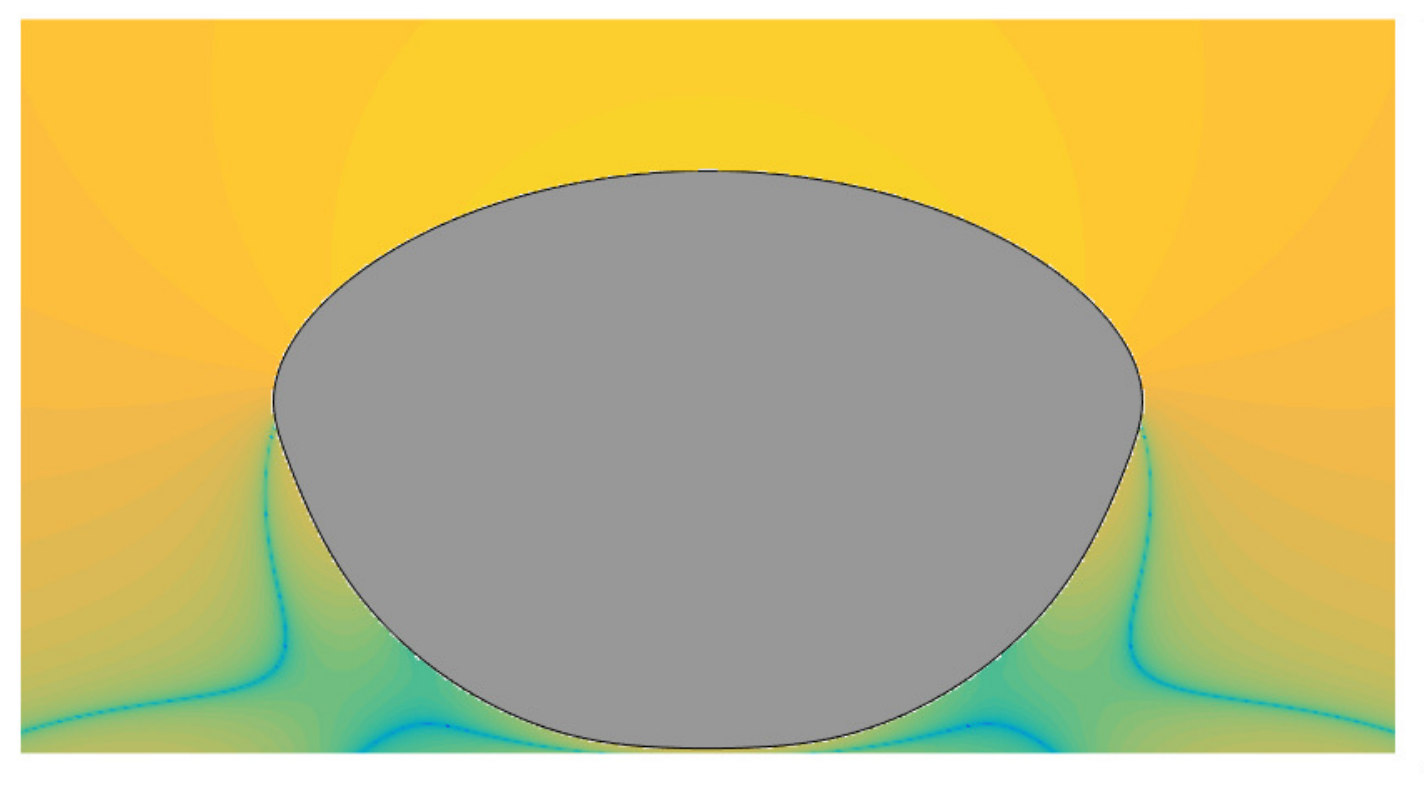}
\caption{}
\end{subfigure}
\begin{subfigure}[b]{0.32\textwidth}
\includegraphics[height = 0.53\textwidth]{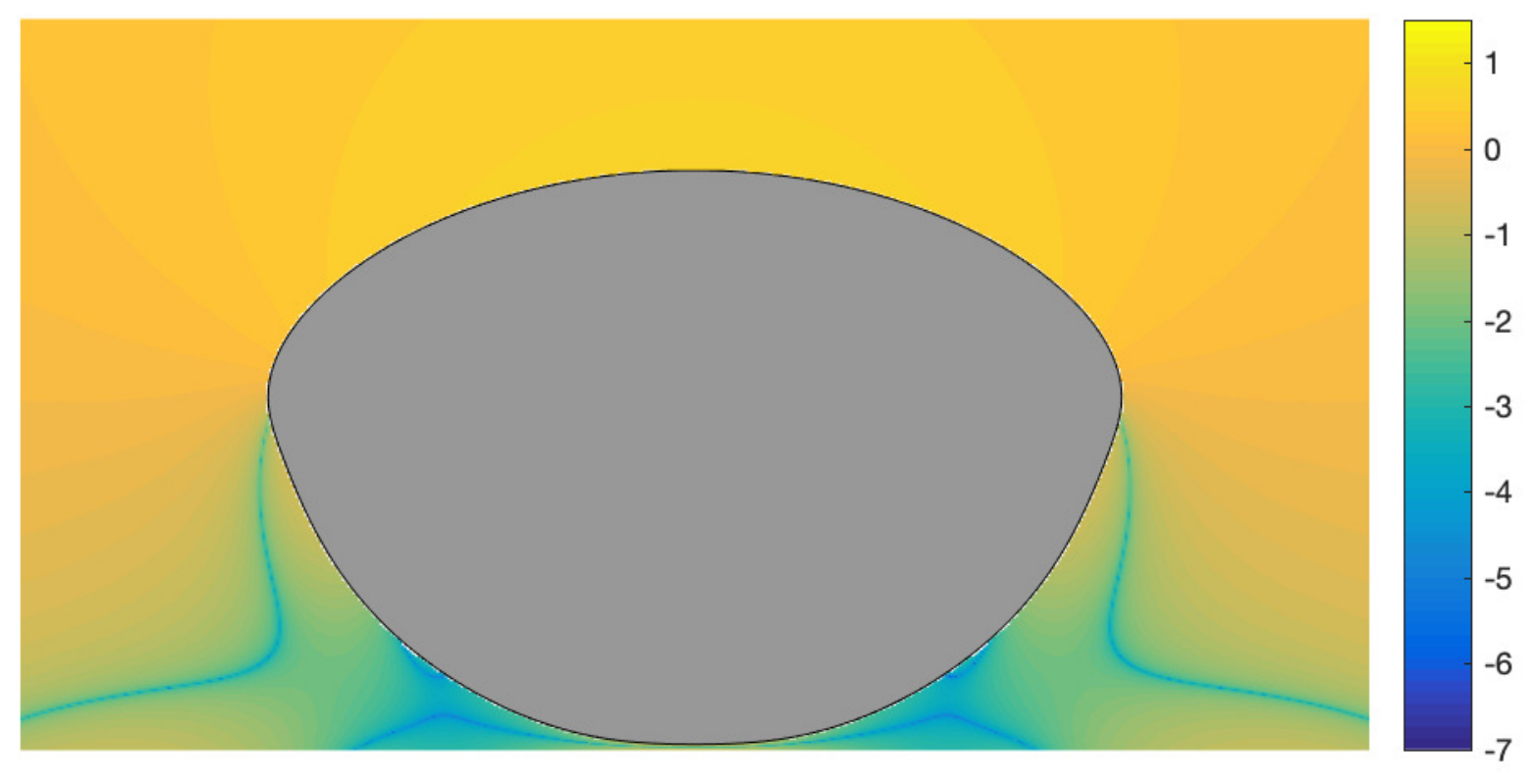}
\caption{}
\end{subfigure}
\caption{\label{fig:NearWall_vort} The vorticity of the fluid with a
single body eroding at time $t=0.1$. The initial distance from the body
to the solid wall are: (a) $h$, (b) $h/2$, and (c) $h/10$.}
\end{center}
\end{figure}

In figure~\ref{fig:NearWall}, we superimpose the eroding body's shape at
equispaced time steps.  For all three initial configurations, the shear
stress is positive for all time, but varies over several orders of
magnitude.  Therefore, we color the eroding body's boundary with the
logarithm of the shear stress.  Since the shear stress is always
positive, the erosion rate is smooth and corners do not develop.
However, in the top half of the body, there is a sudden increase in the
shear stress, and this leads to a region of high curvature.  This
behavior is also present in three dimensions~\citep[see figure
7(c)]{mit-spa2017}.  The biggest difference between the two- and
three-dimensional results is the presence of a recirculation zone.  In
three dimensions, there is no recirculation between the solid wall and
the spherical body~\citep{cha-feu2003}, but recirculation is possible in
two dimensions~\citep{chw-wu1975, hig1985}.  To visualize the flow, we
plot the vorticity of the final time step from figure~\ref{fig:NearWall}
in figure~\ref{fig:NearWall_vort}.  In these examples, a small
recirculation zone, both in size and magnitude, is present in the region
where the vorticity is smallest.


\subsection{20 Bodies at a Medium Porosity}
\label{sec:Eroding20}
\begin{figure}
\begin{center}
  \includegraphics[height=0.227\textwidth]{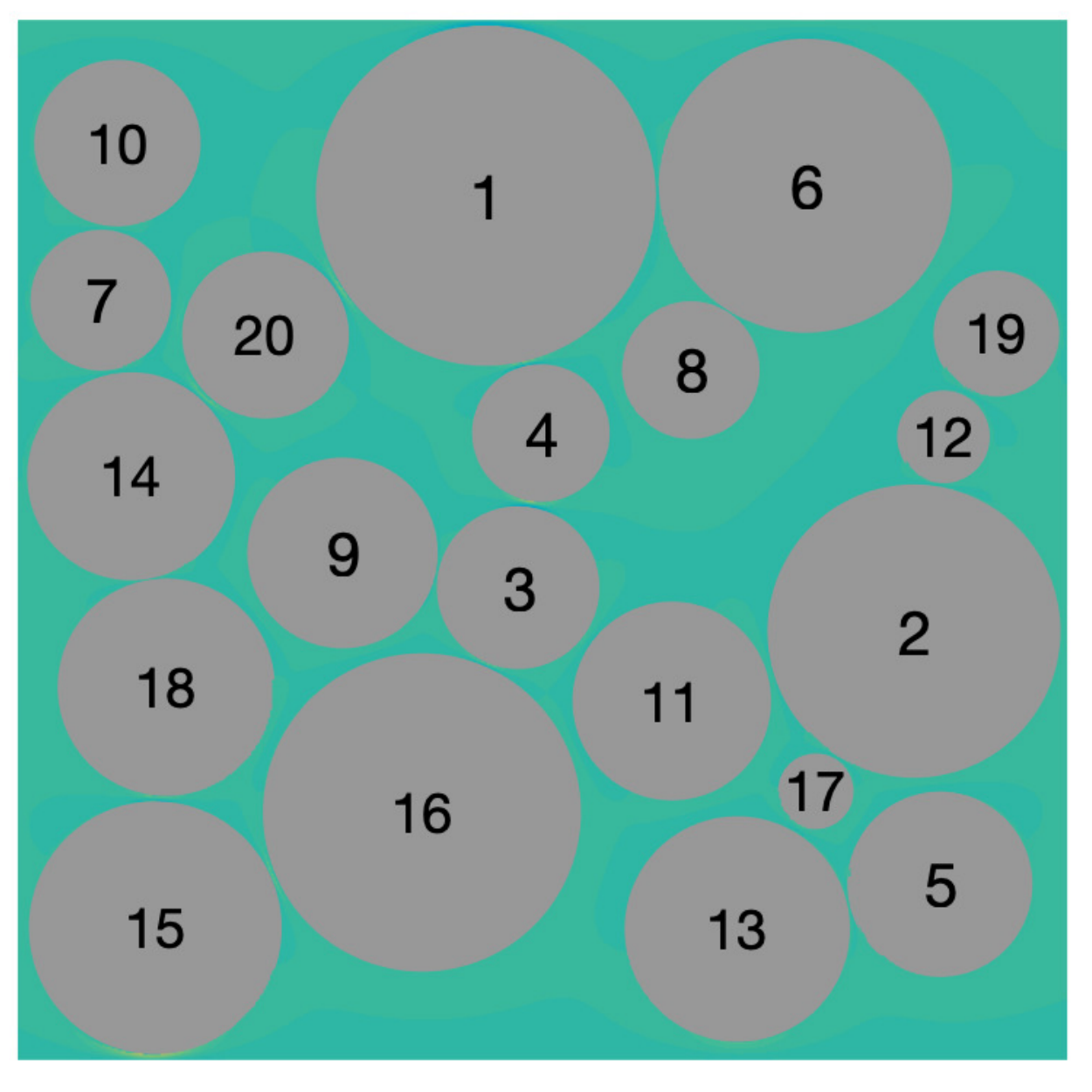}
  \includegraphics[height=0.227\textwidth]{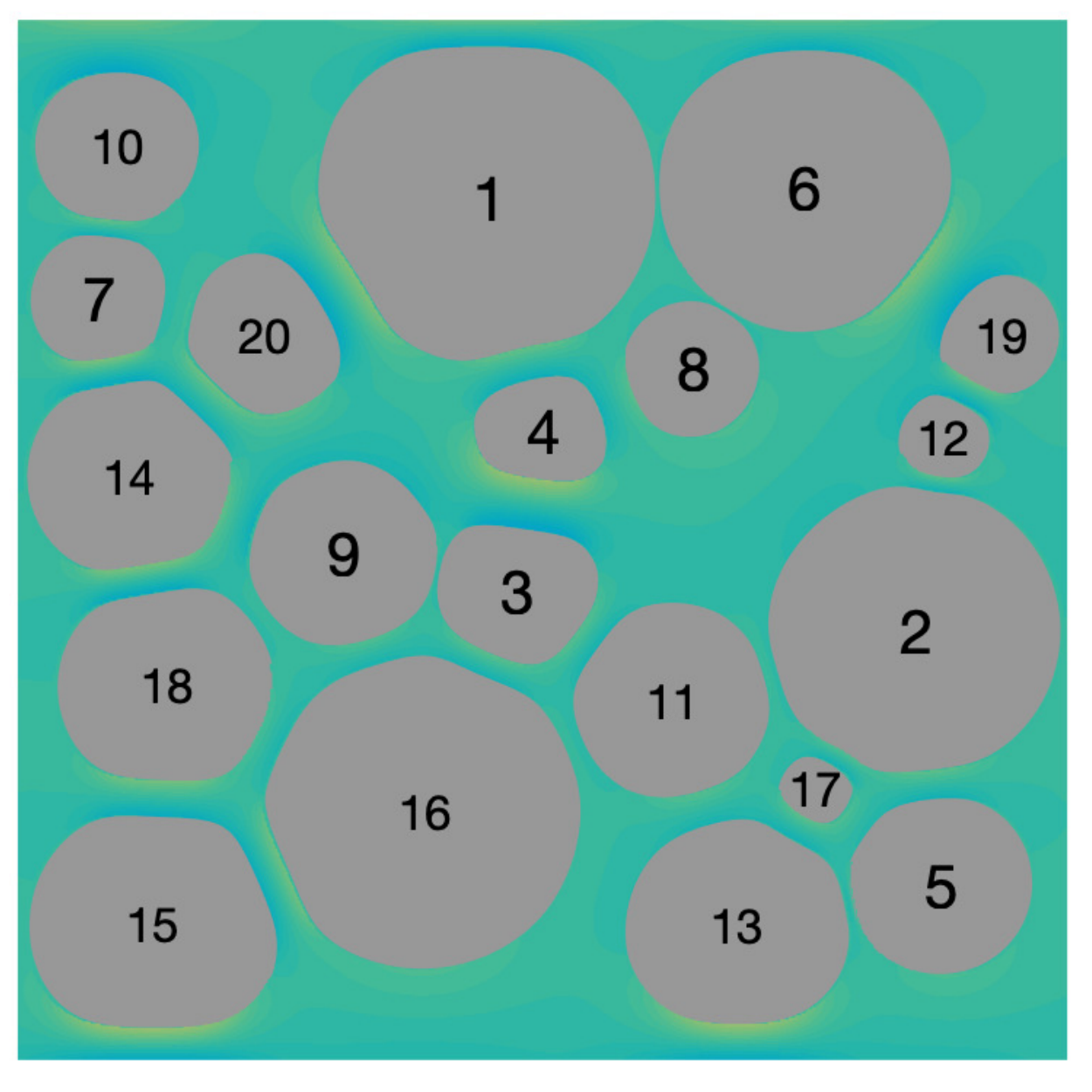}
  \includegraphics[height=0.227\textwidth]{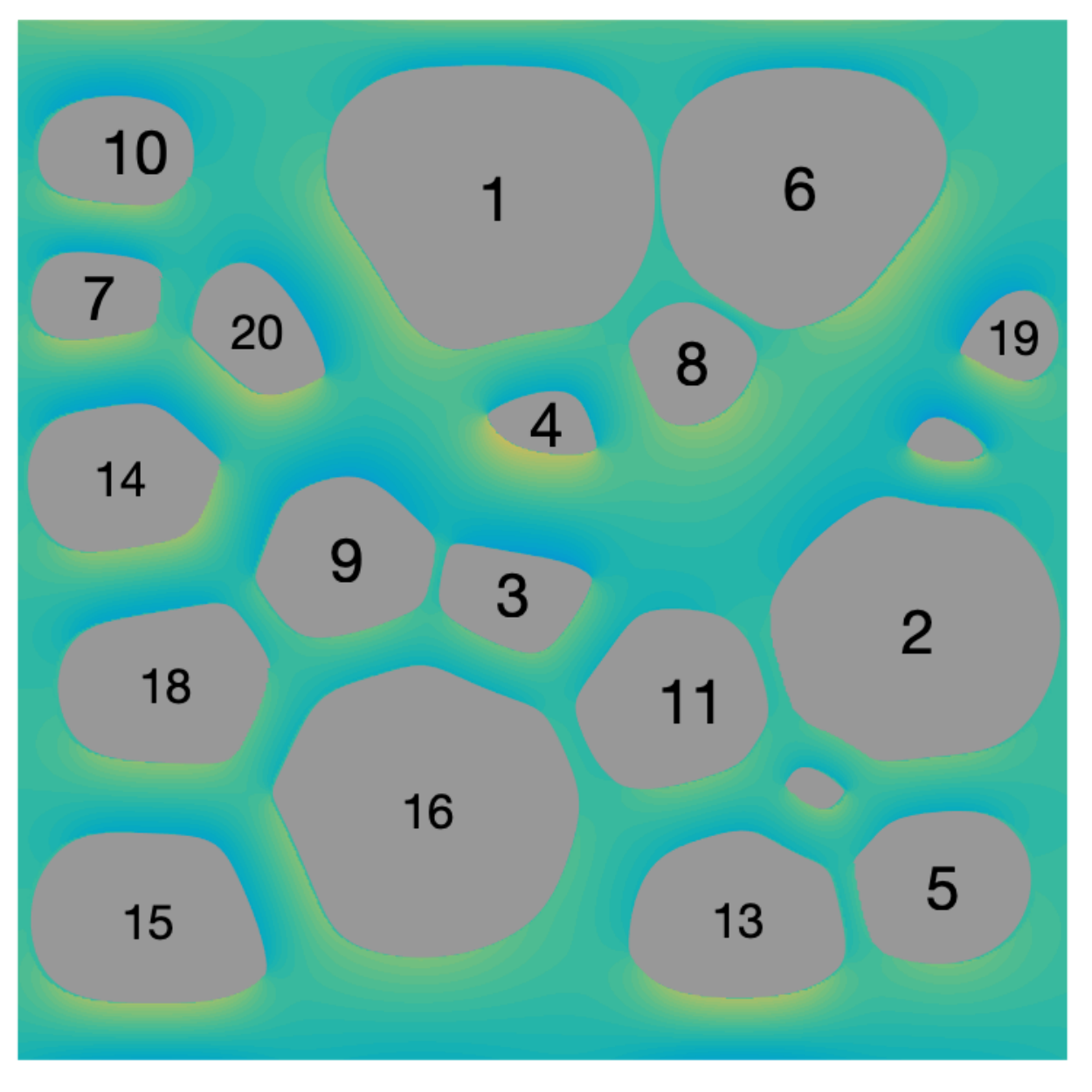}
  \includegraphics[height=0.227\textwidth]{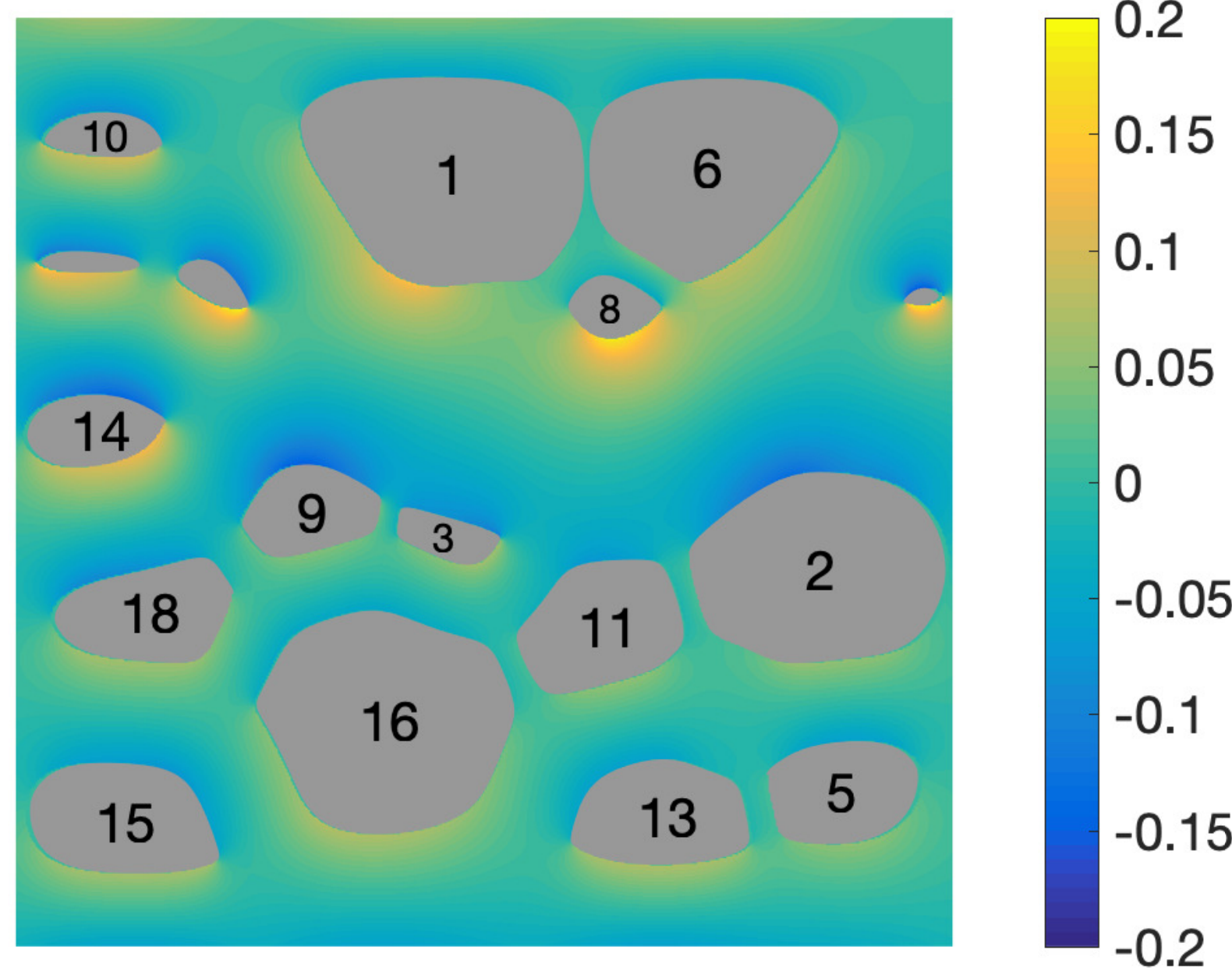}
\caption{\label{fig:Eroding20vort} 20 bodies eroding in a
Hagen-Poiseuille flow.  The four snapshots are equispaced in time, and
the color is the fluid vorticity. In the fourth frame, bodies 4, 12, and
17 have vanished, and bodies 7, 19, 20 have almost vanished.}
\end{center}
\end{figure}

We consider 20 eroding grains with the Hagen-Poiseuille flow $\UU(\xx) =
U \left(1-y^2,0 \right)$ imposed on $\Gamma$. The flow rate $U$ is
chosen so that the average pressure drop from $x=-2$ to $x=2$ is held
fixed at 8. Therefore, $U=1$ once all the grains have vanished.  The
vorticity and grain configuration at four equispaced times are shown in
figure~\ref{fig:Eroding20vort}.  Initially, several of the grains are
closer to the outer wall than the $5h$ threshold required for the
trapezoid rule to achieve machine precision.  In particular, the
distance between bodies 1, 6, 13, and 15 and the outer wall is $1.3h$,
$2.9h$, $2.8h$, and $1.3h$, respectively, where $h$ is the arclength
spacing of the outer wall $\Gamma$.  In addition, the distance between
several pairs of eroding bodies, including 1 \& 6, 3 \& 9, 6 \& 8, and
14 \& 18, is too small to be resolve with the trapezoid rule.  By using
the Barycentric quadrature rule, the interaction between these
nearly-touching bodies is resolved to the desired accuracy, and erosion
can be simulated until all the bodies have vanished.

Erosion causes the some of the pore sizes to quickly grow, and flat
faces develop along the regions of near contact.  This qualitative
behavior is seen in figure~\ref{fig:Eroding20vort} between bodies 3 \&
4, 15 \& 16, and was also observed in previous work~\citep{qua-moo2018}.
However, by resolving the interaction between bodies that are much
closer together, we observe that very little erosion occurs between
certain pairs of bodies, at least initially.  For instance the opening
between bodies 1 \& 6, 3 \& 9, and 5 \& 13 grow much slower than the
opening between bodies 15 \& 16.  A common feature of the pores that
grow slowly is that they are nearly perpendicular to the main flow
direction, resulting in a small erosion rate.

\begin{figure}
\begin{subfigure}[b]{0.5\textwidth}
\includegraphics*[height = 0.7\linewidth]{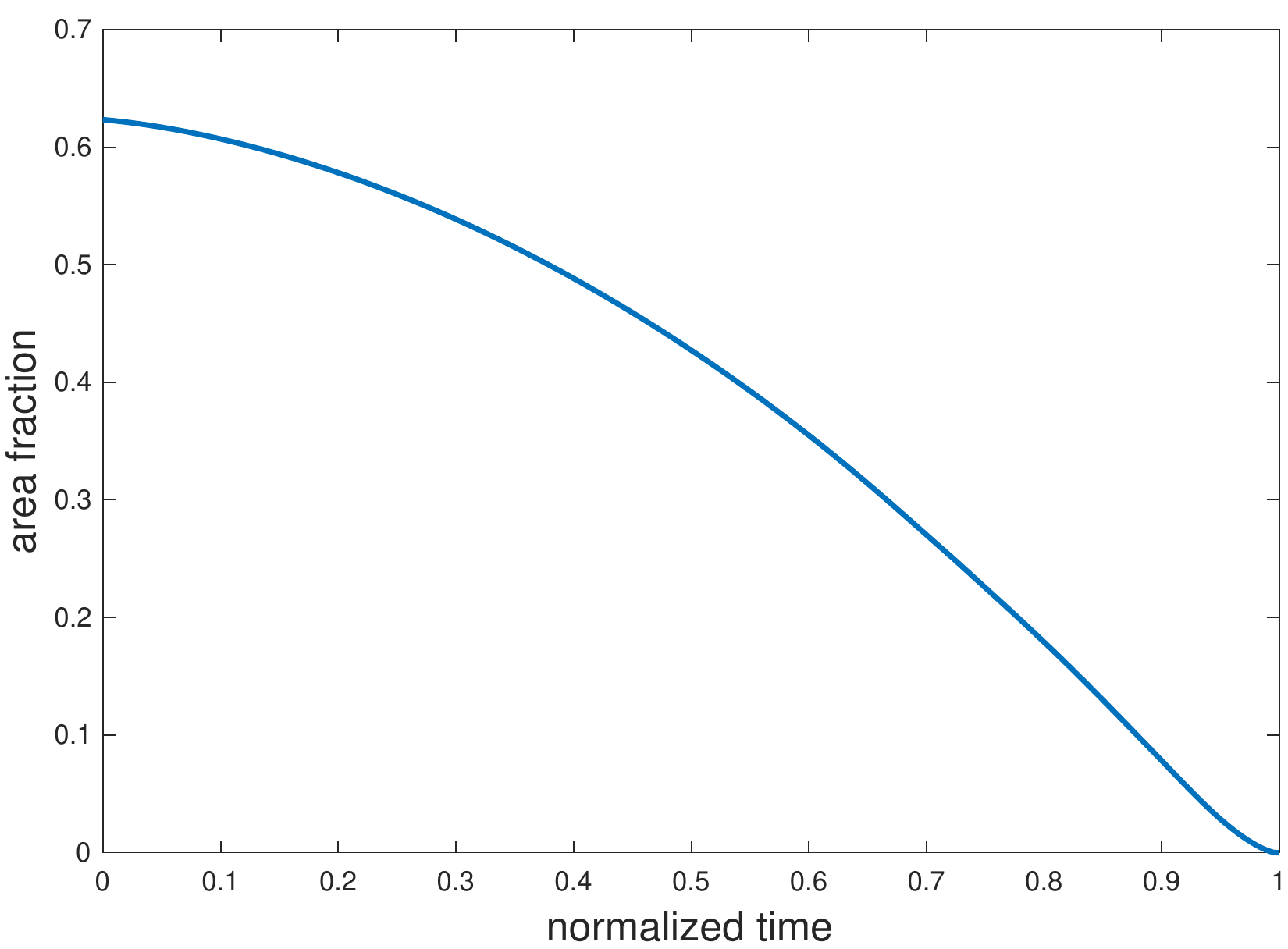}
\caption{}
\end{subfigure}
\begin{subfigure}[b]{0.5\textwidth}
\includegraphics*[height = 0.7\linewidth]{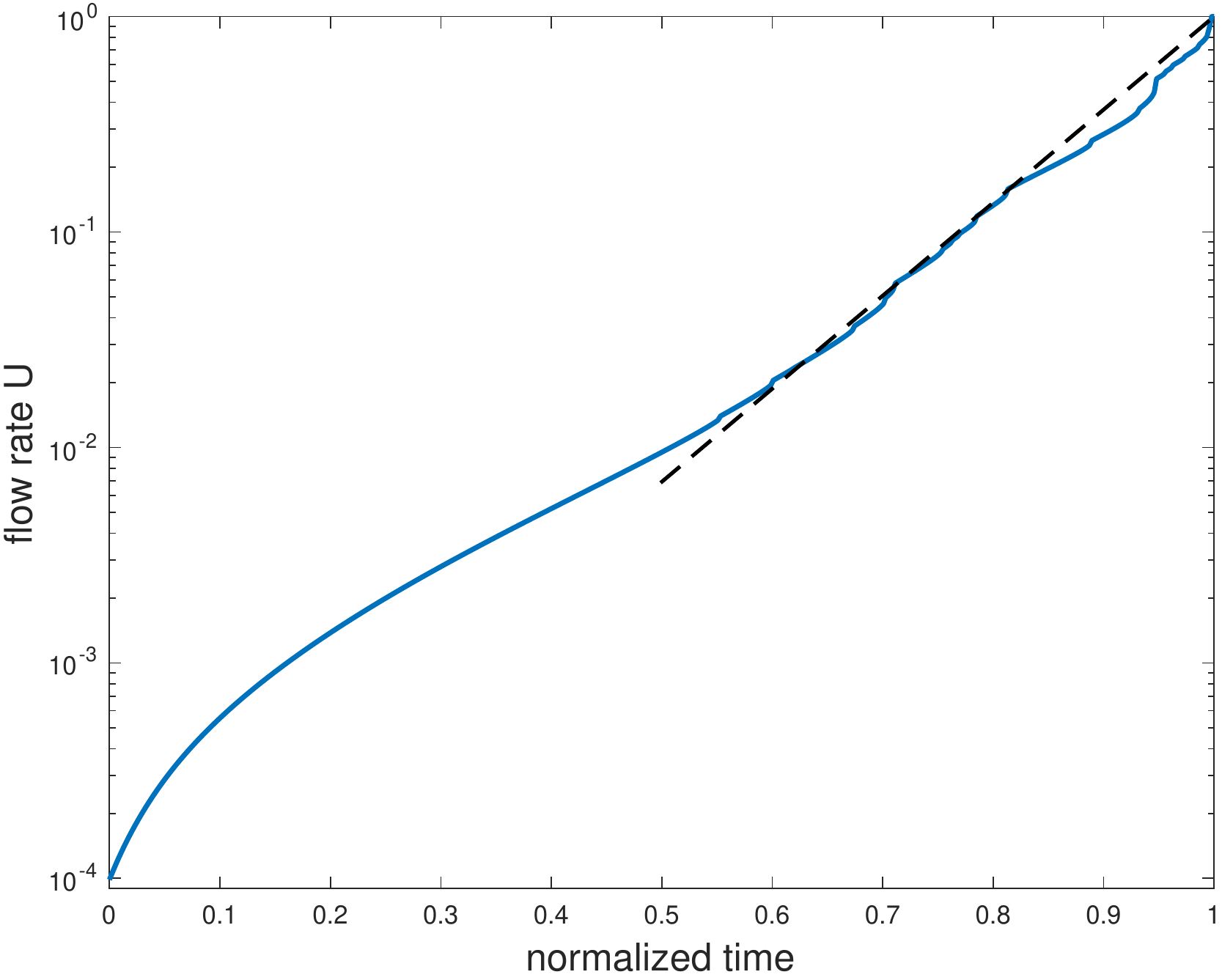}
\caption{}
\end{subfigure}
\caption{\label{fig:Eroding20flowrate}(a) The time-dependent area
fraction of a geometry with 20 eroding bodies. (b) The time-dependent
flow rate, $U$, for a fixed pressure drop across the channel.  The flow
rate is initially small, but it eventually increases as an exponential
law (dashed line) towards the maximum flow rate $U=1$.}
\end{figure}

We next analyze the effect of erosion on the area fraction and the flow
rate.  In figure~\ref{fig:Eroding20flowrate}(a), we plot the area
fraction as a function of normalized time.  The general trend of the
area fraction resembles our previous work~\citep[see][figure
10(a)]{qua-moo2018}, but with a larger initial area fraction.  In
figure~\ref{fig:Eroding20flowrate}(b), we plot the flow rate $U$
required to maintain a constant pressure drop across the channel.
Again, the trend of $U$ resembles that of our previous
work~\citep[see][figure 10(b)]{qua-moo2018}, except that the initial
flow rate is an order of magnitude smaller because of the larger initial
area fraction.  Starting around normalized time $0.2$,
figure~\ref{fig:Eroding20flowrate}(b) is roughly linear which indicates
that the flow rate can be written as an exponential law.  The line of
best fit is $U \approx \exp(9.94(t-t_f)/t_f)$
which is the dashed line in figure~\ref{fig:Eroding20flowrate}(b).

\begin{figure}
\begin{center}
\includegraphics[width = 0.32 \textwidth]{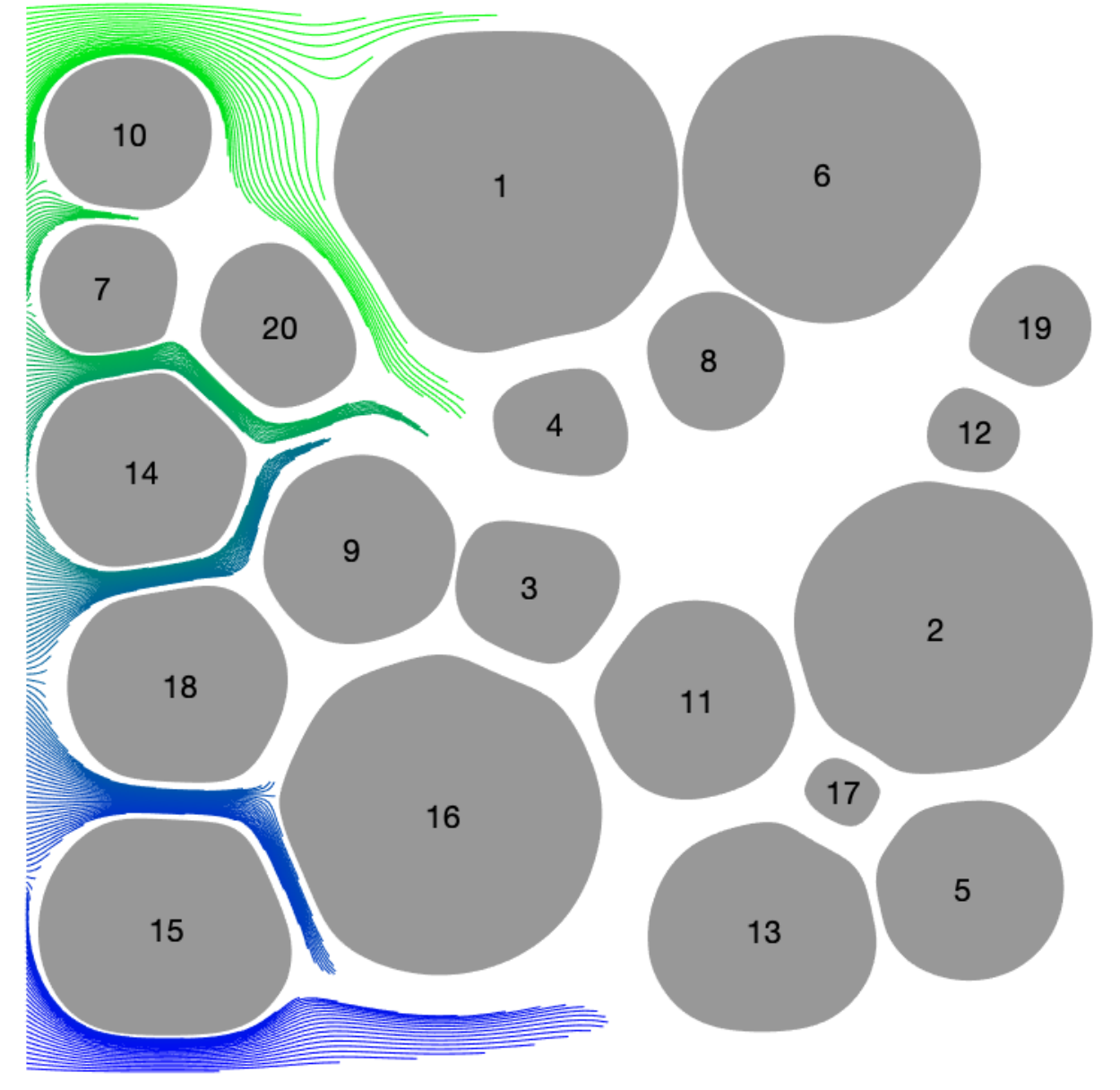}
\includegraphics[width = 0.32 \textwidth]{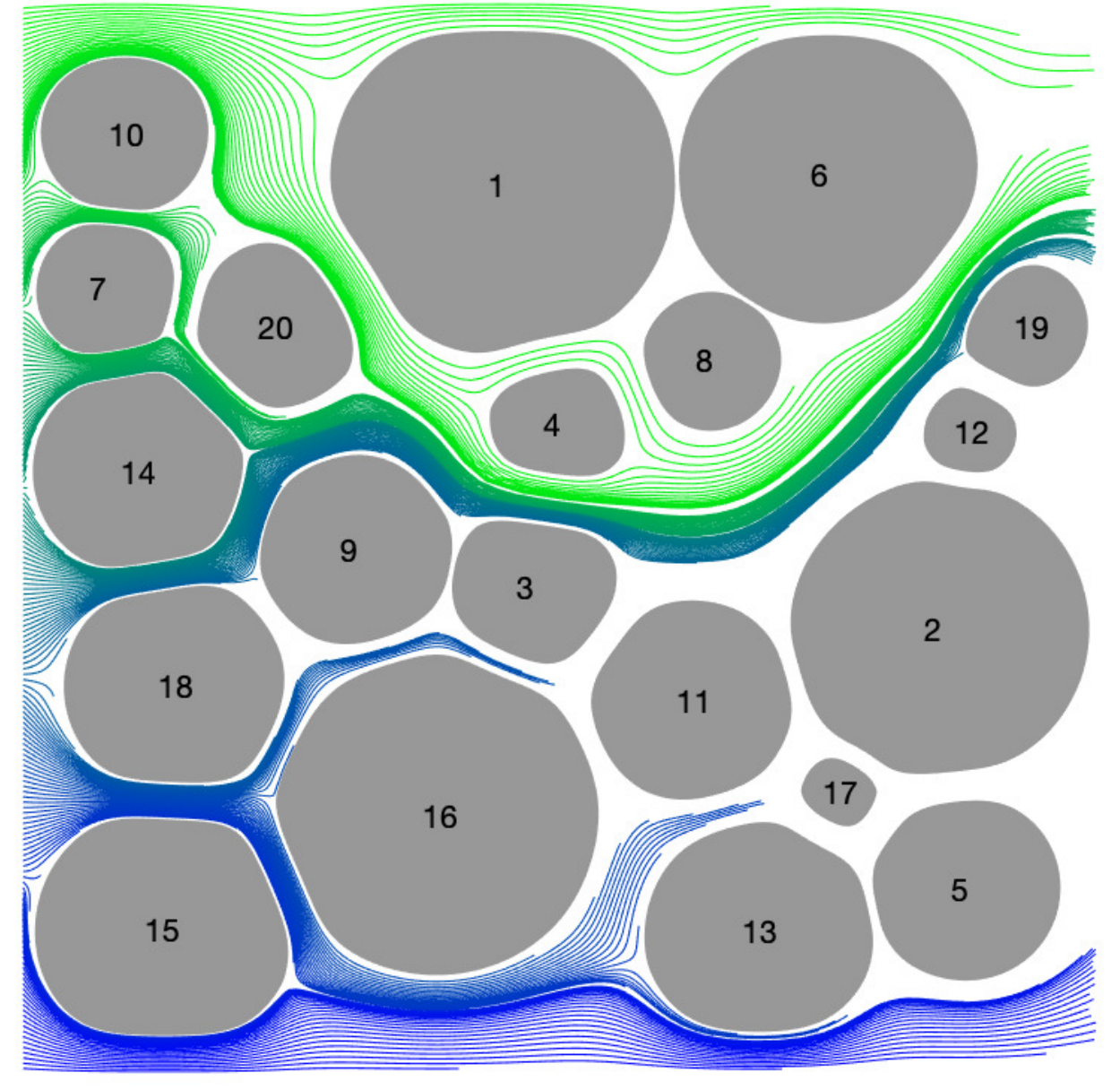}
\includegraphics[width = 0.32 \textwidth]{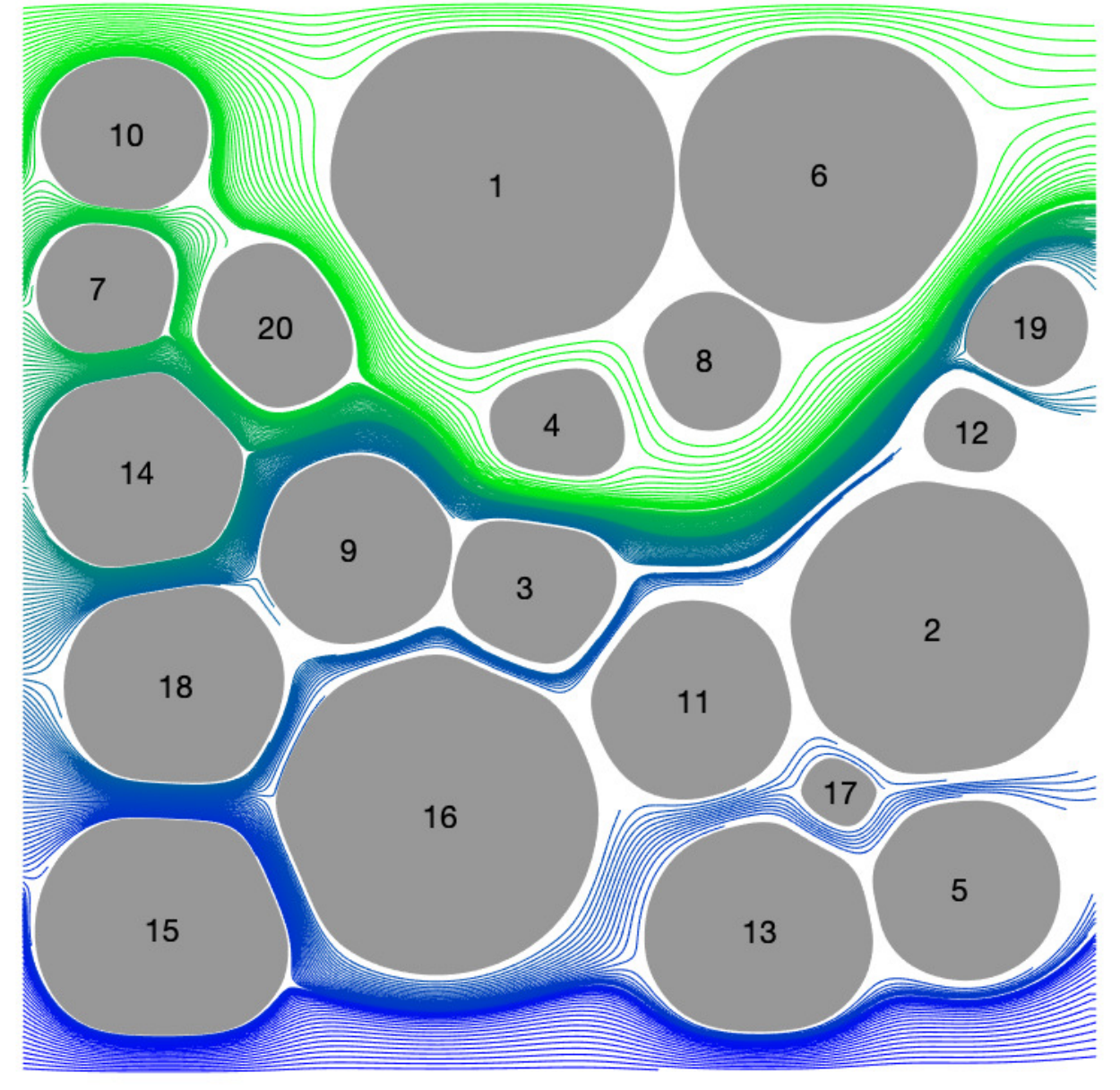}\\
\includegraphics[width = 0.32 \textwidth]{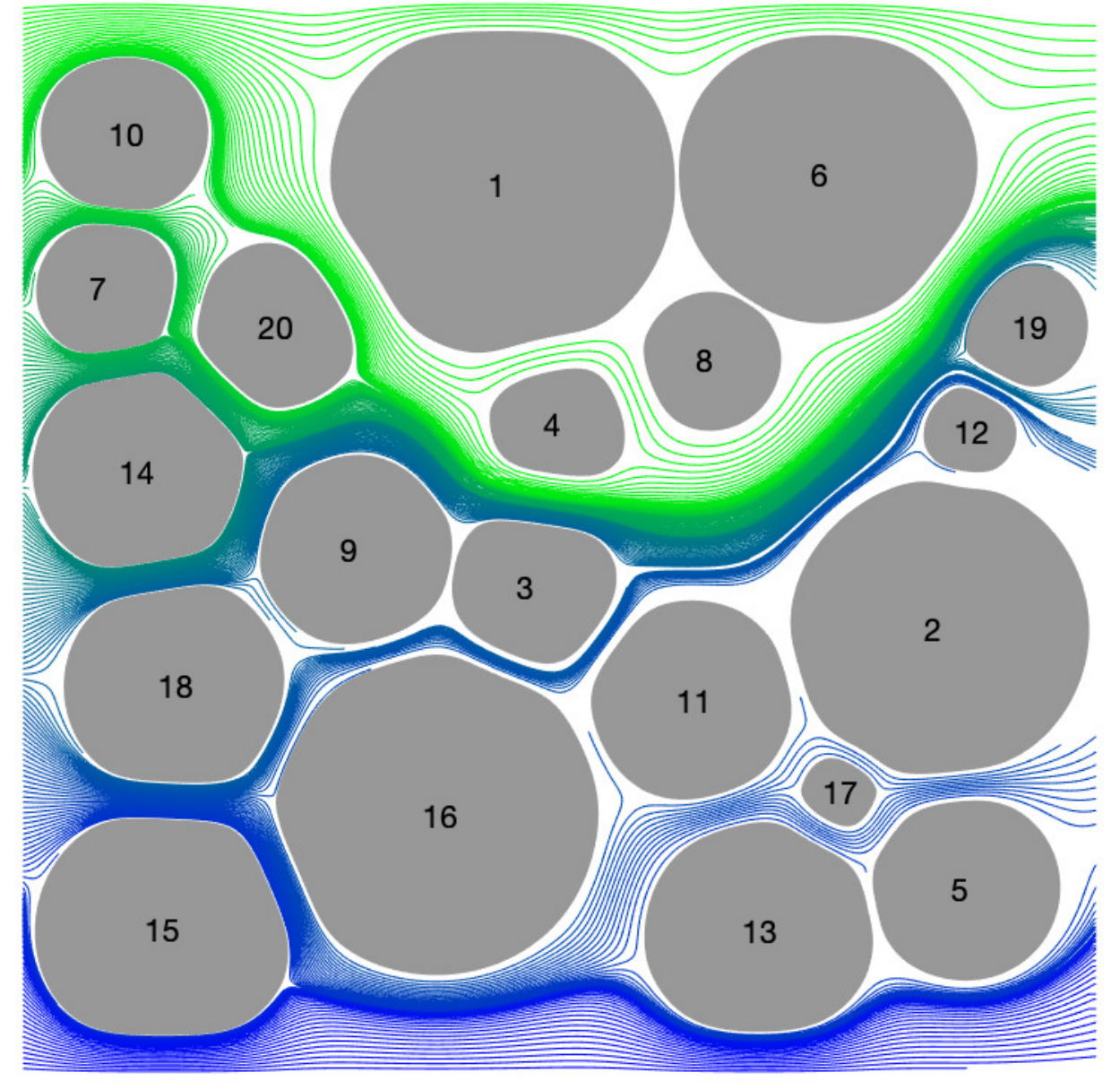}
\includegraphics[width = 0.32 \textwidth]{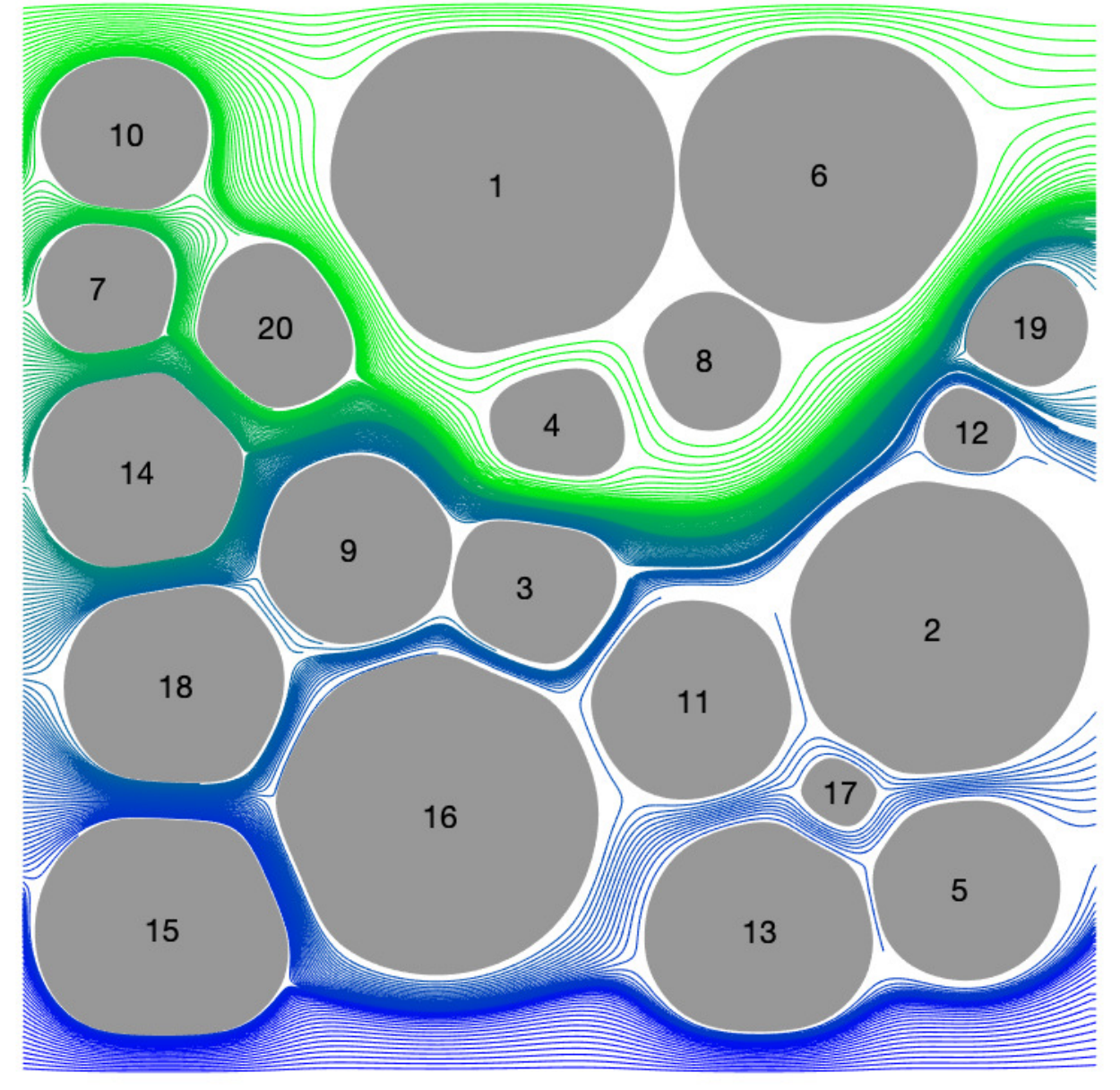}
\includegraphics[width = 0.32 \textwidth]{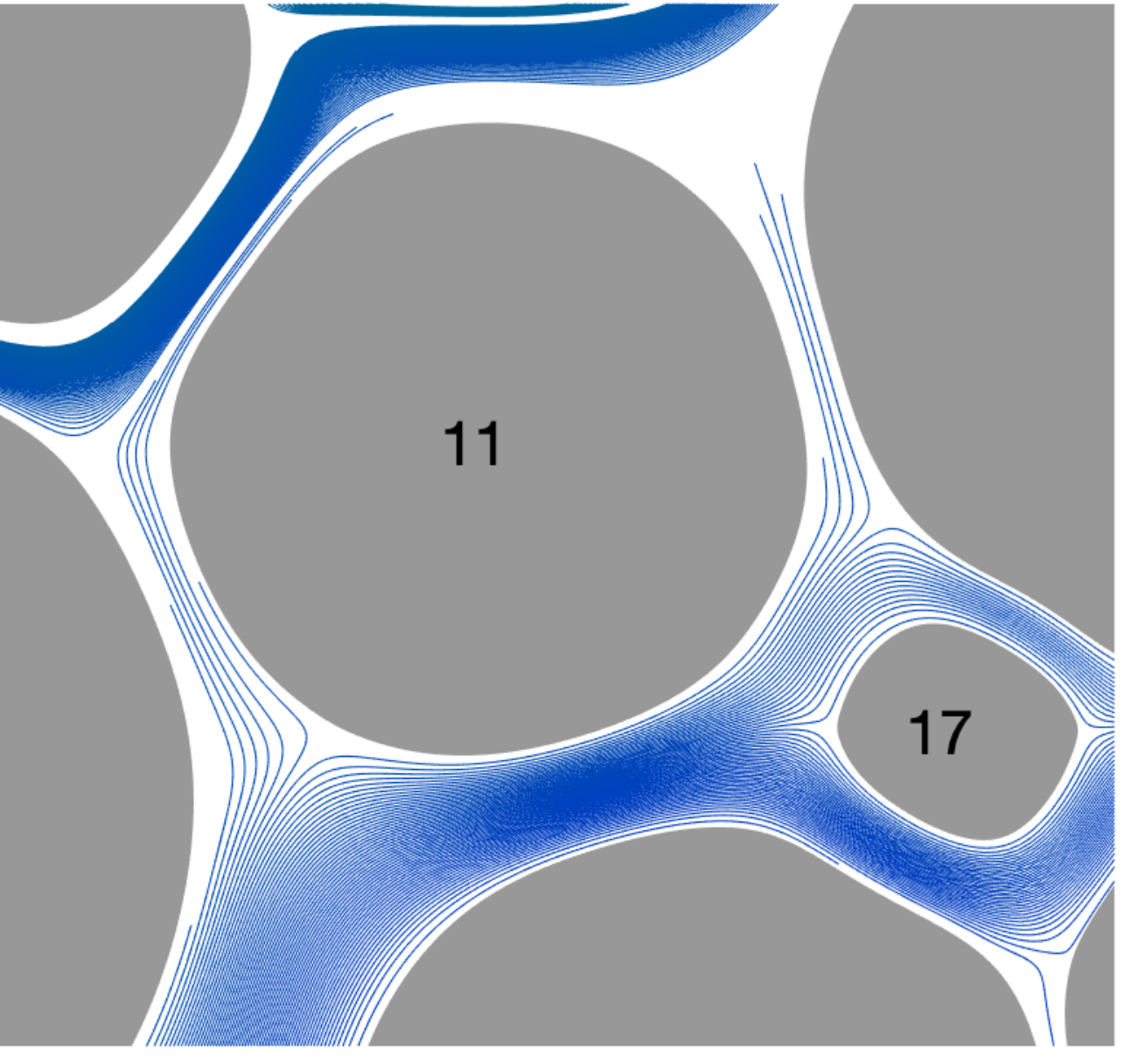}
\caption{\label{fig:Eroding20tracer} 200 streamlines in the second
geometry from figure~\ref{fig:Eroding20vort}. The streamlines are
initially equispaced at $(-1,y)$, where $y \in (-1,1)$. The first five
snapshots are equispaced in time.  The bottom right frame is a
magnification of the fifth snapshot, but with additional streamlines.}
\end{center}
\end{figure}

In figure~\ref{fig:Eroding20vort}, we observe that erosion creates a
network of channels from the inlet to the outlet where the velocity and
vorticity, and therefore erosion rate, are much larger relative to other
regions.  These channels can be further visualized  with the
streamlines.  In figure~\ref{fig:Eroding20tracer}, we freeze the
geometry at the second snapshot from figure~\ref{fig:Eroding20vort} and
plot 200 streamlines that are initially equispaced along $(-1,y)$, where
$y \in (-1,1)$.  The streamlines are shown at five different times, and
the final plot is a zoom in of the lower right quadrant of the fifth
time step, but with additional streamlines.  Since we use a high-order
quadrature rule and time stepping method, we resolve streamlines that
come very close to the eroded bodies.  There are three clear regions
where the streamlines are most concentrated, corresponding to the
regions of highest velocity.  Two of these regions are located between
the bodies and the solid walls at $y=\pm 1$, and the third cuts through
the porous region with the upper part of the channel formed by bodies 1,
4, 6, and 8.  Since the flow is fastest in these regions, the shearing
is largest, and this causes the channels to continue to open fastest as
observed in figure~\ref{fig:Eroding20vort}.

\begin{figure}
\begin{subfigure}[b]{0.45\textwidth}
\begin{subfigure}[b]{\textwidth}
\includegraphics*[width =\linewidth]{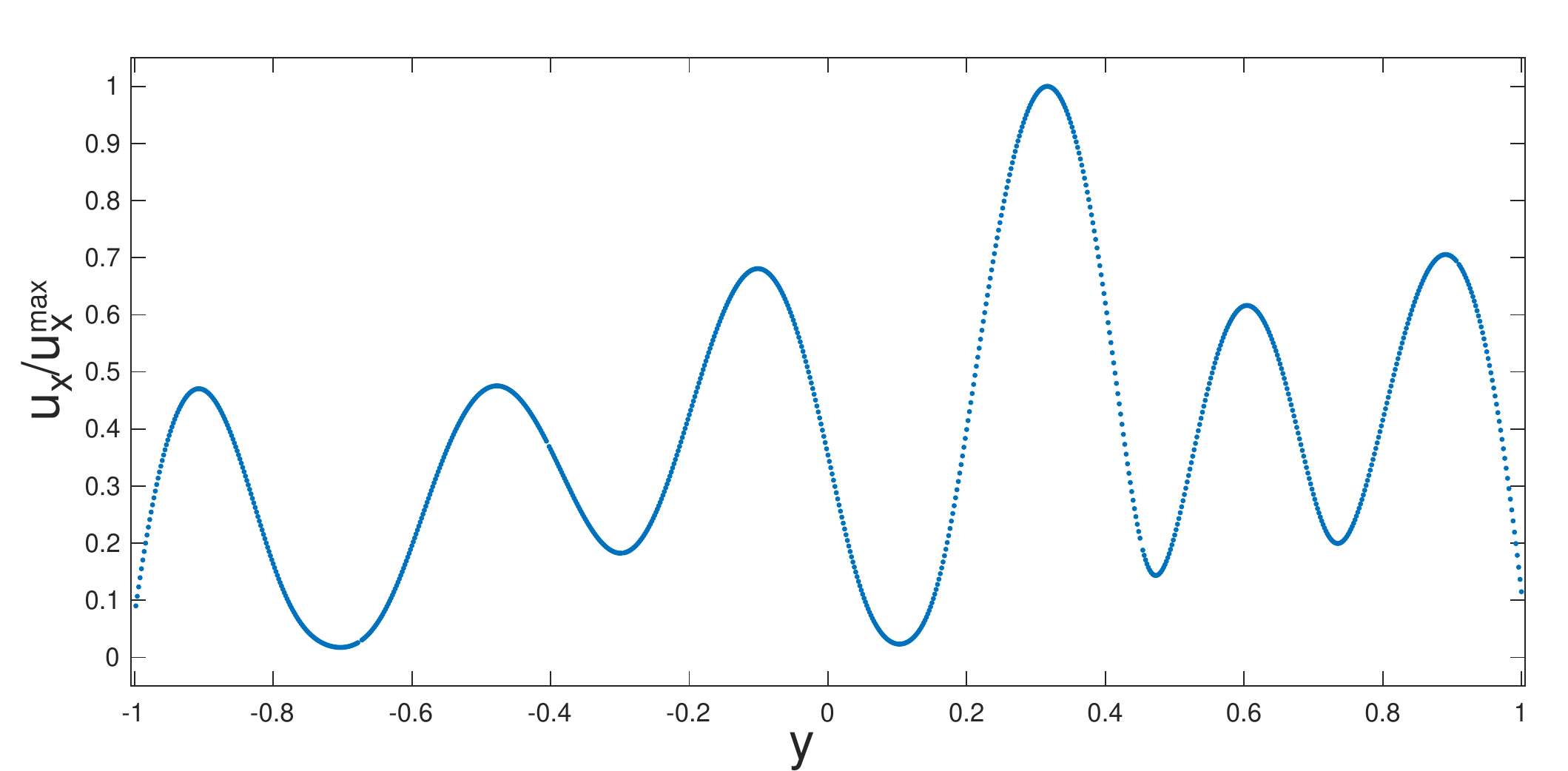}
\caption{}
\end{subfigure}
\begin{subfigure}[b]{\textwidth}
\includegraphics*[width =0.97\linewidth]{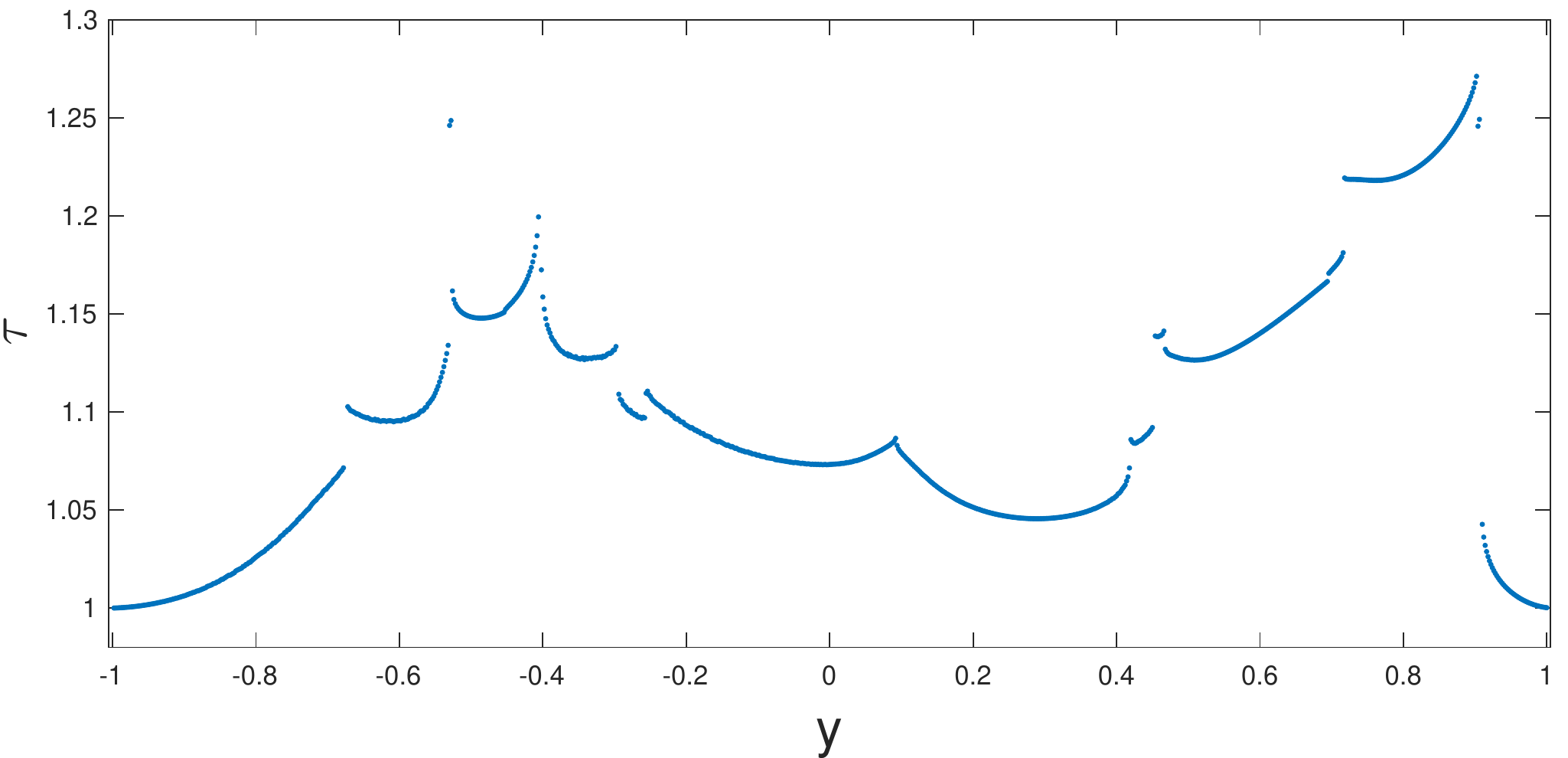}
\caption{}
\end{subfigure}
\end{subfigure}
\begin{subfigure}[b]{0.5\textwidth}
\includegraphics*[width =\linewidth]{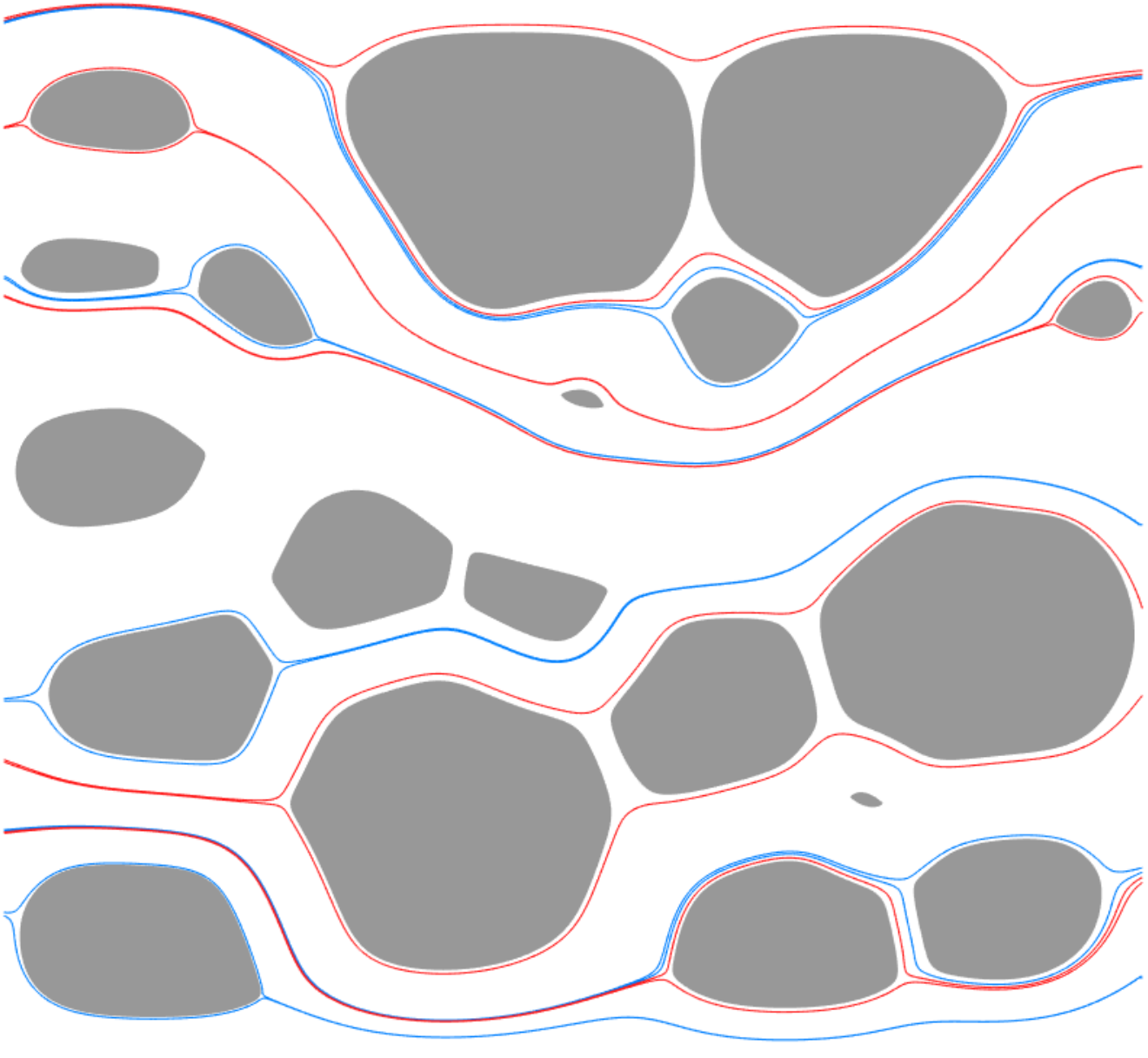}
\caption{}
\end{subfigure}
\caption{\label{fig:Eroding20tort} The local tortuosity of a porous
geometry initialized with 20 grains after eroding to a porosity of
62.9\%.  (a) The $x$-component of the velocity at the inlet,
$u_1(-1,y)$, normalized by its maximum velocity of $2.98 \times
10^{-3}$. (b) The local tortuosity $\tau(y)$ on the cross section $x =
-1$. (c) The streamlines resulting in the ten largest differences of
local tortuosity between neighboring streamlines.  Neighboring
streamlines have the same color.}
\end{figure}

Next, we use the $N_p = 1000$~\citep{bel-sal-rin1992} streamlines to
compute the tortuosity of the eroding geometry.  To compute the
tortuosity, we require the velocity at the inlet $x=-1$.  These
normalized velocities are plotted in figure~\ref{fig:Eroding20tort}(a)
for the eroded geometry at porosity $\phi = 62.9\%$
(figure~\ref{fig:Eroding20tort}(c)).  The velocities are similar those
of~\citet[see figure 4(a)]{mat-kha-koz2008}, except that our
cross-section, by construction, does not cut through any of the grains.
Next, in figure~\ref{fig:Eroding20tort}(b), we plot the local
tortuosity~\eqref{eqn:localTort} by calculating the relative length of
each streamline as it traverses the channel from $x=-1$ to $x=1$.  The
local tortuosity ranges from 1 to 1.27, meaning that one of the
streamlines is 27\% longer than it would have been if the grains were
absent.  The average streamline is 9.79\% longer or equivalently the
tortuosity of the geometry is $1.098$.  Again, comparing the local
tortuosity to~\citet[see figure 4(b)]{mat-kha-koz2008}, the results are
qualitatively similar. However, since our initial cross-section does not
cut through the grains, the local tortuosity does not have any gaps.
Discontinuities in local tortuosity occur when nearby streamlines
diverge to circumvent a grain.  In figure~\ref{fig:Eroding20tort}(c), we
plot pairs of streamlines associated with the ten largest jumps in the
local tortuosity, with each pair of corresponding streamlines plotted in
the same color.

\begin{figure}
\begin{subfigure}[b]{0.5\textwidth}
\includegraphics*[height = 0.8\linewidth]{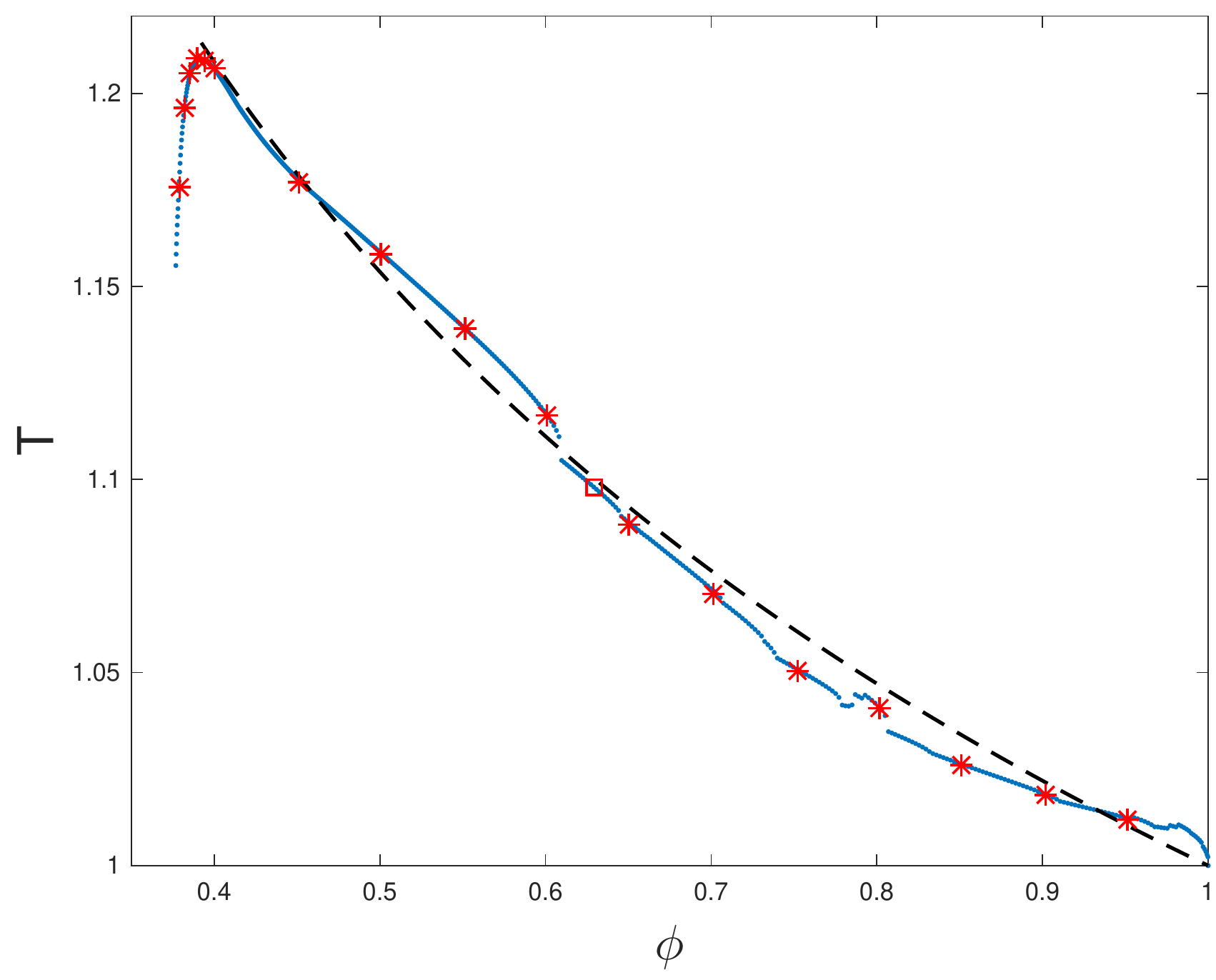}
\caption{}
\end{subfigure}
\begin{subfigure}[b]{0.5\textwidth}
\includegraphics*[height = 0.8\linewidth]{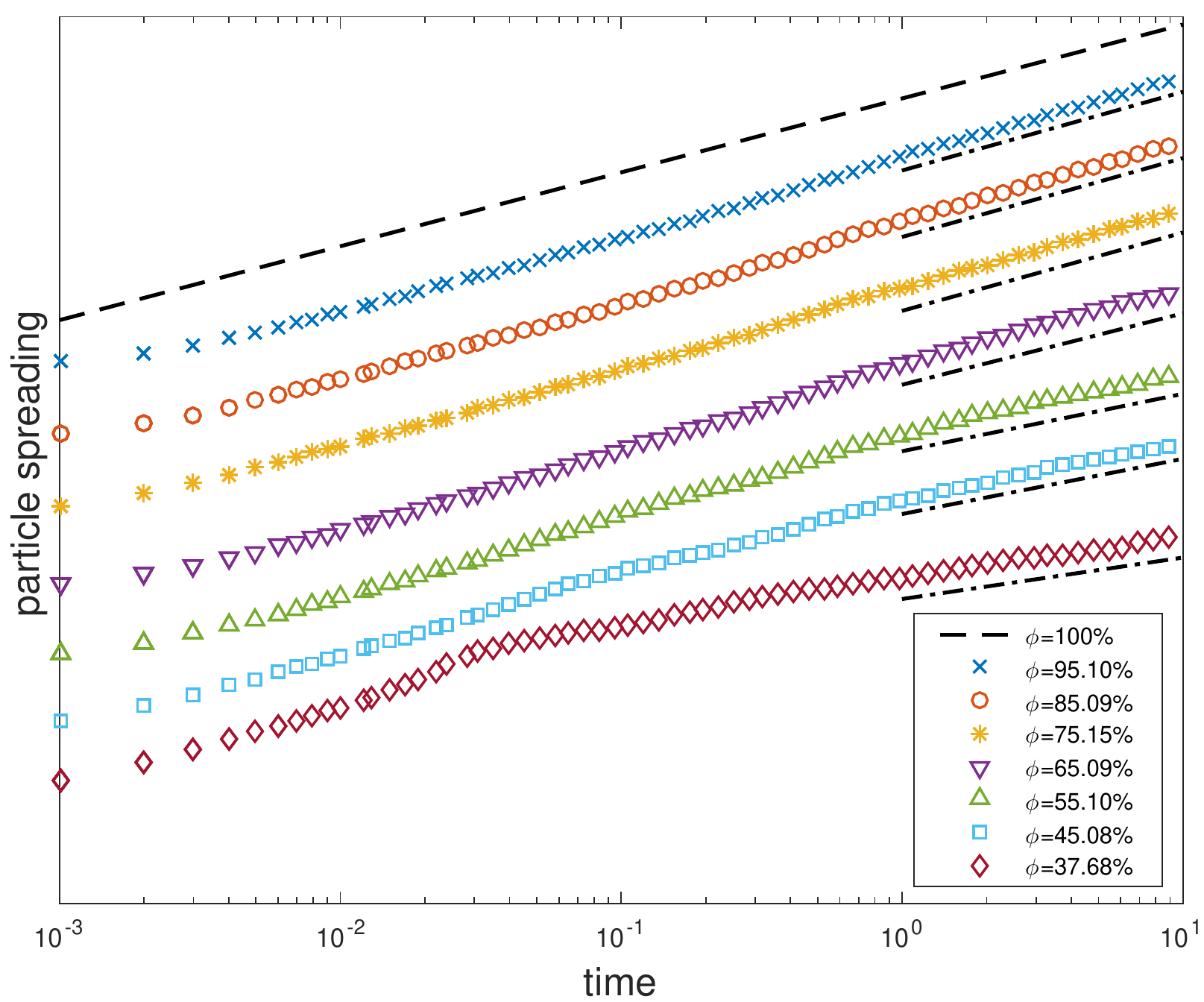}
\caption{}
\end{subfigure}
\caption{\label{fig:Eroding20Transport} (a) The tortuosity of an eroding
geometry initialized with 20 grains.  The tortuosity is calculated using
the Eulerian method~\eqref{eqn:tortuosity2} (blue dots) and Lagrangian
method~\eqref{eqn:tortuosity1} (red stars).  The red square corresponds
to the geometry in figure~\ref{fig:Eroding20tort}(c).  The dashed line
is the line of best fit $\widehat{T}(\phi)=\phi^{-p}$ with $p=0.2064$.
(b) The temporal evolution of $\sigma_\lambda$ at seven porosities.  The
dashed line has slope one and corresponds to ballistic dispersion.
Asymptotically, the spreading is super-dispersive with $\sigma_\lambda
\sim t^\alpha$, $\alpha \in (1/2,1)$.  The dashed-dotted lines of best
fit have slopes $\alpha = 1.06$ ($\phi=95.10\%$), $\alpha = 1.07$
($\phi=85.09\%$), $\alpha = 1.06$ ($\phi=75.15\%$), $\alpha = 0.97$
($\phi=65.09\%$), $\alpha = 0.78$ ($\phi=55.10\%$), $\alpha = 0.75$
($\phi=45.08\%$), and $\alpha = 0.56$ ($\phi=37.68\%$).  Values greater
than 1 result from using a least-squares fit for the tails of the
particle spreading.}
\end{figure}

In figure~\ref{fig:Eroding20Transport}(a), we plot the tortuosity as a
function of the porosity. The initial porosity is $\phi = 37.68\%$, and
the initial tortuosity is $T = 1.16$.  The tortuosity is computed with
both the length of the streamlines~\eqref{eqn:tortuosity1} (red stars)
and using the spatial average of the velocity on an Eulerian
grid~\eqref{eqn:tortuosity2} (blue marks).  The red square corresponds
to the porosity of the geometry in figure~\ref{fig:Eroding20tort}(c).
The two tortuosity formulas give similar results, and any discrepancy
can be accounted for by slow regions of recirculation and from applying
quadrature to compute the tortuosity.  As the bodies erode, wide
channels form where streamlines undergo only minor vertical deflections,
and this explains why the tortuosity eventually decreases with porosity.
We computed lines of best fit using the porosity-tortuosity
models~\eqref{eqn:tortuosityModels} and found that the power law
minimizes the error.  The  black dashed line in
figure~\ref{fig:Eroding20Transport}(a) is the line of best fit
$\widehat{T}(\phi) = \phi^{-0.2064}$ with a root-mean-square error of
$5.90 \times 10^{-3}$.  Interestingly, at the low porosities, the
tortuosity initially increases. This increase occurs because in the
absence of erosion (left plot in figure~\ref{fig:Eroding20vort}), many
of the streamlines, such as those initialized between bodies 15 \& 18,
only perform minor deflections to pass through the narrow regions,
albeit, very slowly.  However, as erosion starts to open the channels,
the streamlines deflect into the fast regions, such as the region above
body 11, and this increases the amount of vertical deflection, and
therefore the tortuosity. While this increase in tortuosity is
interesting, in the next two examples we will see that the tortuosity
does not initially increase.

We next use streamlines to investigate the temporal evolution of the
particle spreading $\sigma_\lambda$.  The spreading is computed for
seven geometries of different porosities that are formed during the
erosion process (figure~\ref{fig:Eroding20Transport}(b)).  So that the
spreading reaches a statistical equilibrium, we use the reinsertion
algorithm described in section~\ref{sec:dispersion} to form sufficiently
long trajectories.  For all the reported porosities, the particle
dispersion exhibits two distinct power law regimes.  Initially, the
dispersion is ballistic ($\sigma_\lambda \sim t$) since individual fluid
particles have not yet explored enough space to significantly alter
their velocity.  However, once the particles have been subjected to a
range of velocities, their dispersion slows, and we observe
super-dispersive ($\sigma_\lambda \sim t^\alpha$, $\alpha \in (1/2,1)$)
behavior over at least one order of magnitude in time.  Before any
erosion takes place, the anomalous dispersion coefficient is $\alpha =
0.56$.  Then, as the grains begin to erode, the dispersion rate grows
towards the ballistic regime that occurs in the absence of grains.  The
monotonic increase in dispersion with respect to the porosity is
explained by the onset of channels where many tracers experience less
variability in their velocities.

\subsection{20 Bodies at a Low Porosity}
\begin{figure}
\begin{center}
  \includegraphics[height=0.227\textwidth]{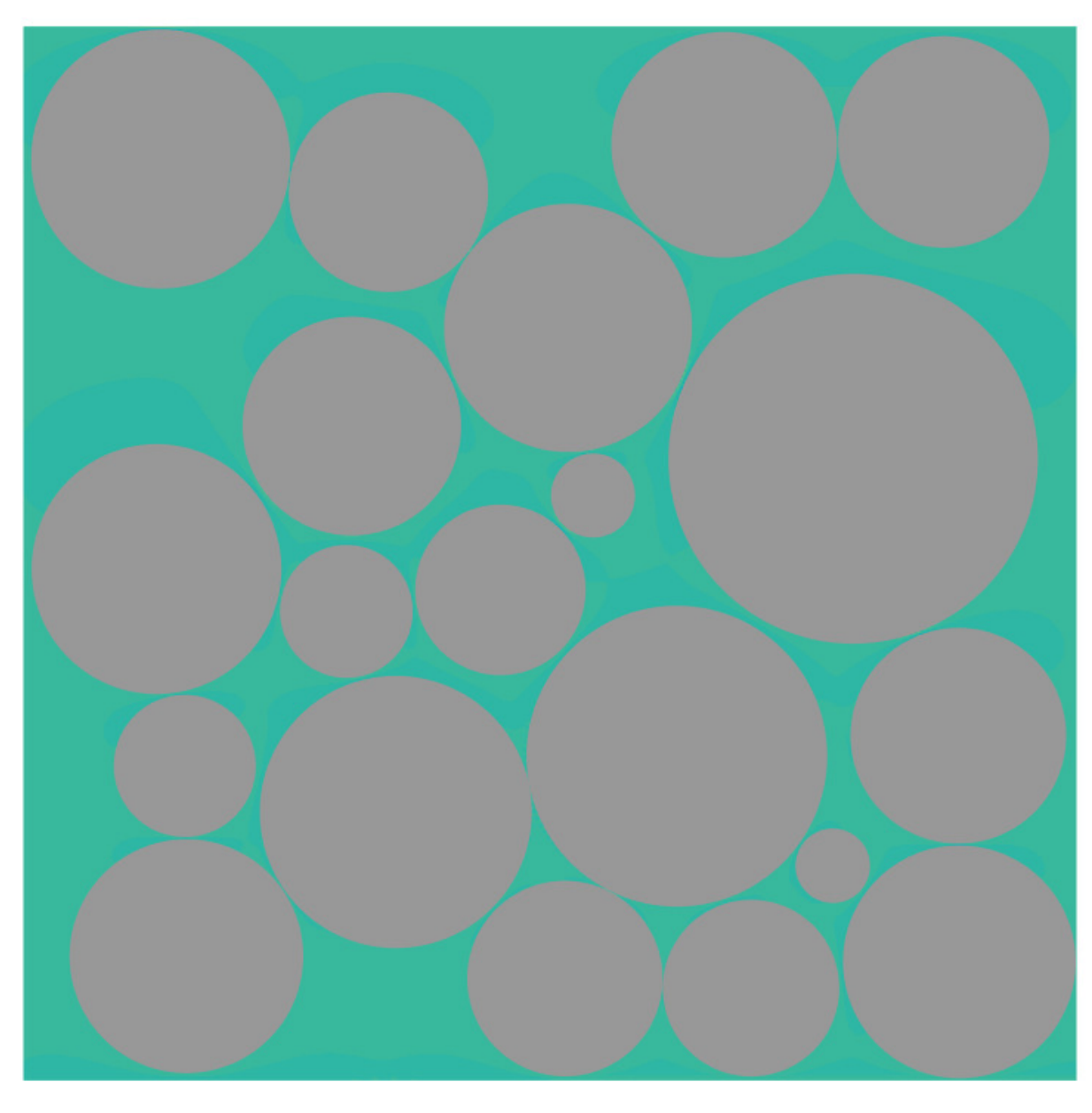}
  \includegraphics[height=0.227\textwidth]{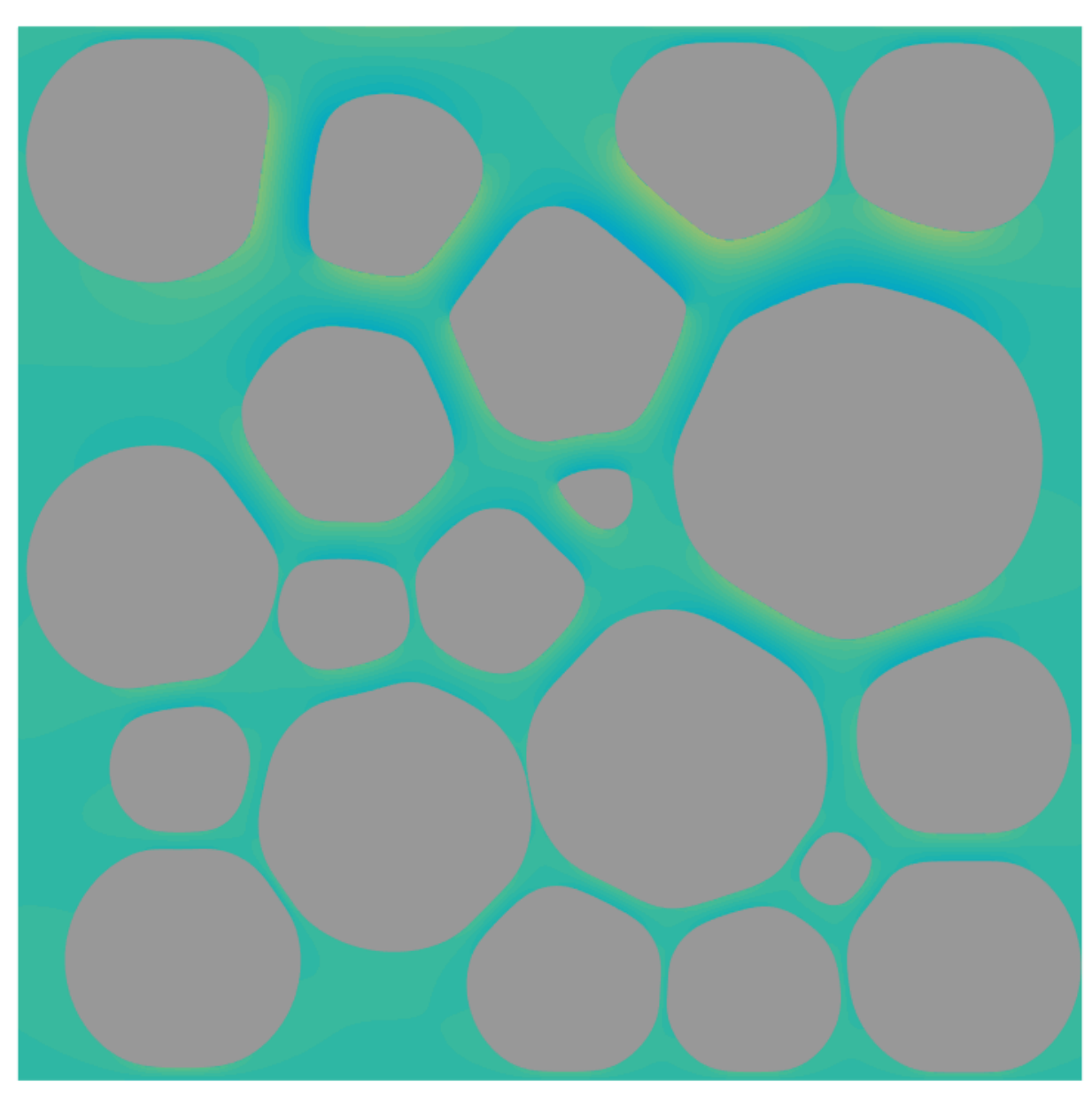}
  \includegraphics[height=0.227\textwidth]{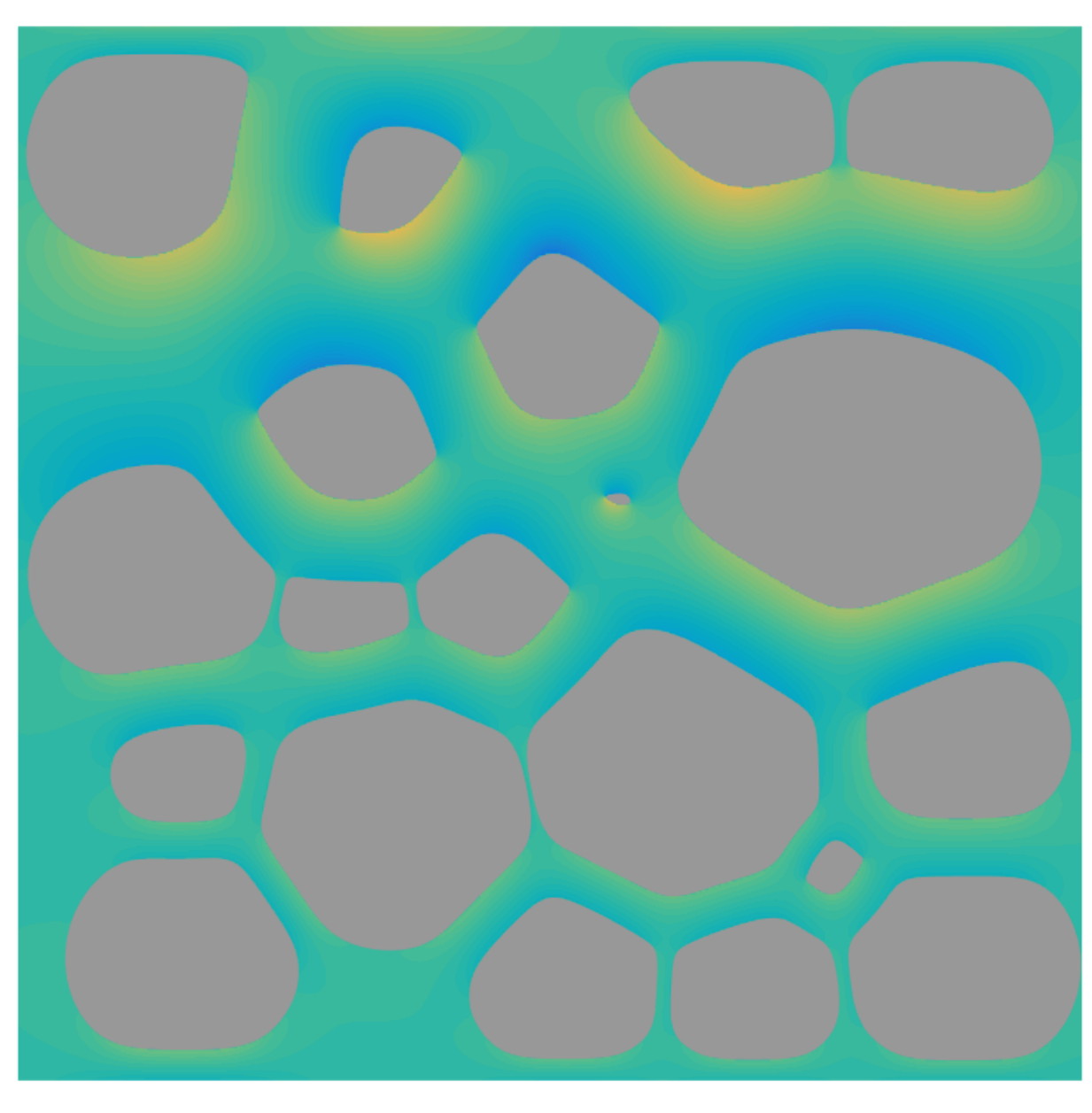}
  \includegraphics[height=0.227\textwidth]{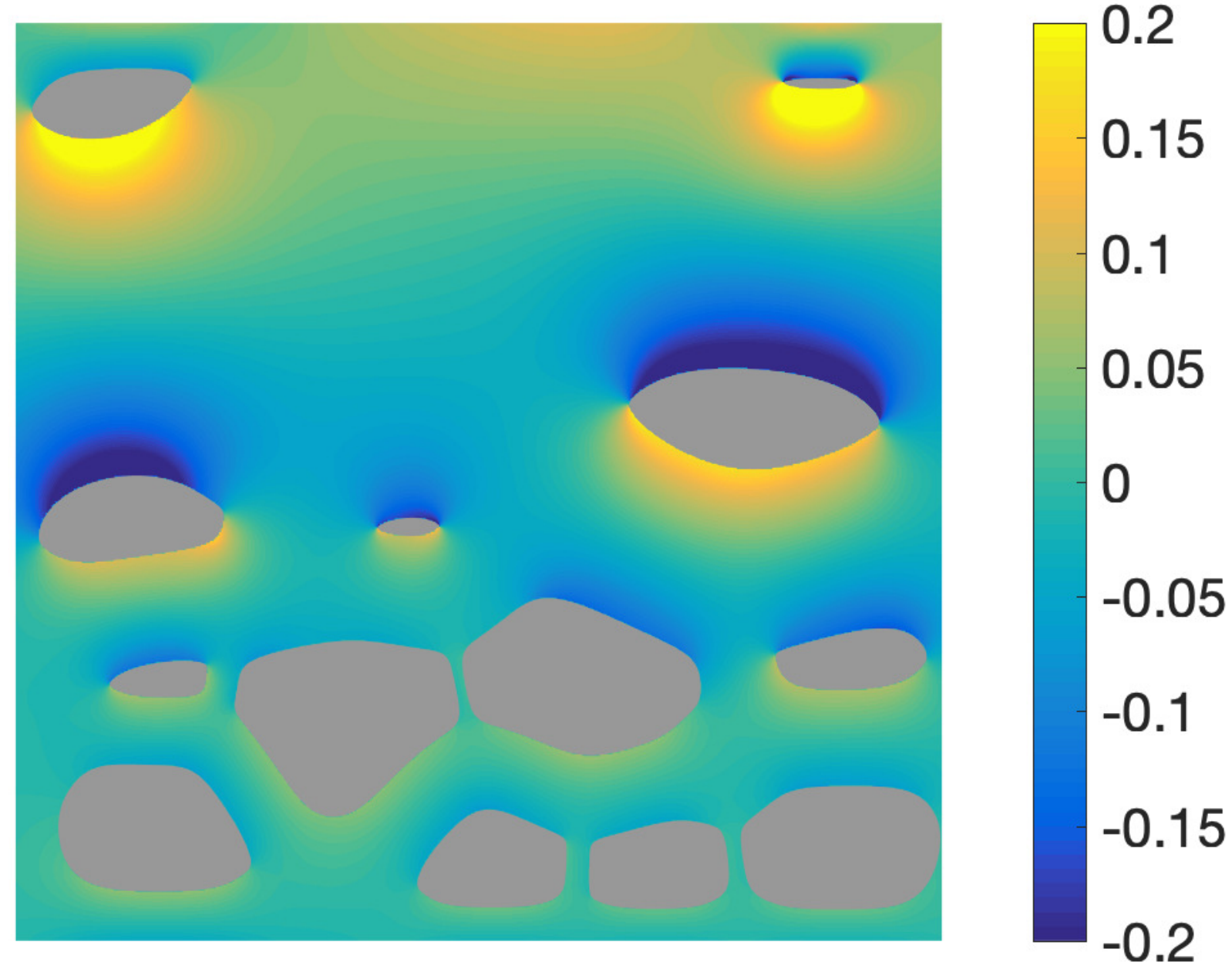}
\end{center}
\caption{\label{fig:ErodingLow20vort} 20 bodies eroding in a
Hagen-Poiseuille flow. The snapshots are equispaced in time, and the
color is the fluid vorticity. In addition to the channels that develop
between the bodies and the solid walls, erosion leads to two main
channels through the geometry---one in the top half and one near the
middle.}
\end{figure}

We consider a second example with 20 eroding bodies, but with a smaller
initial porosity.  In figure~\ref{fig:ErodingLow20vort}, we plot the
eroding geometry and vorticity at four evenly spaced instances in time.
Initially, the smallest distance between pairs of bodies is $3.29 \times
10^{-4}$, and the smallest distance between the bodies and solid wall is
$4.50 \times 10^{-3}$.  At these distances, a resolution of
approximately $N_\iin = 27,000$ and $N_\out = 18,000$ discretization
points is required to satisfy the $5h$ threshold needed for the
trapezoid rule to achieve machine precision.

\begin{figure}
\begin{subfigure}[b]{0.5\textwidth}
\includegraphics*[height = 0.8\linewidth]{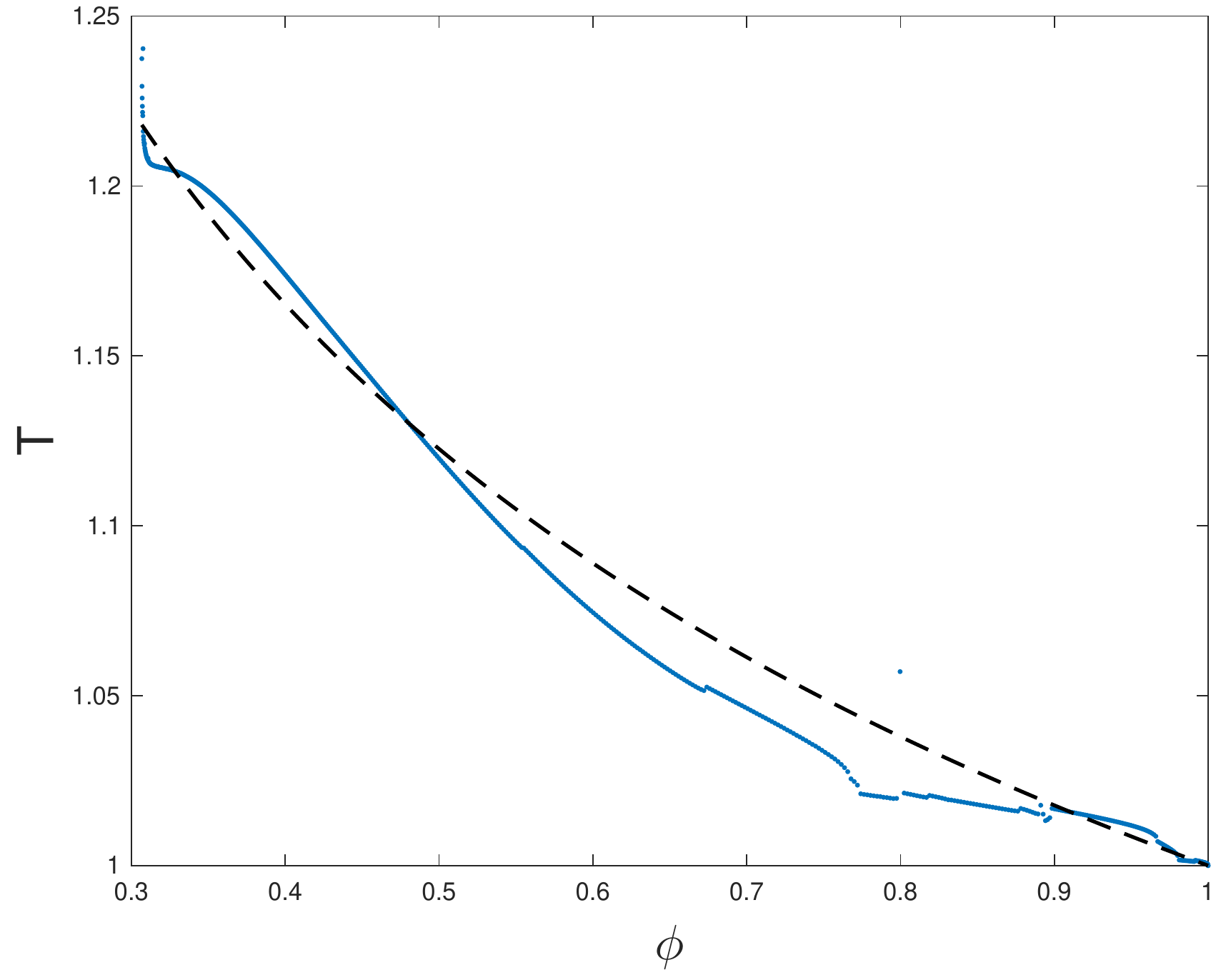}
\caption{}
\end{subfigure}
\begin{subfigure}[b]{0.5\textwidth}
\includegraphics*[height=0.8\linewidth]{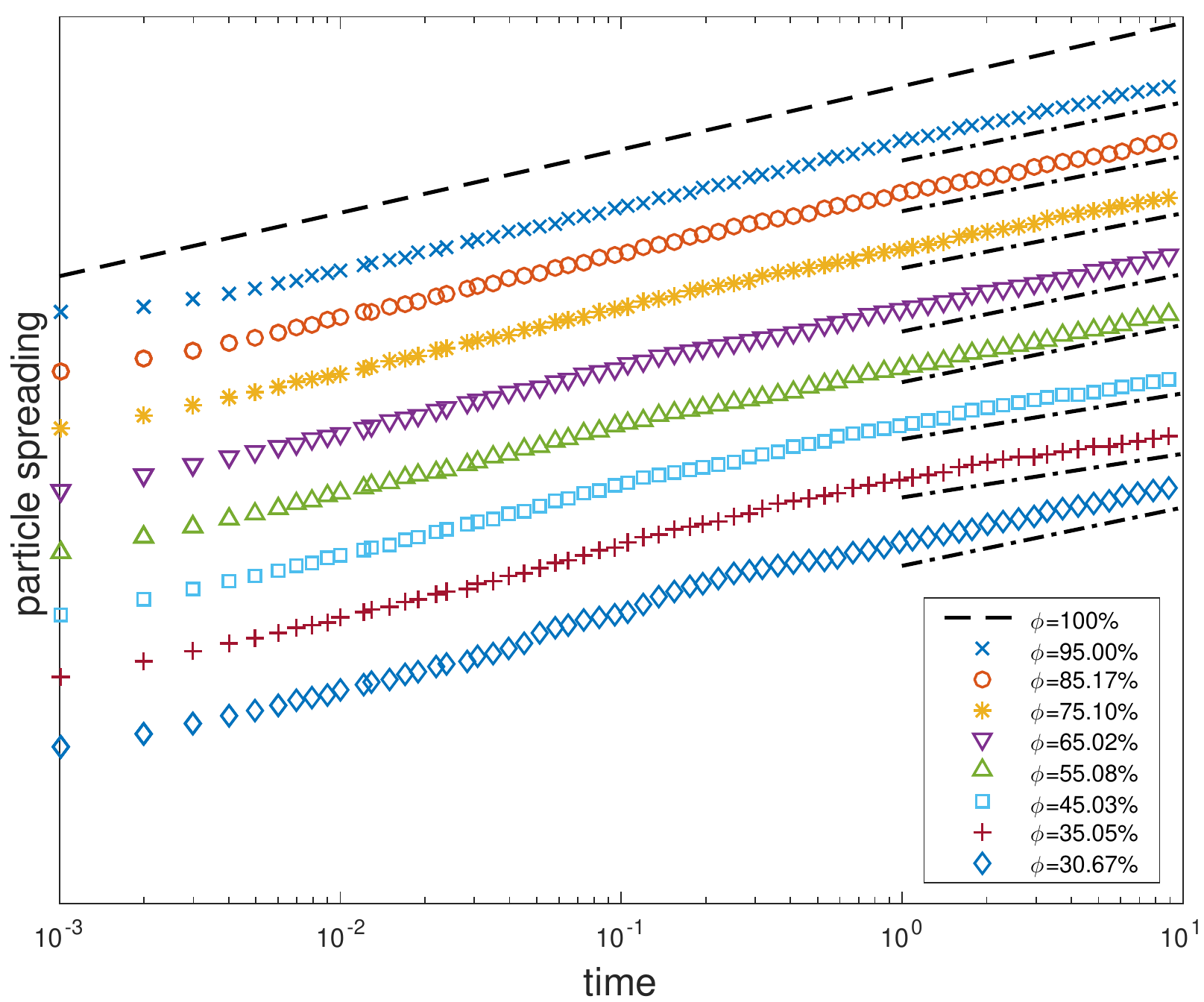}
\caption{}
\end{subfigure}
\caption{\label{fig:ErodingLow20Transport} (a) The tortuosity  of an
eroding geometry initialized with 20 grains.  When compared to the last
example, the bodies are initially much closer together, and the porosity
is smaller.  The dashed line is the line of best fit
$\widehat{T}(\phi)=\phi^{-p}$ with $p=0.1669$.  (b) The temporal
evolution of $\sigma_\lambda$ at eight porosities.  The dashed line has
slope one and corresponds to ballistic motion. Asymptotically, the
spreading is super-dispersive with $\sigma_\lambda \sim t^{\alpha}$,
$\alpha \in (1/2,1)$.  The dashed-dotted lines of best fit have slopes
$0.92$ ($\phi=95.00\%$), $0.87$ ($\phi=85.17\%$), $0.87$
($\phi=75.10\%$), $0.91$ ($\phi=65.02\%$), $0.91$ ($\phi=55.08\%$),
$0.72$ ($\phi=45.03\%$), $0.69$ ($\phi=35.05\%$), and $0.92$
($\phi=30.67\%$).}
\end{figure}

We compute the tortuosity using the Eulerian
method~\eqref{eqn:tortuosity2} at each time step. The initial porosity
is $\phi = 30.67\%$ and the initial tortuosity is $T = 1.24$.  In
figure~\ref{fig:ErodingLow20Transport}(a), we plot the tortuosity with
respect to the porosity (blue) and the line of best fit (black) using
the power law $\widehat{T}(\phi) = \phi^{-0.1669}$. This model
outperforms the other three models in
equation~\eqref{eqn:tortuosityModels}, and its root-mean-squared error
is $1.13 \times 10^{-2}$.  As grains erode, there is an increase in the
number of streamlines that take a nearly direct path through the
geometry, and this decreases the tortuosity. However, the channelization
effect of erosion results in an increase in the tortuosity since the
length of many of the streamlines increases when they deflect from a
high porosity region (low pressure) to a low porosity region (high
pressure). For this example, we see that the net effect is a decrease in
the tortuosity for all time.

In figure~\ref{fig:ErodingLow20Transport}(b), we plot the temporal
evolution of the particle spreading $\sigma_\lambda$.  As in the last
example, we analyze the spreading at several different porosities and we
use the reinsertion algorithm described in section~\ref{sec:dispersion}.
For all the porosities, the dispersion is much closer to ballistic when
compared to the results in figure~\ref{fig:Eroding20Transport}.
However, there are still clear transitions from ballistic dynamics to
asymptotic super-dispersive spreading.  In contrast to the higher
porosity initial condition (section~\ref{sec:Eroding20}), at early times
the erosion results in a decrease in the dispersion rate. In particular,
after the first 5\% of the bodies have eroded, the particle spreading
transitions from $\sigma_\lambda \sim t^{0.92}$ to $\sigma_\lambda \sim
t^{0.69}$.  To explain this behavior, recall that anomalous dispersion
is caused by tracers spending time in both the fast and slow regimes.
Since the initial configuration has a reasonably uniform velocity (see
figure~\ref{fig:ErodingLow20vort}), albeit a small one, the dispersion
is nearly ballistic. However, as the geometry erodes, the flow becomes
more intermittent, and this results in an increased anomalous dispersion
rate~\citep{dea-leb-den-tar-bol-dav2013}.  Then, as the bodies continue
to erode, the geometry channelizes, and most tracers are transported
with a large velocity through the channels, again resulting in a nearly
ballistic motion~\citep{sie-ili-pri-riv-gua2019}.

\subsection{100 eroding bodies}
\label{sec:Eroding100}
\begin{figure}
\includegraphics[height =0.227\linewidth]{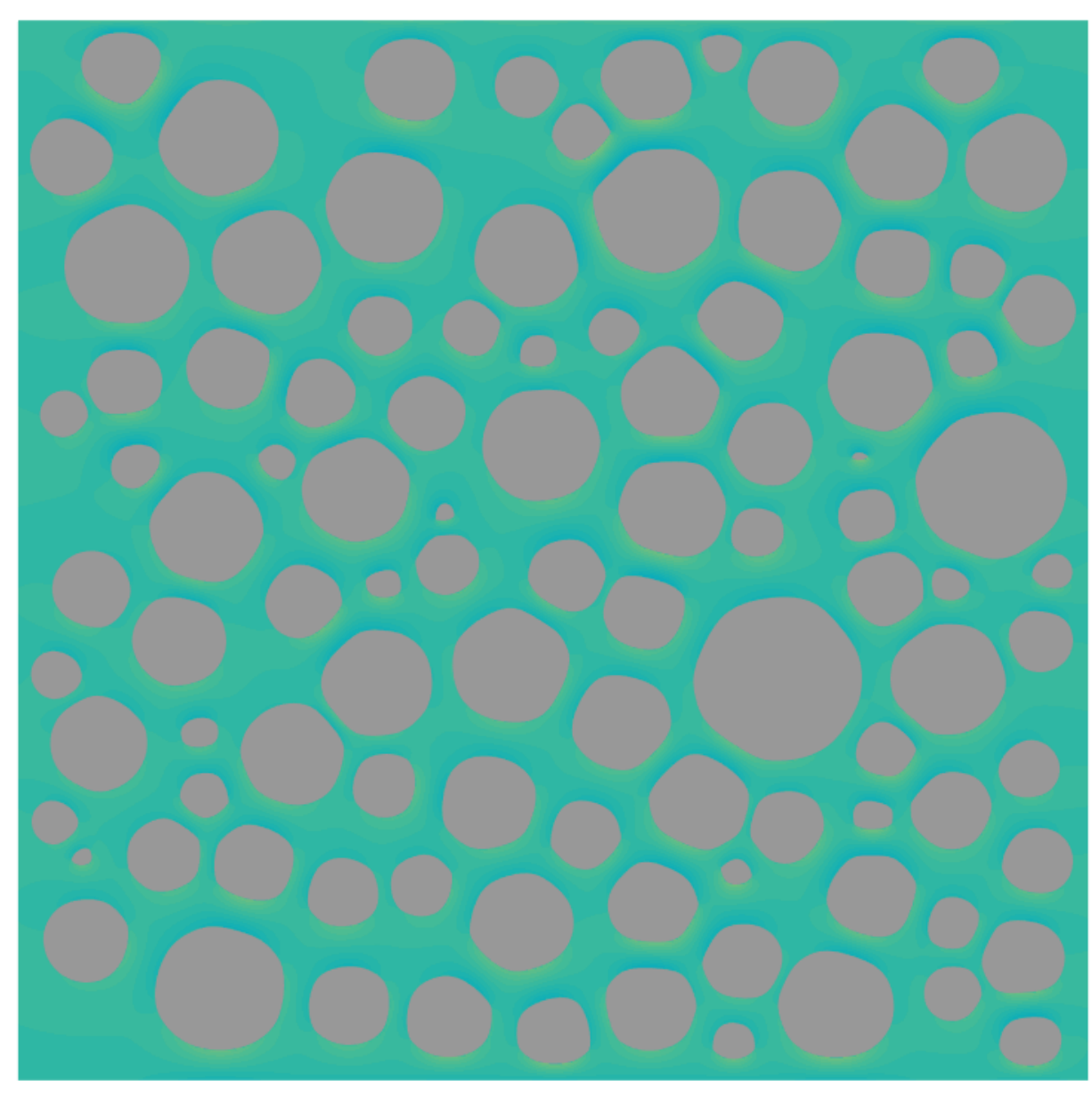}
\includegraphics[height =0.227\linewidth]{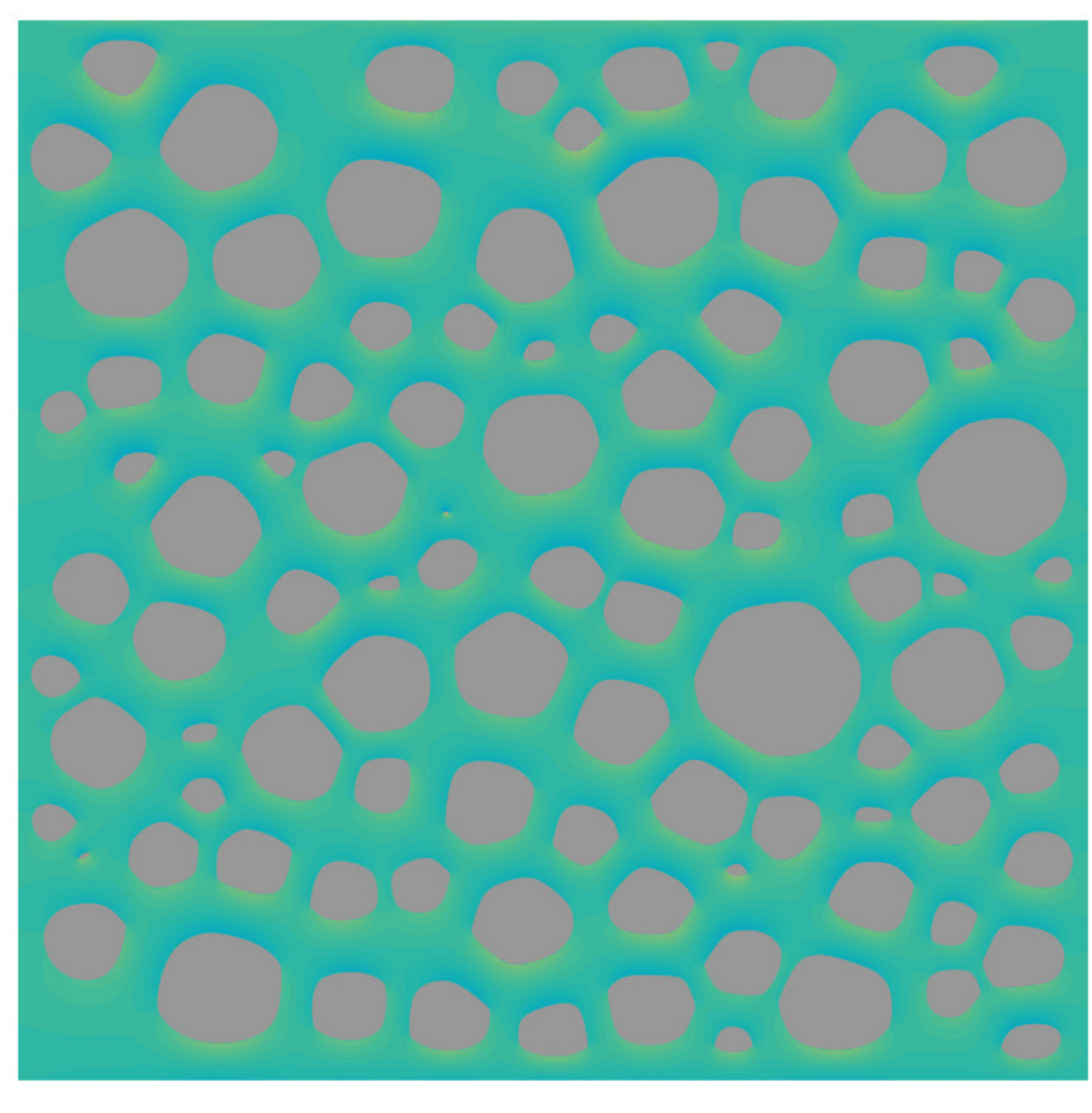}
\includegraphics[height =0.227\linewidth]{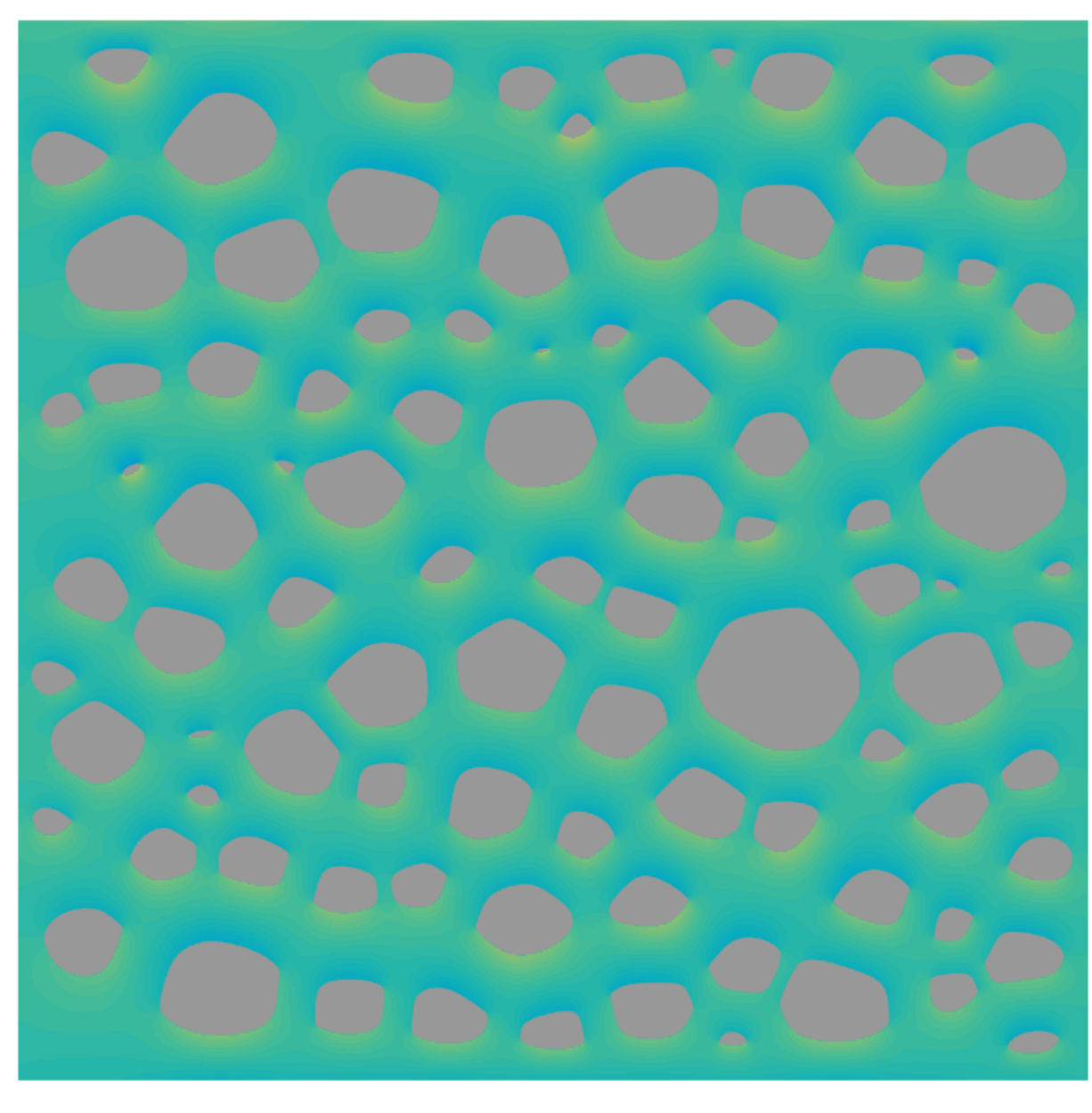}
\includegraphics[height =0.227\linewidth]{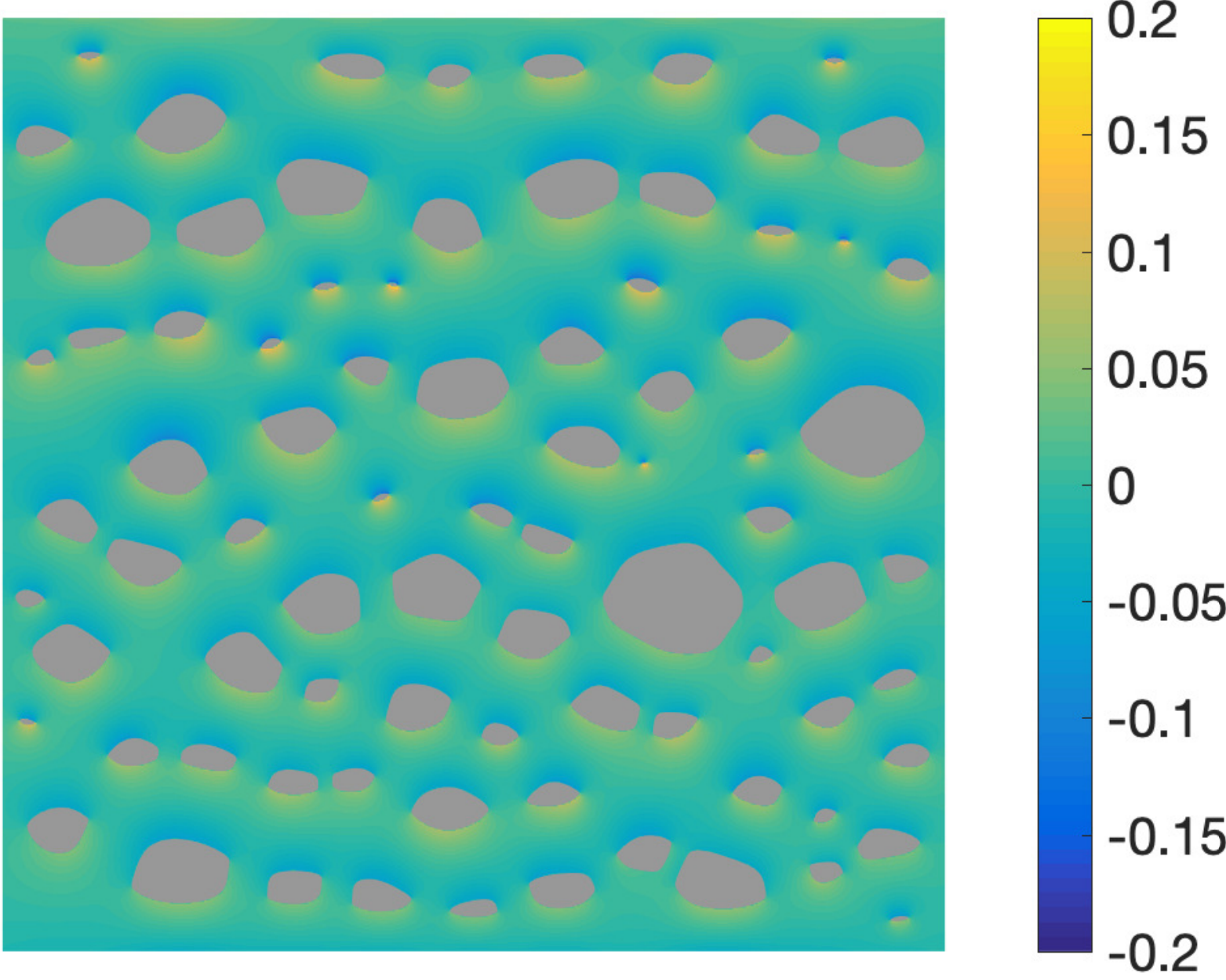}
\caption{\label{fig:Eroding100vort} The erosion of 100 nearly touching
grains in a Hagen-Poiseuille flow. The four snapshots are evenly spaced
in time, and the color is the fluid vorticity.  Because of the large
number of bodies, erosion creates many channels connecting the inlet to
the outlet.}
\end{figure}

\begin{figure}
\begin{subfigure}[b]{0.45\textwidth}
\begin{subfigure}[b]{\textwidth}
\includegraphics*[width =\linewidth]{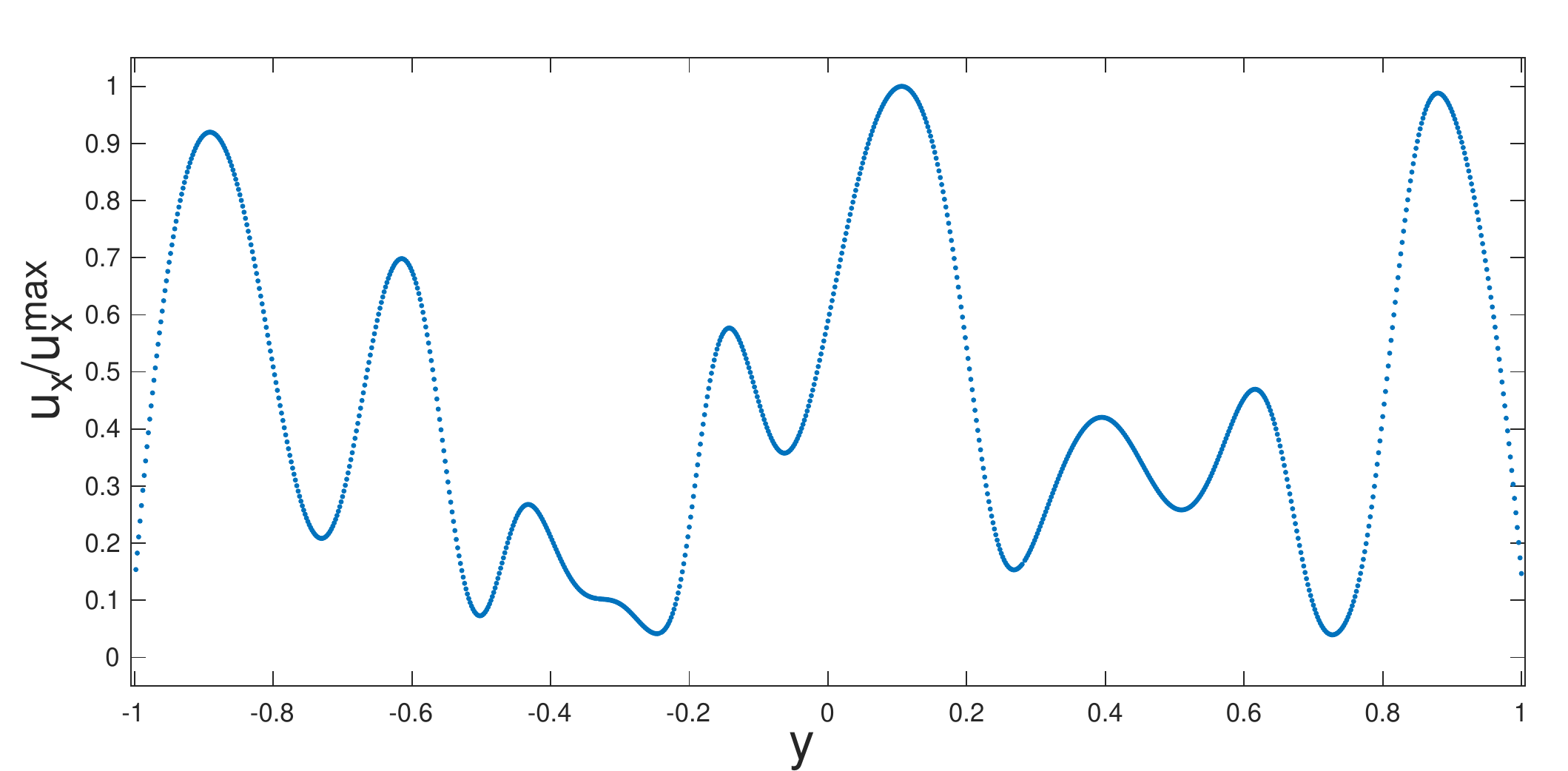}
\caption{}
\end{subfigure}
\begin{subfigure}[b]{\textwidth}
\includegraphics*[width =\linewidth]{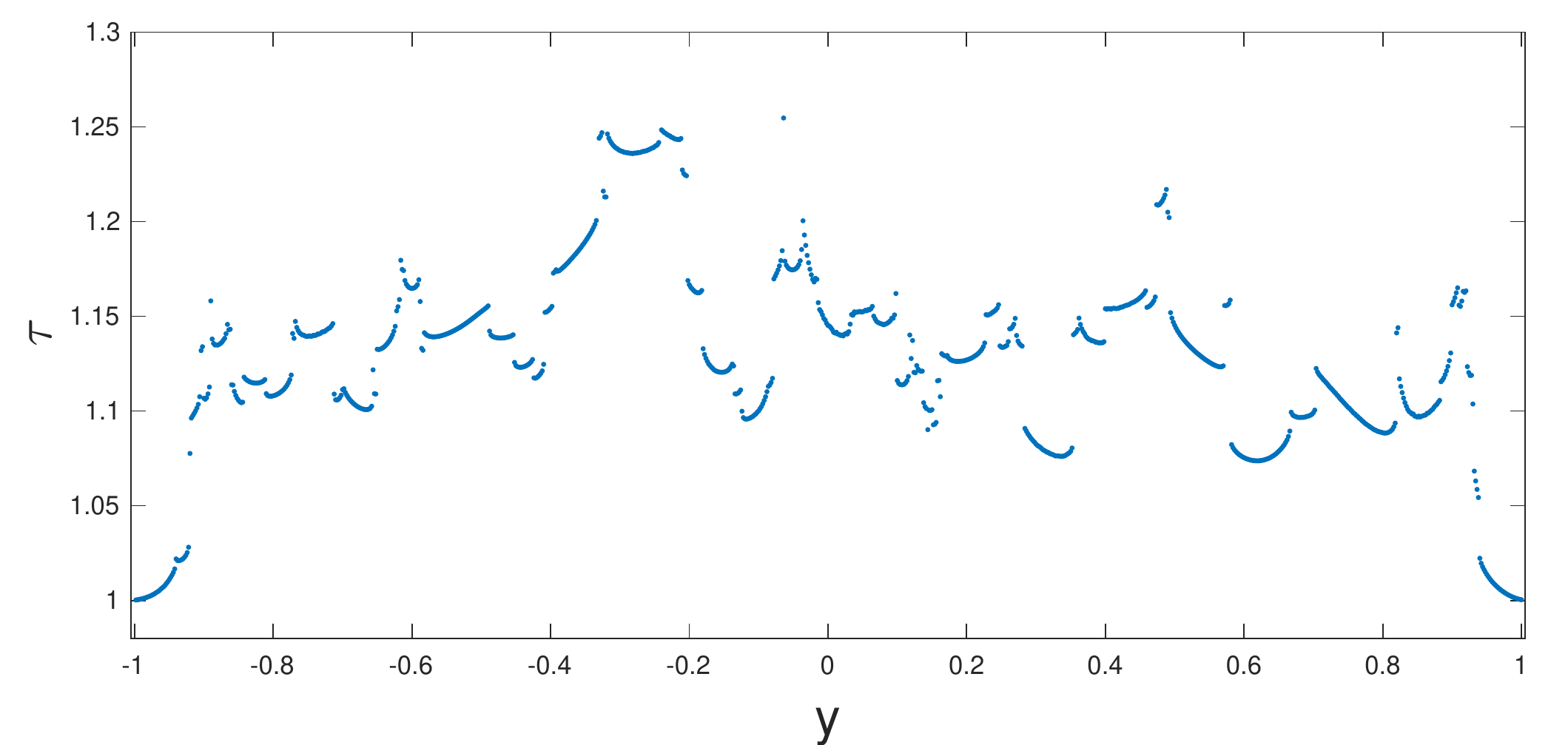}
\caption{}
\end{subfigure}
\end{subfigure}
\begin{subfigure}[b]{0.5\textwidth}
\includegraphics[width = \textwidth]{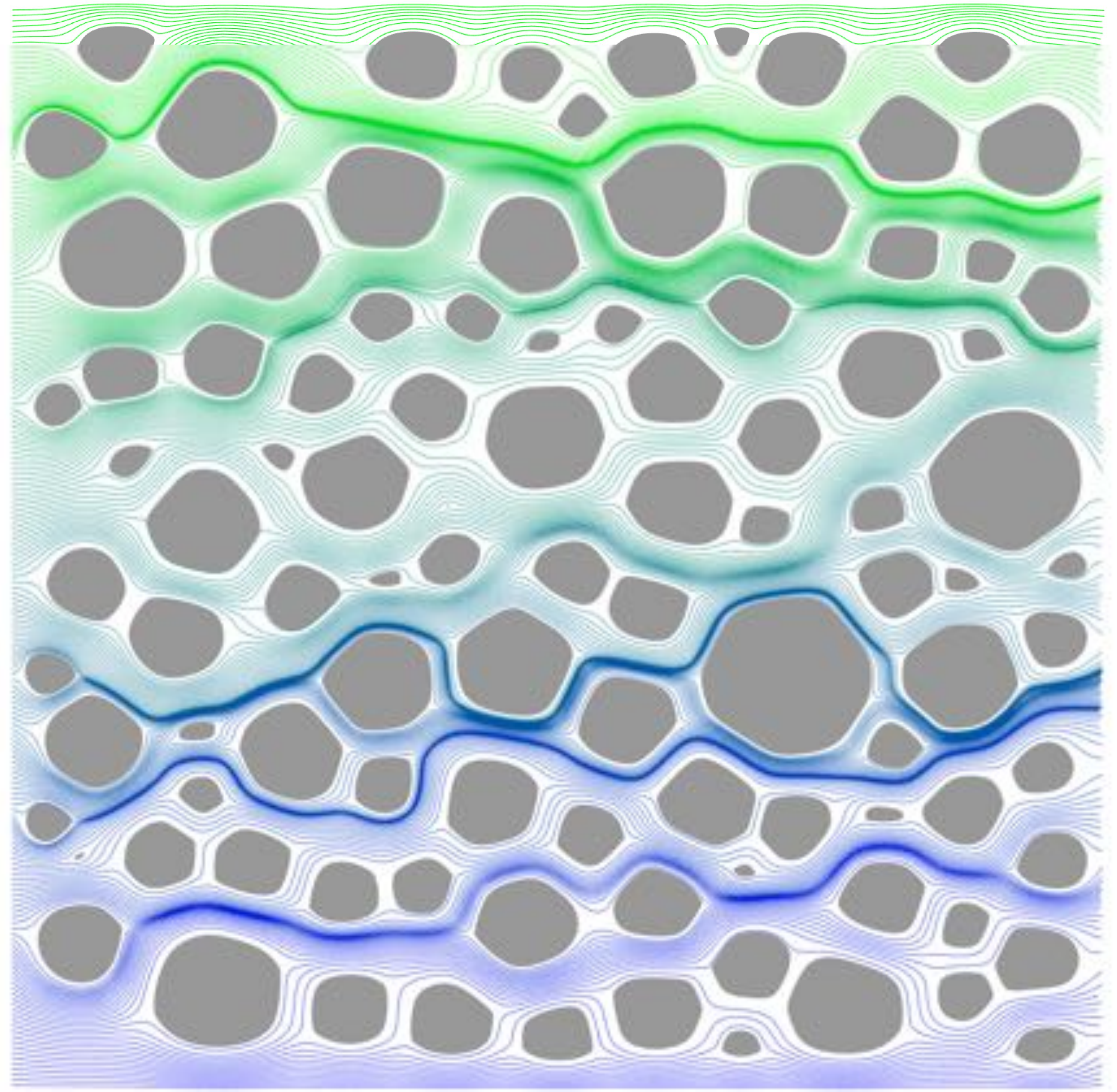}
\caption{}
\end{subfigure}
\caption{\label{fig:Eroding100tort} The local tortuosity of a porous
geometry initialized with 100 grains after eroding to a porosity of
62.98\%.  (a) The $x$-component of the velocity at the inlet, $u_1(-1,
y)$, normalized by its maximum velocity $u_{max}=3.90 \times 10^{-4}$.
Note that this maximum velocity is about a order of magnitude smaller
than the 20 body example in figure~\ref{fig:Eroding20tort}.  (b) The
local tortuosity $\tau(y)$ on the cross section $x = -1$. Compared to
figure~\ref{fig:Eroding20tort}, this example has more small bodies, and
this results in more discontinuities in the local tortuosity.  (c) The
trajectories of 200 tracers initialized at $x = -1$.}
\end{figure}

As a final example, we consider 100 eroding bodies with an initial
porosity near 50\%.  Snapshots of the configurations and vorticity are
in figure~\ref{fig:Eroding100vort}. We compute the tortuosity using both
the Lagrangian~\eqref{eqn:tortuosity1} and Eulerian
methods~\eqref{eqn:tortuosity2}. Therefore, we compute and plot the
normalized velocity at $N_p = 1000$ points along the inlet $x=-1$ in
figure~\ref{fig:Eroding100tort}(a) for the eroded geometry at porosity
$\phi = 62.98\%$ (figure~\ref{fig:Eroding100tort}(c)).  The initial
velocity of the tracers is qualitatively similar to the 20 body example
(figure~\ref{fig:Eroding20tort}(a)), except with additional oscillations
because of the additional grains.  In
figure~\ref{fig:Eroding100tort}(b), we plot the local tortuosity by
finding the length of the streamlines as they pass from $x=-1$ to $x=1$.
Compared to figure~\ref{fig:Eroding20tort}(b), the local tortuosity is
much more discontinuous.  These discontinuities can be explained by
examining the trajectories of tracers in
figure~\ref{fig:Eroding100tort}(c).  Here, there are many instances of
nearby streamlines that are deflected apart from one another as they
tend to a stagnation point in the flow, and this results in trajectories
with significantly different lengths.  At this porosity, one of the
tracers travels 25.5\% farther than it would have if the bodies had been
absent, and the average tracer travelled 12\% farther resulting in a
tortuosity of $T = 1.12$.

\begin{figure}
\begin{subfigure}[b]{0.5\textwidth}
\includegraphics*[height = 0.8\linewidth]{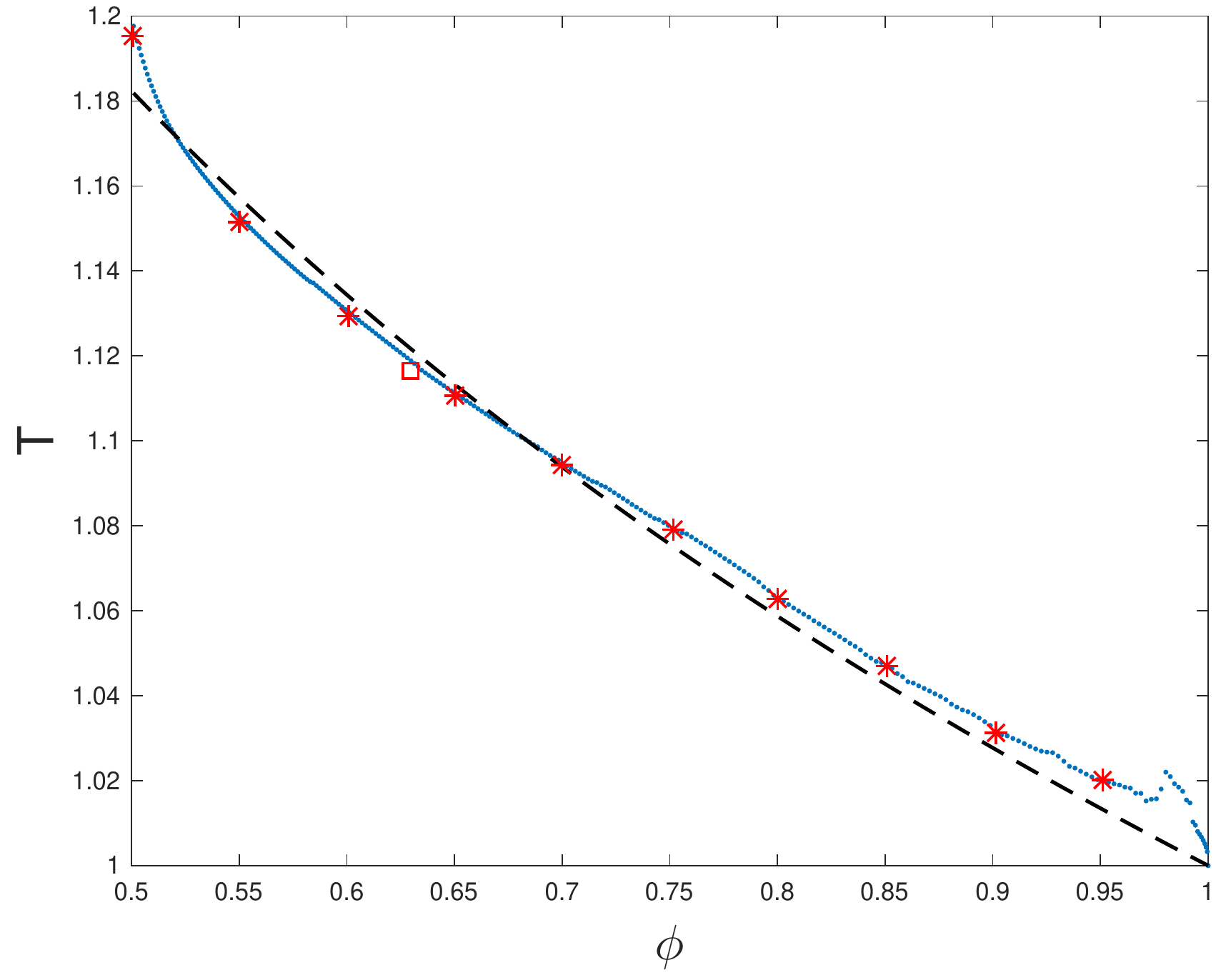}
\caption{}
\end{subfigure}
\begin{subfigure}[b]{0.5\textwidth}
\includegraphics*[height=0.8\linewidth]{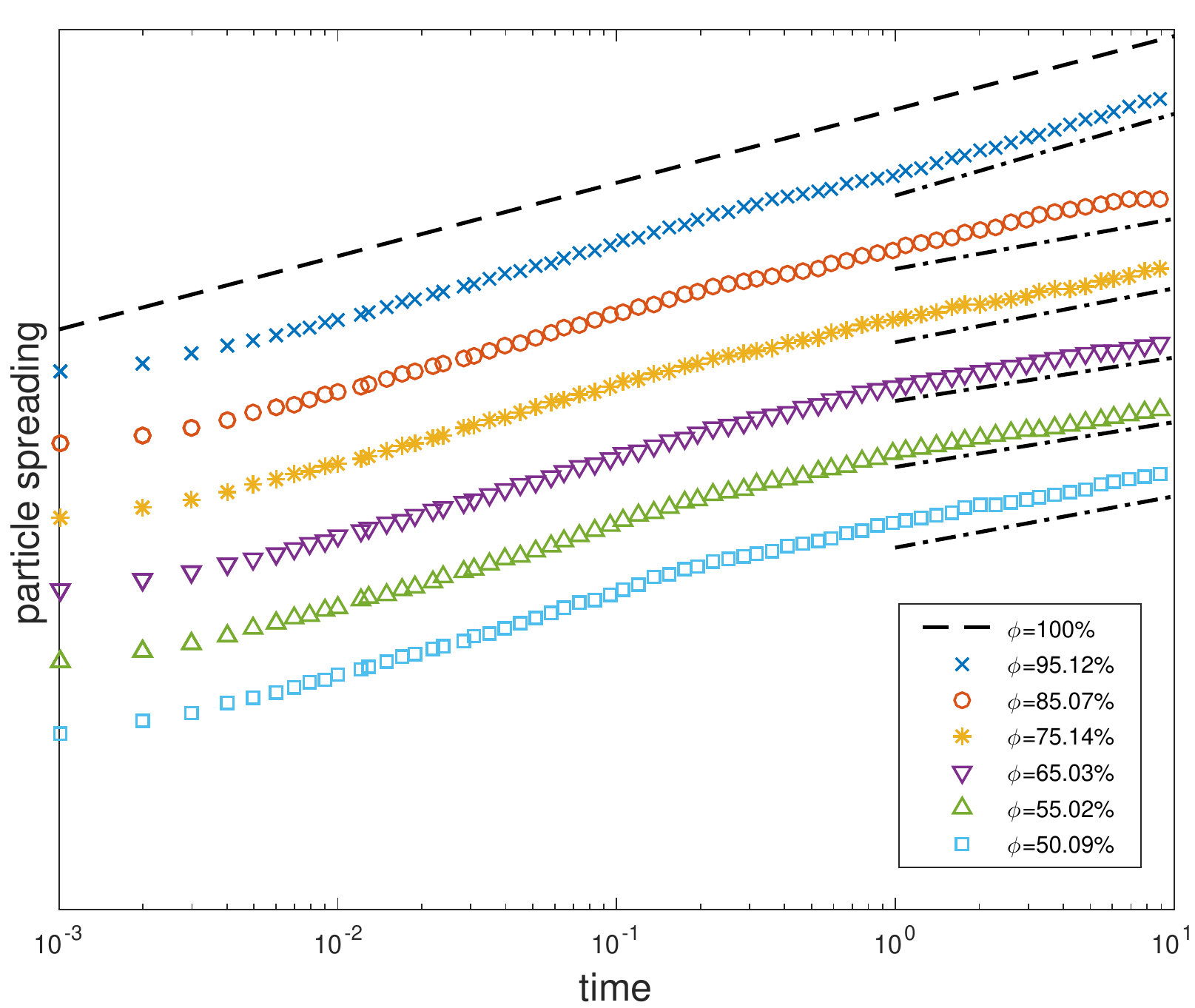}
\caption{}
\end{subfigure}
\caption{\label{fig:Eroding100Transport} (a) The tortuosity of an
eroding geometry initialized with 100 grains.  The tortuosity is
calculated using the Eulerian method~\eqref{eqn:tortuosity2} (blue dots)
and Lagrangian method~\eqref{eqn:tortuosity1} (red stars).  The dashed
line is the line of best fit $\widehat{T}(\phi)=\phi^{-p}$ with
$p=0.2459$. (b) The temporal evolution of $\sigma_\lambda$ at six
porosities.  The dashed line has slope one and corresponds to ballistic
dispersion. Asymptotically, the spreading becomes super-dispersive with
$\sigma_\lambda \sim t^{\alpha}$, $\alpha \in (1/2,1)$.  The
dashed-dotted lines of best fit have slopes $1.11$ ($\phi=95.12\%$),
$0.68$ ($\phi=85.07\%$), $0.73$ ($\phi=75.14\%$), $0.59$
($\phi=65.03\%$), $0.61$ ($\phi=55.02\%$), and $0.70$ ($\phi=50.09\%$).}
\end{figure}

In figure~\ref{fig:Eroding100Transport}(a), we plot the tortuosity as a
function of the porosity.  The initial geometry has a porosity of $\phi
= 50.09\%$ and the tortuosity is $T = 1.20$.  The tortuosity is computed
with both the length of the streamlines~\eqref{eqn:tortuosity1} (red
stars) and using the spatial average of the velocity on an Eulerian
grid~\eqref{eqn:tortuosity2} (blue marks).  The red square corresponds
to the porosity of the geometry in figure~\ref{fig:Eroding100tort}(c).
Again, the two tortuosity formulas give similar results.  For this
geometry, the tortuosity decreases monotonically at almost all
porosities.  However, the tortuosity undergoes a sudden increase near
the end of the simulation, and we have observed this behavior in other
examples.  The increase is caused by a single small body near the middle
of the channel being completely eroded.  While this results in
straighter streamlines, therefore reducing $\lambda$, the horizontal
flow, $u_1(y)$, increases since there is no longer a no-slip boundary,
and this increases the tortuosity.  We also compute the lines of best
fit using the porosity-tortuosity models~\eqref{eqn:tortuosityModels}.
The black dashed line in figure~\ref{fig:Eroding100Transport}(a) is the
line of best fit $\widehat{T}(\phi) = \phi^{-0.2459}$, with a
root-mean-square error of $5.50 \times 10^{-3}$.  We note a slightly
better root-mean-square error of $5.20 \times 10^{-3}$ is possible with
the model $\widehat{T}(\phi) = 1 - 0.2631 \ln(\phi)$.

In figure~\ref{fig:Eroding100Transport}(b), we plot the temporal
evolution of the particle spreading $\sigma_\lambda$ at six different
porosities. Again, we initially observe ballistic motion (black dashed
line), and then super-dispersion.  Similar to the example in
figure~\ref{fig:ErodingLow20Transport}(b), the asymptotic anomalous
dispersion rate is not growing with the porosity.  Therefore, it appears
that the dispersion rate in an eroding geometry depends not only the
porosity, but also the location and shape of the bodies.  Finally, at
the highest porosity, anomalous dispersion is only observed briefly in
the time interval $(0.5,1)$, and then transitions back to a ballistic
regime.  Since the bodies are so small at this high porosity, after
reinsertion, the streamline is not significantly deflected by any of the
bodies, and this results in a ballistic regime.

Finally, we investigate the effect of erosion on pore sizes.  The
distribution of the pore sizes is directly related to the distribution
of the velocity, and thus effects the tortuosity~\citep{den-ica-hid2018}
and anomalous dispersion~\citep{dea-qua-bir-jua2018}. In addition, the
pore sizes are used in network models~\citep{bry-mel-cad1993,
bry-kin-mel1993}.  As described in section~\ref{sec:throats}, we use a
Delaunay triangulation to define neighboring eroding bodies, and we
compute the pore size by finding the closest distance between all
neighboring bodies.  Instead of computing the Delaunay configuration at
each time step, which would result in new definitions for the pores at
each time step, we only compute a new Delaunay triangulation when a
grain completely erodes.  Once all pore sizes are computed, we analyze
their distribution as a function of the porosity.  

\begin{figure}
\begin{subfigure}[b]{0.33\textwidth}
\includegraphics*[width =\linewidth]{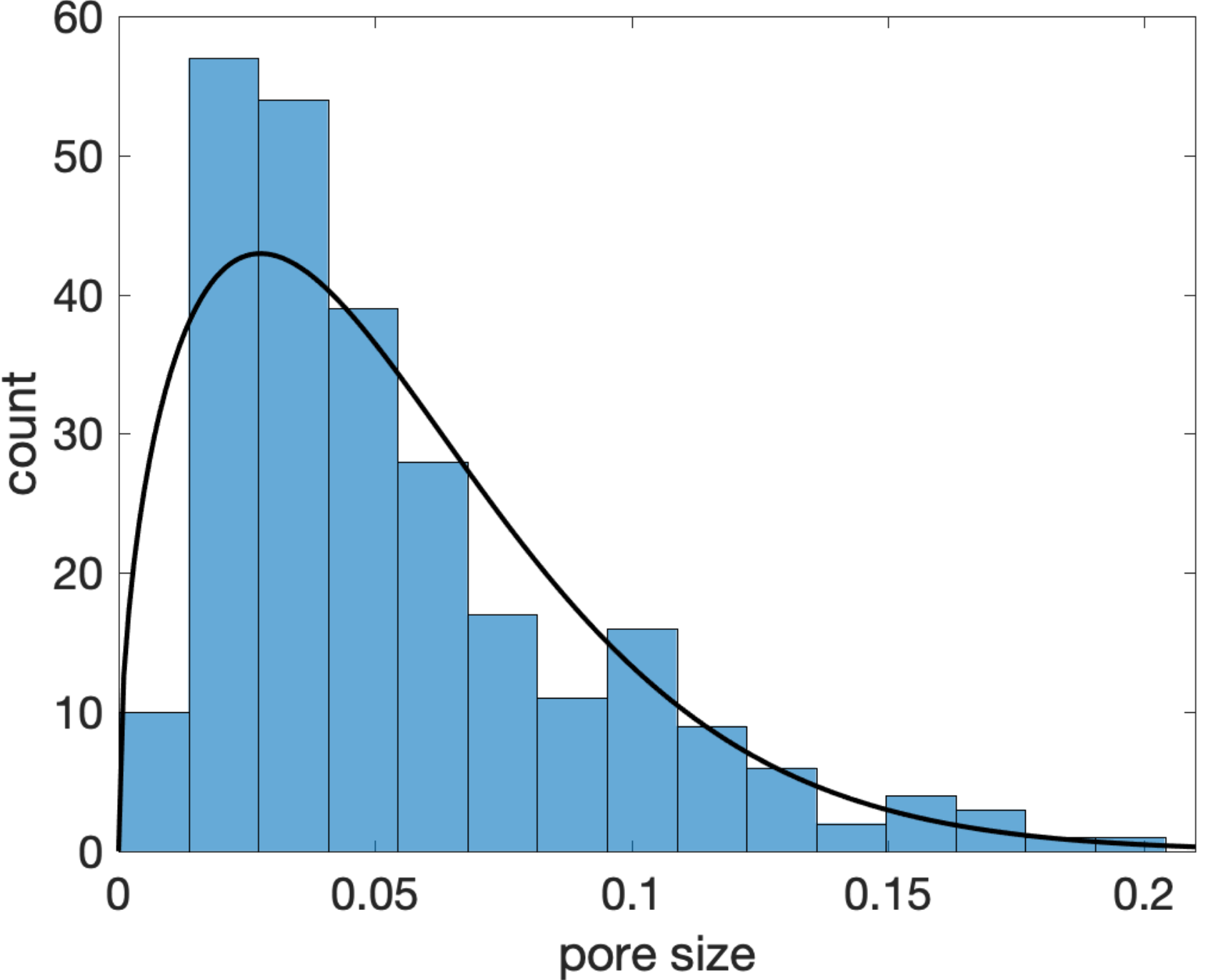}
\caption{100 bodies, 258 pores}
\end{subfigure}%
\begin{subfigure}[b]{0.33\textwidth}
\includegraphics*[width =\linewidth]{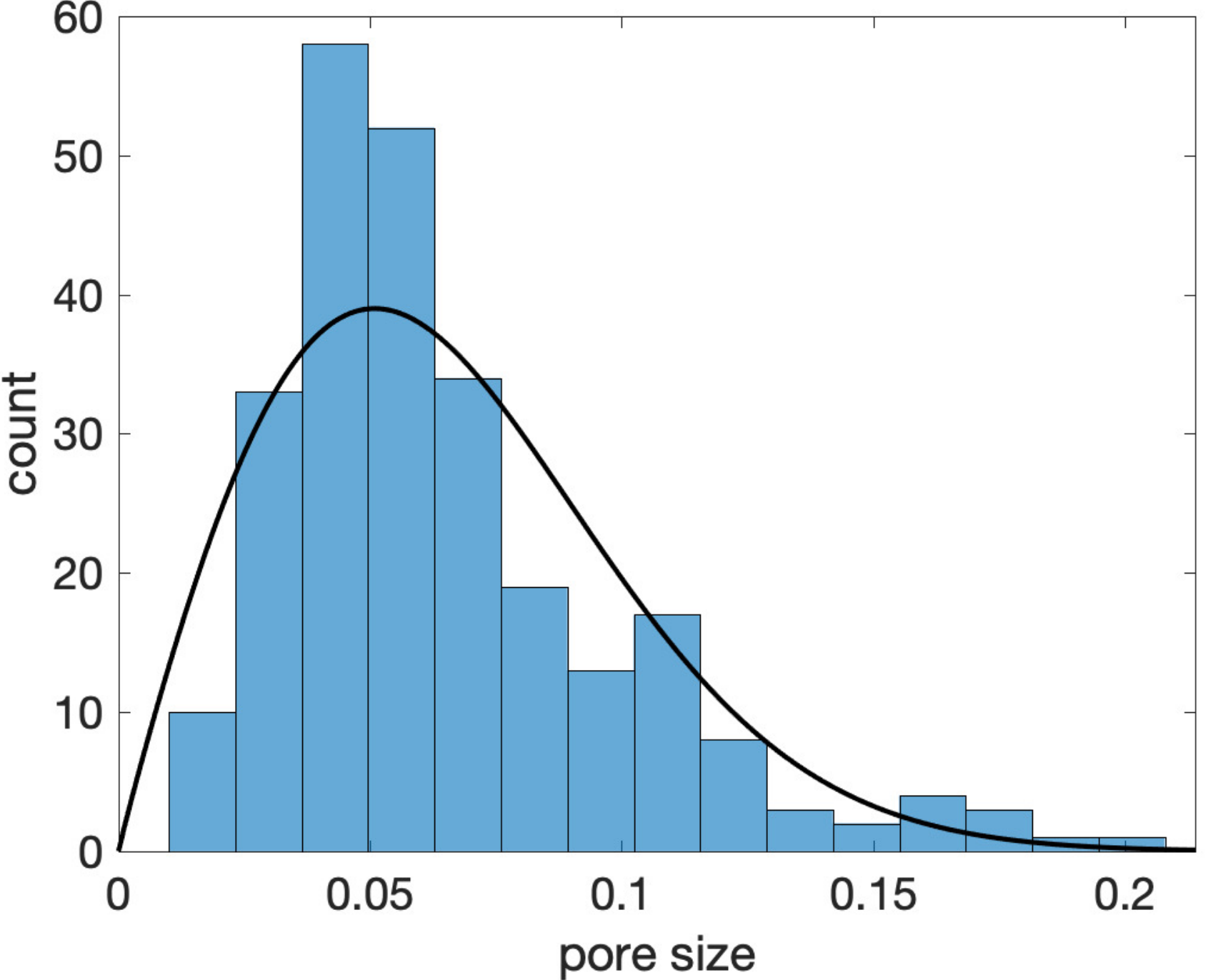}
\caption{100 bodies, 258 pores}
\end{subfigure}%
\begin{subfigure}[b]{0.33\textwidth}
\includegraphics*[width =\linewidth]{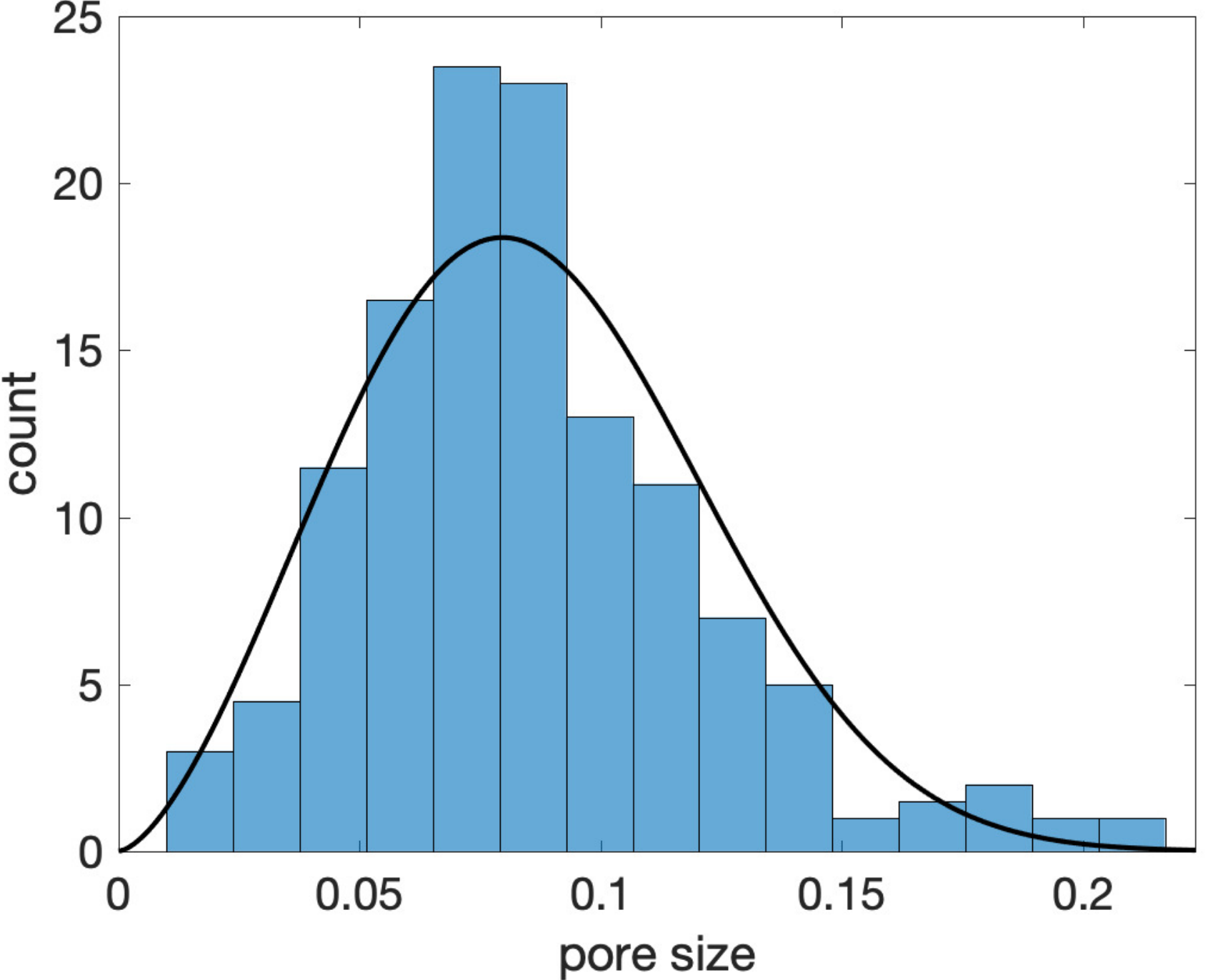}
\caption{97 bodies, 249 pores}
\end{subfigure}
\begin{subfigure}[b]{0.33\textwidth}
\includegraphics*[width =\linewidth]{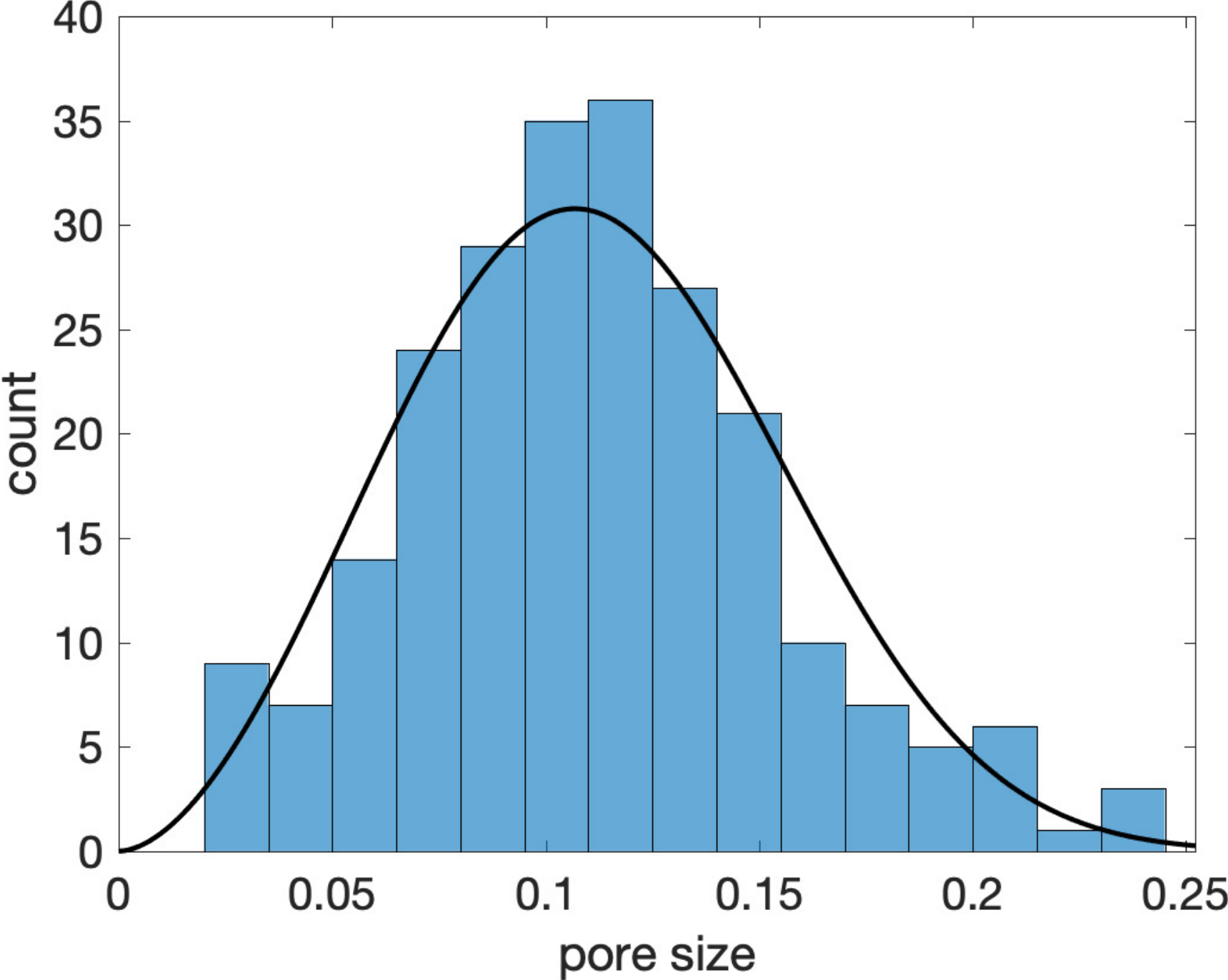}
\caption{92 bodies, 234 pores}
\end{subfigure}%
\begin{subfigure}[b]{0.33\textwidth}
\includegraphics*[width =\linewidth]{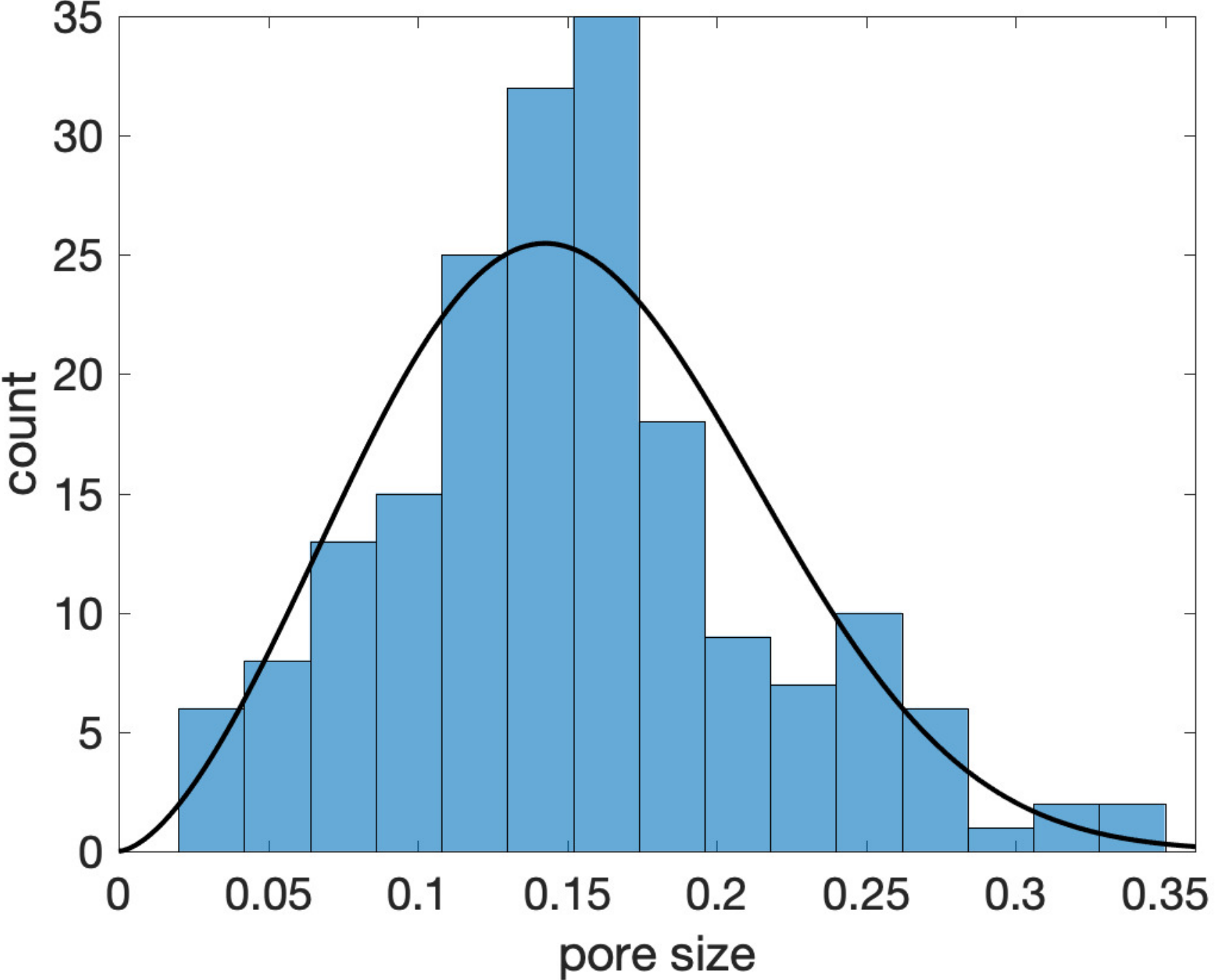}
\caption{75 bodies, 189 pores}
\end{subfigure}%
\begin{subfigure}[b]{0.33\textwidth}
\includegraphics*[width =\linewidth]{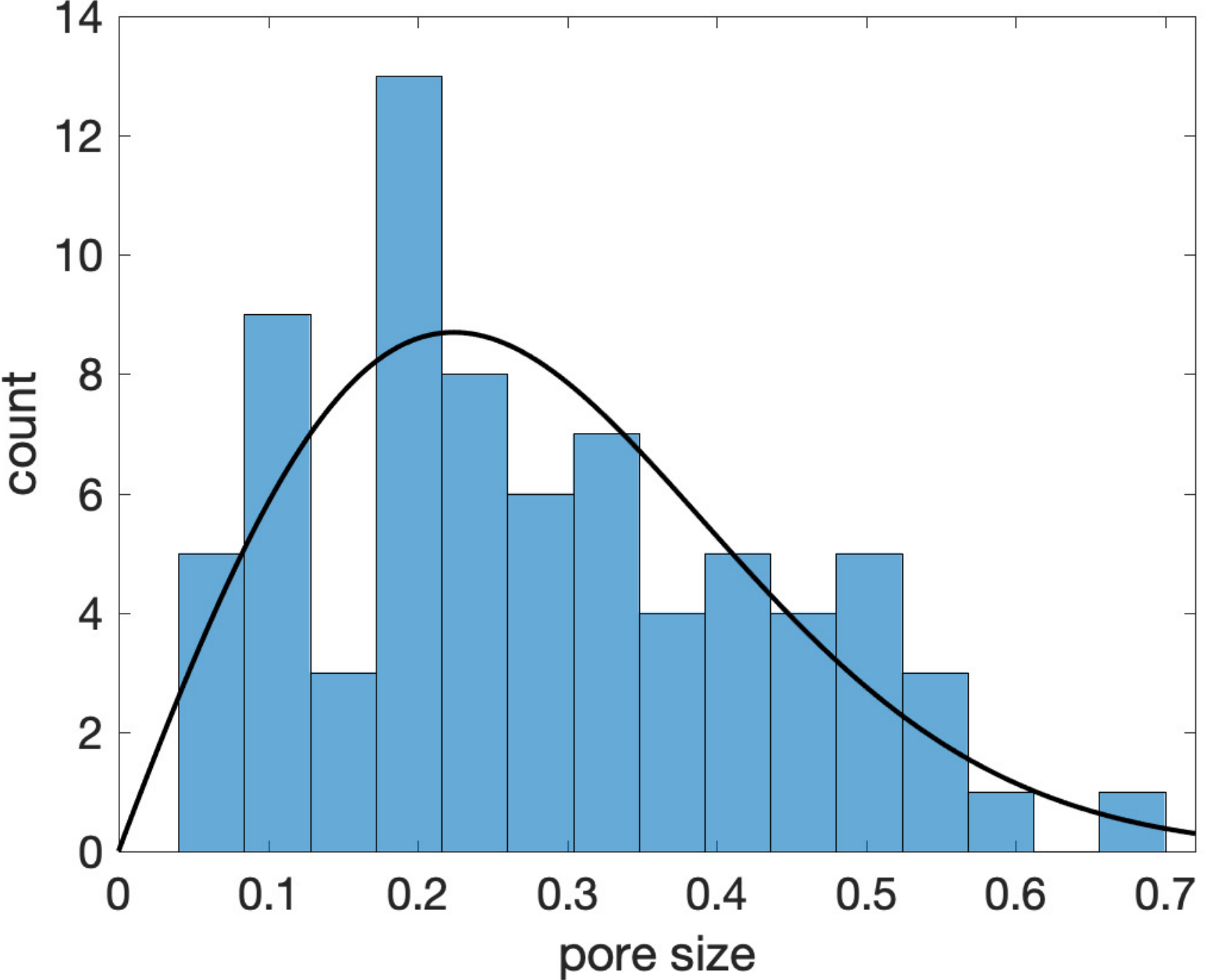}
\caption{34 bodies, 74 pores}
\end{subfigure}
\caption{\label{fig:Eroding100gap_hist} The pore sizes of 100 eroding
bodies at six porosities. The black curves are Weibull distributions
whose first and second moments agree with the data. The porosities and
Weibull distribution parameters $(k,\lambda)$ at each of the porosities
are: (a) $\phi = 50.09\%$ and $(k,\lambda)=(1.4650,0.0605)$; (b) $\phi =
55.02\%$ and $(k,\lambda)=(1.9485,0.0737)$; (c) $\phi = 65.03\%$ and
$(k,\lambda)=(2.5682,0.0965)$; (d) $\phi = 75.14\%$ and
$(k,\lambda)=(2.7771,0.1255)$; (e) $\phi = 85.07\%$ and
$(k,\lambda)=(2.6235,0.1713)$; (f) $\phi = 95.12\%$ and
$(k,\lambda)=(1.9840,0.3194)$.}
\end{figure}

\begin{figure}
\begin{subfigure}[b]{0.5\textwidth}
\includegraphics*[height = 0.8\linewidth]{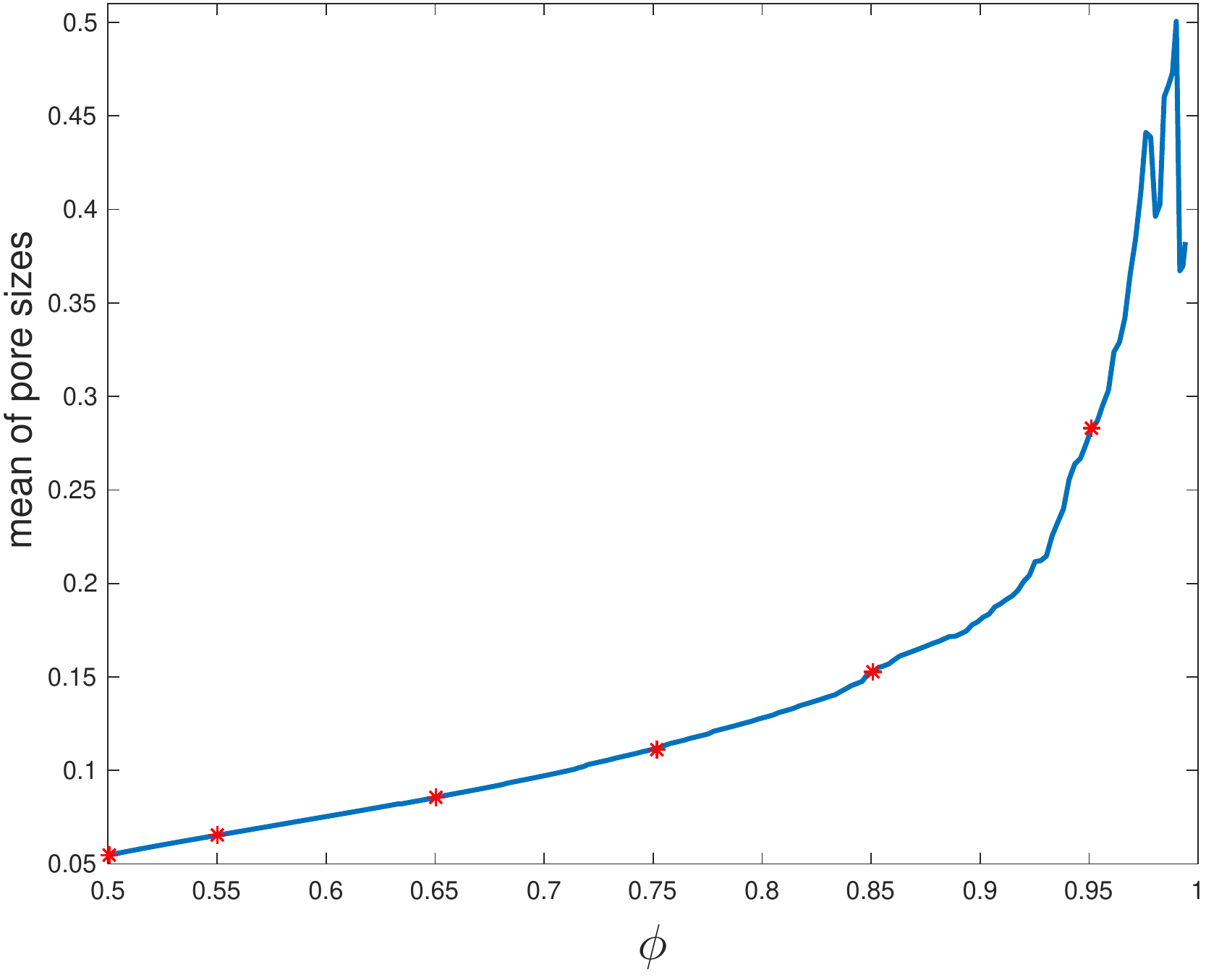}
\caption{}
\end{subfigure}
\begin{subfigure}[b]{0.5\textwidth}
\includegraphics*[height=0.8\linewidth]{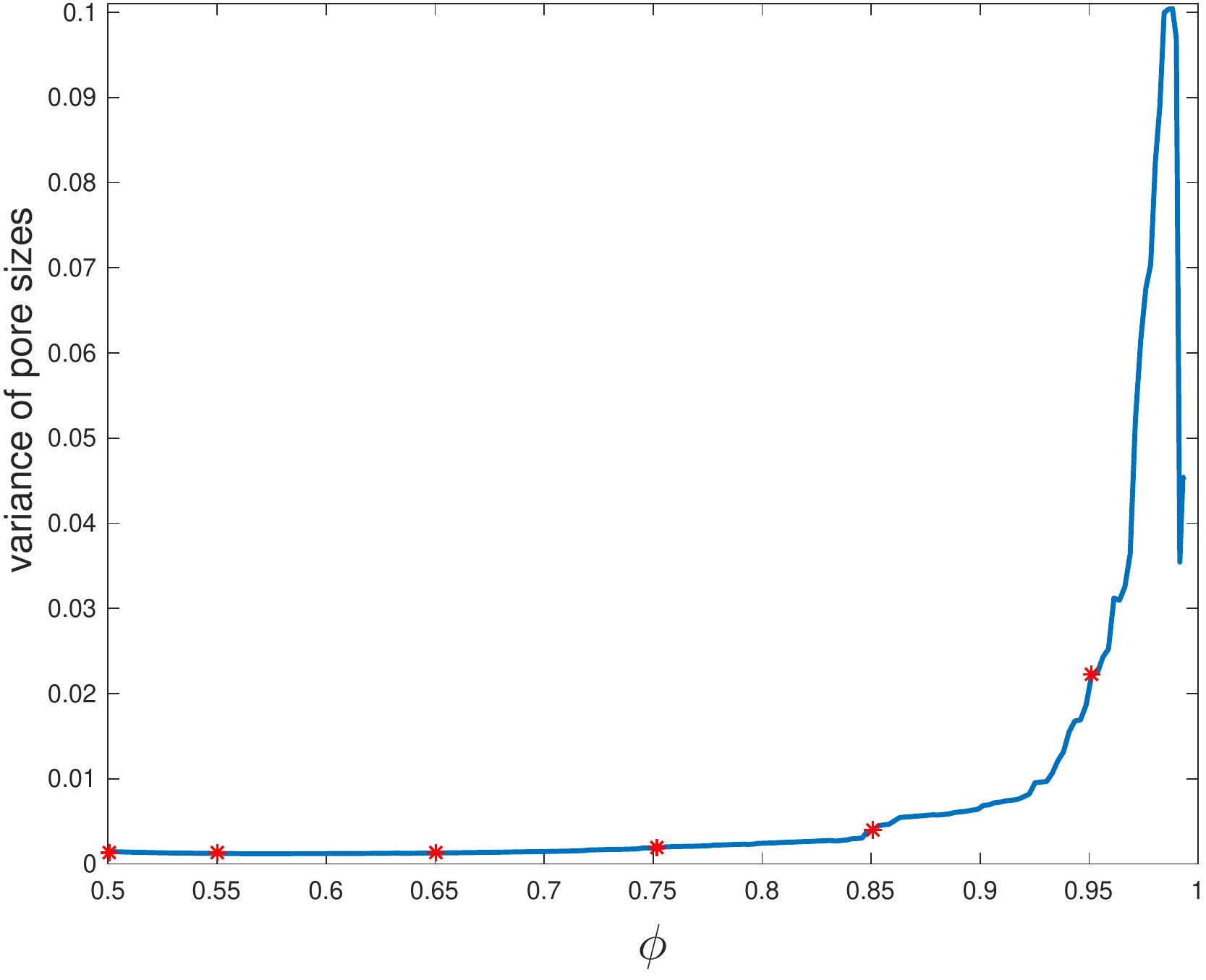}
\caption{}
\end{subfigure}
\caption{\label{fig:Eroding100gap_mean_var} The effect of erosion on (a)
the mean and (b) the variance of the pore sizes. The geometry initially
contains 100 eroding bodies. The distributions of the pore sizes in
figure~\ref{fig:Eroding100gap_hist} are indicated by the red stars.}
\end{figure}

In figure~\ref{fig:Eroding100gap_hist}, we plot histograms of the pore
pore sizes at six porosities throughout the erosion process. We
superimpose the Weibull distribution~\citep{ioa-cha1993} with the same
first two moments as the data.  The parameters of the distribution,
$(k,\lambda)$, are included in the caption of
figure~\ref{fig:Eroding100gap_hist}.  In
figure~\ref{fig:Eroding100gap_mean_var}, we plot the mean and variance
of the pore sizes as a function of the porosity.  Interestingly, for
porosities less than $\phi = 85\%$, the mean pore size grows linearly
and the variance remains nearly flat. Since a channelized geometry has
large variance, this indicates that channelization is less prevalent at
low porosities.

\section{Conclusions}
\label{sec:conclusions}
As a continuation of our previous work~\citep{qua-moo2018}, we have
simulated dense suspensions and characterized transport in viscous
eroding porous media. This is accomplished by using high-order time
stepping methods and a new quadrature methods to solve a BIE formulation
of the Stokes equations.  By using these numerical methods, we are able
to perform stable simulations of erosion with $N = O(100)$
discretization points, while the trapezoid rule would require $O(10^5)$
discretization points.

The transport is characterized in terms of tortuosity and anomalous
dispersion. While the local tortuosity agrees qualitatively with other
works~\citep{mat-kha-koz2008}, the tortuosity of eroded geometries
cannot be completely described in terms of the porosity. In particular,
we observe that for certain configurations, the tortuosity transiently
increases, even though the porosity always increases due to erosion. We
also observe super-dispersive spreading, and the rate of dispersion
significantly depends not only on the porosity, but also the number of
eroding bodies and their distribution.

To further our understanding of erosion, we are examining other bulk and
statistical properties of an eroding porous media. In this work, we
provide results for the pore throat sizes which affect the anomalous
dispersion rate~\citep{dea-qua-bir-jua2018}.  At a later date, we will
report results on the  development of anisotropic effects and the
distributions of grain sizes, shapes, and opening angles.

As a long term goal, we plan to include the inertial effects and other
transport models. Including inertia requires an integral equation
formulation of the Navier-Stokes equations, which is an active area of
research with promising directions recently
proposed~\citep{gray2019boundary, kli-ask-kro2019}.  Regarding other
transport models, this would involve a diffusive term to consider the
transport of heat or a contaminant. Forming high-fidelity simulations of
such an advection-diffusion equation can be accomplished by using time
splitting methods and recent work on heat solvers in complex
geometries~\citep{fry-kro-tor2019}.

\paragraph{\bf Acknowledgments} BQ and NM were supported by Florida
State University startup funds and Simons Foundation Mathematics and
Physical Sciences-Collaboration Grants for Mathematicians 527139 and
524259.

\bibliographystyle{jfm}


\begin{thebibliography}{84}
\expandafter\ifx\csname natexlab\endcsname\relax\def\natexlab#1{#1}\fi
\def\au#1{#1} \def\ed#1{#1} \def\yr#1{#1}\def\at#1{#1}\def\jt#1{\textit{#1}}
  \def\bt#1{#1}\def\bvol#1{\textbf{#1}} \def\vol#1{#1} \def\pg#1{#1}
  \def\publ#1{#1}\def\arxiv#1{#1}\def\org#1{#1}\def\st#1{\textit{#1}}

\bibitem[Alim {\em et~al.\/}(2017)Alim, Parsa, Weitz \&
  Brenner]{ali-par-wei-bre2017}
{\sc \au{Alim, K.}, \au{Parsa, S.}, \au{Weitz, D.~A.} \& \au{Brenner, M.~P.}}
  \yr{2017}  \at{{Local Pore Size Correlations Determine Flow Distributions in
  Porous Media}}.  \jt{Physical Review Letters}  \bvol{119},  \pg{144501}.

\bibitem[Allen(2019)]{allen2019sde}
{\sc \au{Allen, E.~J.}} \yr{2019}  \at{{An SDE model for deterioration of rock
  surfaces}}.  \jt{Stochastic Analysis and Applications}  \pg{pp. 1--16}.

\bibitem[Alley {\em et~al.\/}(2002)Alley, Healy, LaBaugh \&
  Reilly]{all-hea-lab-rei2002}
{\sc \au{Alley, W.~M.}, \au{Healy, R.~W.}, \au{LaBaugh, J.~W.} \& \au{Reilly,
  T.~E.}} \yr{2002}  \at{Flow and storage in groundwater systems}.
  \jt{Science}  \bvol{296}~(5575),  \pg{1985--1990}.

\bibitem[Amin {\em et~al.\/}(2019)Amin, Huang, Hu, Zhang \&
  Ristroph]{amin2019role}
{\sc \au{Amin, K.}, \au{Huang, J.~M.}, \au{Hu, K.~J.}, \au{Zhang, J.} \&
  \au{Ristroph, L.}} \yr{2019}  \at{The role of shape-dependent flight
  stability in the origin of oriented meteorites}.  \jt{Proceedings of the
  National Academy of Sciences}  \bvol{116}~(33),  \pg{16180--16185}.

\bibitem[de~Anna {\em et~al.\/}(2013)de~Anna, Borgne, Dentz, Tartakovsky,
  Bolster \& Davy]{dea-leb-den-tar-bol-dav2013}
{\sc \au{de~Anna, P.}, \au{Borgne, T.~Le}, \au{Dentz, M.}, \au{Tartakovsky,
  A.~M.}, \au{Bolster, D.} \& \au{Davy, P.}} \yr{2013}  \at{{Flow
  Intermittency, Dispersion, and Correlated Continuous Time Random Walks in
  Porous Media}}.  \jt{Physical Review Letters}  \bvol{110}~(18),  \pg{184502}.

\bibitem[de~Anna {\em et~al.\/}(2018)de~Anna, Quaife, Biros \&
  Juanes]{dea-qua-bir-jua2018}
{\sc \au{de~Anna, P.}, \au{Quaife, B.}, \au{Biros, G.} \& \au{Juanes, R.}}
  \yr{2018}  \at{{Prediction of velocity distribution from pore structure in
  simple porous media}}.  \jt{Physical Review Fluids}  \bvol{2}~(12),
  \pg{124103}.

\bibitem[Baker \& Shelley(1986)]{bak-she1986}
{\sc \au{Baker, G.~R.} \& \au{Shelley, M.~J.}} \yr{1986}  \at{Boundary integral
  techniques for multi-connected domains}.  \jt{Journal of Computational
  Physics}  \bvol{64}~(1),  \pg{112--132}.

\bibitem[Barnett {\em et~al.\/}(2015)Barnett, Wu \&
  Veerapaneni]{bar-wu-vee2015}
{\sc \au{Barnett, A.}, \au{Wu, B.} \& \au{Veerapaneni, S.}} \yr{2015}
  \at{Spectrally-accurate quadratures for evaluation of layer potentials close
  to the boundary for the 2d stokes and laplace equations}.  \jt{SIAM Journal
  on Scientific Computing}  \bvol{37}~(4),  \pg{B519--B542}.

\bibitem[Barnett(2014)]{bar2014}
{\sc \au{Barnett, A.~H.}} \yr{2014}  \at{{Evaluation of layer potentials close
  to the boundary for Laplace and Helmholtz problems on analytic planar
  domains}}.  \jt{SIAM Journal on Scientific Computing}  \bvol{36}~(2),
  \pg{A427--A451}.

\bibitem[Beale \& Lai(2001)]{bea-lai2001}
{\sc \au{Beale, J.T.} \& \au{Lai, M.-C.}} \yr{2001}  \at{{A Method for
  Computing Nearly Singular Integrals}}.  \jt{SIAM Journal on Numerical
  Analysis}  \bvol{38}~(6),  \pg{1902--1925}.

\bibitem[Beale {\em et~al.\/}(2016)Beale, Ying \& Wilson]{bea-yin-wil2016}
{\sc \au{Beale, J.~T.}, \au{Ying, W.} \& \au{Wilson, J.~R.}} \yr{2016}  \at{{A
  Simple Method for Computing Singular or Nearly Singular Integrals on Closed
  Surfaces}}.  \jt{Communications in Computational Physics}  \bvol{20}~(3),
  \pg{733--753}.

\bibitem[Bear(1972)]{bea1972}
{\sc \au{Bear, J.}} \yr{1972} {\em Dynamics of Fluids in Porous Media\/}.
  \publ{New York: Dover}.

\bibitem[Beckermann \& Viskanta(1988)]{bec-vis1998}
{\sc \au{Beckermann, C.} \& \au{Viskanta, R.}} \yr{1988}  \at{Natural
  convection solid/liquid phase change in porous media}.  \jt{International
  journal of heat and mass transfer}  \bvol{31}~(1),  \pg{35--46}.

\bibitem[Bellin {\em et~al.\/}(1992)Bellin, Salandin \&
  Rinaldo]{bel-sal-rin1992}
{\sc \au{Bellin, A.}, \au{Salandin, P.} \& \au{Rinaldo, A.}} \yr{1992}
  \at{{Simulation of Dispersion in Heterogeneous Porous Formations: Statistics,
  First-Order Theories, Convergence of Computations}}.  \jt{Water Resources
  Research}  \bvol{28}~(9),  \pg{2211--2227}.

\bibitem[Berhanu {\em et~al.\/}(2012)Berhanu, Petroff, Devauchelle, Kudrolli \&
  Rothman]{berhanu2012shape}
{\sc \au{Berhanu, M.}, \au{Petroff, A.}, \au{Devauchelle, O.}, \au{Kudrolli,
  A.} \& \au{Rothman, D.~H.}} \yr{2012}  \at{Shape and dynamics of seepage
  erosion in a horizontal granular bed}.  \jt{Physical Review E}
  \bvol{86}~(4),  \pg{041304}.

\bibitem[Berkowitz \& Scher(2001)]{ber-sch2001}
{\sc \au{Berkowitz, B.} \& \au{Scher, H.}} \yr{2001}  \at{{The Role of
  Probabilistic Approaches to Transport Theory in Heterogeneous Media}}.
  \jt{Transport in Porous Media}  \bvol{42},  \pg{241--263}.

\bibitem[Berkowitz {\em et~al.\/}(2000)Berkowitz, Scher \&
  Silliman]{ber-sch-sil2000}
{\sc \au{Berkowitz, B.}, \au{Scher, H.} \& \au{Silliman, S.~E.}} \yr{2000}
  \at{Anomalous transport in laboratory-scale, heterogeneous porous media}.
  \jt{Water Resources Research}  \bvol{36}~(1),  \pg{149--158}.

\bibitem[Bijeljic \& Blunt(2006)]{bij-blu2006}
{\sc \au{Bijeljic, B.} \& \au{Blunt, M.~J.}} \yr{2006}  \at{Pore-scale modeling
  and continuous time random walk analysis of dispersion in porous media}.
  \jt{Water Resources Research}  \bvol{42}~(1).

\bibitem[Borgne {\em et~al.\/}(2011)Borgne, Dentz, Davy, Bolster, Carrera,
  de~Dreuzy \& Bour]{leb-den-dav-bol-car-dec-bou2011}
{\sc \au{Borgne, T.~Le}, \au{Dentz, M.}, \au{Davy, P.}, \au{Bolster, D.},
  \au{Carrera, J.}, \au{de~Dreuzy, J.-R.} \& \au{Bour, O.}} \yr{2011}
  \at{{Persistence of incomplete mixing: A key to anomalous transport}}.
  \jt{Physical Review E}  \bvol{84},  \pg{015301}.

\bibitem[Borgne {\em et~al.\/}(2007)Borgne, de~Dreuzy, Davy \&
  Bour]{leb-ded-dav-bou2007}
{\sc \au{Borgne, T.~Le}, \au{de~Dreuzy, J.{-}R.}, \au{Davy, P.} \& \au{Bour,
  O.}} \yr{2007}  \at{Characterization of the velocity field organization in
  heterogeneous media by conditional correlation}.  \jt{Water Resources
  Research}  \bvol{43}.

\bibitem[Bryant {\em et~al.\/}(1993{\natexlab{{\em a\/}}})Bryant, King \&
  Mellor]{bry-kin-mel1993}
{\sc \au{Bryant, S.~L.}, \au{King, P.~R.} \& \au{Mellor, D.~W.}}
  \yr{1993{\natexlab{{\em a\/}}}}  \at{Network model evaluation of permeability
  and spatial correlation in a real random sphere packing}.  \jt{Transport in
  Porous Media}  \bvol{11}~(1),  \pg{53--70}.

\bibitem[Bryant {\em et~al.\/}(1993{\natexlab{{\em b\/}}})Bryant, Mellor \&
  Cade]{bry-mel-cad1993}
{\sc \au{Bryant, S.~L.}, \au{Mellor, D.~W.} \& \au{Cade, C.~A.}}
  \yr{1993{\natexlab{{\em b\/}}}}  \at{Physically representative network models
  of transport in porous media}.  \jt{AIChE Journal}  \bvol{39}~(3),
  \pg{387--396}.

\bibitem[Carman(1937)]{car1937}
{\sc \au{Carman, P.~C.}} \yr{1937}  \at{Fluid flow through granular beds}.
  \jt{Transactions of the Institution of Chemical Engineers}  \bvol{15},
  \pg{150--166}.

\bibitem[Chaoui \& Feuillebois(2003)]{cha-feu2003}
{\sc \au{Chaoui, M.} \& \au{Feuillebois, F.}} \yr{2003}  \at{Creeping flow
  around a sphere in a shear flow close to a wall}.  \jt{Quarterly Journal of
  Mechanics and Applied Mathematics}  \bvol{56}~(3),  \pg{381--410}.

\bibitem[Cho {\em et~al.\/}(2019)Cho, Lu, Howard, Adams \& Datta]{cho2019crack}
{\sc \au{Cho, H.~J.}, \au{Lu, N.~B.}, \au{Howard, M.~P.}, \au{Adams, R.~A.} \&
  \au{Datta, S.~S.}} \yr{2019}  \at{Crack formation and self-closing in
  shrinkable, granular packings}.  \jt{Soft Matter} .

\bibitem[Chwang \& Wu(1975)]{chw-wu1975}
{\sc \au{Chwang, A.~T.} \& \au{Wu, T.~Y.{-}T.}} \yr{1975}  \at{{Hydromechanics
  of low-Reynolds-number flow. Part 2. Singularity method for Stokes flows}}.
  \jt{Journal of Fluid Mechanics}  \bvol{67}~(4),  \pg{787--815}.

\bibitem[Cushman {\em et~al.\/}(1995)Cushman, Hu \& Deng]{cus-hu-den1995}
{\sc \au{Cushman, J.~H.}, \au{Hu, B.~X.} \& \au{Deng, F.{-}W.}} \yr{1995}
  \at{{Nonlocal reactive transport with physical and chemical
  heterogeneity:Localization errors}}.  \jt{Water Resources Research}
  \bvol{31}~(9),  \pg{2219--2237}.

\bibitem[Cvetkovic {\em et~al.\/}(1996)Cvetkovic, Cheng \&
  Wen]{cve-che-wen1996}
{\sc \au{Cvetkovic, V.}, \au{Cheng, H.} \& \au{Wen, X.{-}H.}} \yr{1996}
  \at{{Analysis of nonlinear effects on tracer migration in heterogeneous
  aquifers using Lagrangian travel time statistics}}.  \jt{Water Resources
  Research}  \bvol{32}~(6),  \pg{1671--1680}.

\bibitem[Dagan(1987)]{dag1987}
{\sc \au{Dagan, G.}} \yr{1987}  \at{{Theory of Solute Transport by
  Groundwater}}.  \jt{Annual Review of Fluid Mechanics}  \bvol{19},
  \pg{183--215}.

\bibitem[Dardis \& McCloskey(1998)]{dar-mcc1998}
{\sc \au{Dardis, O.} \& \au{McCloskey, J.}} \yr{1998}  \at{Permeability
  porosity relationships from numerical simulations of fluid flow}.
  \jt{Geophysical Research Letters}  \bvol{25}~(9),  \pg{1471--1474}.

\bibitem[Dentz {\em et~al.\/}(2011)Dentz, Borgne, Englert \&
  Bijeljic]{den-leb-eng-bij2011}
{\sc \au{Dentz, M.}, \au{Borgne, T.~Le}, \au{Englert, A.} \& \au{Bijeljic, B.}}
  \yr{2011}  \at{{Mixing, spreading and reaction in heterogeneous media: A
  brief review}}.  \jt{Journal of Contaminant Hydrology}  \bvol{120},
  \pg{1--17}.

\bibitem[Dentz {\em et~al.\/}(2004)Dentz, Cortis, Scher \&
  Berkowitz]{den-cor-sch-ber2004}
{\sc \au{Dentz, M.}, \au{Cortis, A.}, \au{Scher, H.} \& \au{Berkowitz, B.}}
  \yr{2004}  \at{Time behavior of solute transport in heterogeneous media:
  transition from anomalous to normal transport}.  \jt{Advances in Water
  Resources}  \bvol{27},  \pg{55–173}.

\bibitem[Dentz {\em et~al.\/}(2018)Dentz, Icardi \& Hidalgo]{den-ica-hid2018}
{\sc \au{Dentz, M.}, \au{Icardi, M.} \& \au{Hidalgo, J.~J.}} \yr{2018}
  \at{Mechanisms of dispersion in a porous medium}.  \jt{Journal of Fluid
  Mechanics}  \bvol{841},  \pg{851--882}.

\bibitem[Duda {\em et~al.\/}(2011)Duda, Koza \& Matyka]{dud-koz-mat2011}
{\sc \au{Duda, A.}, \au{Koza, Z.} \& \au{Matyka, M.}} \yr{2011}  \at{Hydraulic
  tortuosity in arbitrary porous media flow}.  \jt{Physical Review E}
  \bvol{84},  \pg{036319}.

\bibitem[Favier {\em et~al.\/}(2019)Favier, Purseed \&
  Duchemin]{favier2019rayleigh}
{\sc \au{Favier, B.}, \au{Purseed, J.} \& \au{Duchemin, L.}} \yr{2019}
  \at{{Rayleigh--B{\'e}nard convection with a melting boundary}}.  \jt{Journal
  of Fluid Mechanics}  \bvol{858},  \pg{437--473}.

\bibitem[Fryklund {\em et~al.\/}(2019)Fryklund, Kropinski \&
  Tornberg]{fry-kro-tor2019}
{\sc \au{Fryklund, F.}, \au{Kropinski, M.~C.~A.} \& \au{Tornberg, A.{-}K.}}
  \yr{2019}  \at{An integral equation based numerical method for the forced
  heat equation on complex domains}.  \jt{arxiv}  \bvol{1907.08537}.

\bibitem[Gray {\em et~al.\/}(2019)Gray, Jakowski, Moore \&
  Ye]{gray2019boundary}
{\sc \au{Gray, L.~J.}, \au{Jakowski, J.}, \au{Moore, M.~N.~J.} \& \au{Ye, W.}}
  \yr{2019}  \at{{Boundary integral analysis for non-homogeneous,
  incompressible Stokes flows}}.  \jt{Advances in Computational Mathematics}
  \bvol{45}~(3),  \pg{1729--1734}.

\bibitem[Greengard \& Rokhlin(1987)]{gre-rok1987}
{\sc \au{Greengard, L.} \& \au{Rokhlin, V.}} \yr{1987}  \at{{A Fast Algorithm
  for Particle Simulations}}.  \jt{Journal of Computational Physics}
  \bvol{73},  \pg{325--348}.

\bibitem[Hakoun {\em et~al.\/}(2019)Hakoun, Comolli \& Dentz]{hak-com-den2019}
{\sc \au{Hakoun, V.}, \au{Comolli, A.} \& \au{Dentz, M.}} \yr{2019}
  \at{{Upscaling and Prediction of Lagrangian Velocity Dynamics in
  Heterogeneous Porous Media}}.  \jt{Water Resources Research}  \bvol{55}~(5),
  \pg{3976--3996}.

\bibitem[Helsing \& Ojala(2008)]{hel-oja2008a}
{\sc \au{Helsing, J.} \& \au{Ojala, R.}} \yr{2008}  \at{On the evaluation of
  layer potentials close to their sources}.  \jt{Journal of Computational
  Physics}  \bvol{227},  \pg{2899--2921}.

\bibitem[Hewett \& Sellier(2017)]{hewett2017evolution}
{\sc \au{Hewett, J.~N.} \& \au{Sellier, M.}} \yr{2017}  \at{Evolution of an
  eroding cylinder in single and lattice arrangements}.  \jt{Journal of Fluids
  and Structures}  \bvol{70},  \pg{295--313}.

\bibitem[Hewett \& Sellier(2018)]{hewett2018modelling}
{\sc \au{Hewett, J.~N.} \& \au{Sellier, M.}} \yr{2018}  \at{Modelling ripple
  morphodynamics driven by colloidal deposition}.  \jt{Computers \& Fluids}
  \bvol{163},  \pg{54--67}.

\bibitem[Higdon(1985)]{hig1985}
{\sc \au{Higdon, J.~J.~L.}} \yr{1985}  \at{Stokes flow in arbitrary
  two-dimensional domains: shear flow over ridges and cavities}.  \jt{Journal
  of Fluid Mechanics}  \bvol{159},  \pg{195--226}.

\bibitem[Hou {\em et~al.\/}(1994)Hou, Lowengrub \& Shelley]{hou-low-she1994}
{\sc \au{Hou, T.~Y.}, \au{Lowengrub, J.~S.} \& \au{Shelley, M.~J.}} \yr{1994}
  \at{{Removing the Stiffness for Interfacial Flows with Surface Tension}}.
  \jt{Journal of Computational Physics}  \bvol{114},  \pg{312--338}.

\bibitem[Huang {\em et~al.\/}(2015)Huang, Moore \& Ristroph]{mac2015shape}
{\sc \au{Huang, J.~M.}, \au{Moore, M.~N.~J.} \& \au{Ristroph, L.}} \yr{2015}
  \at{Shape dynamics and scaling laws for a body dissolving in fluid flow}.
  \jt{Journal of Fluid Mechanics}  \bvol{765}.

\bibitem[Ioakimidis {\em et~al.\/}(1991)Ioakimidis, Papadakis \&
  Perdios]{ioa-pap-per1991}
{\sc \au{Ioakimidis, N.~I.}, \au{Papadakis, K.~E.} \& \au{Perdios, E.~A.}}
  \yr{1991}  \at{{Numerical Evaluations of Analytic Functions by Cauchy's
  Theorem}}.  \jt{BIT Numerical Mathematics}  \bvol{31}~(2),  \pg{276--285}.

\bibitem[Ioannidis \& Chatzis(1993)]{ioa-cha1993}
{\sc \au{Ioannidis, M.~A.} \& \au{Chatzis, I.}} \yr{1993}  \at{{Network
  Modelling of Pore Structure and Transport Properties of Porous Media}}.
  \jt{Chemical Engineering Science}  \bvol{48}~(5),  \pg{951--972}.

\bibitem[Jambon-Puillet {\em et~al.\/}(2018)Jambon-Puillet, Shahidzadeh \&
  Bonn]{jambon2018singular}
{\sc \au{Jambon-Puillet, E.}, \au{Shahidzadeh, N.} \& \au{Bonn, D.}} \yr{2018}
  \at{Singular sublimation of ice and snow crystals}.  \jt{Nature
  Communications}  \bvol{9}~(1),  \pg{4191}.

\bibitem[Johnson \& Elimelech(1995)]{joh-eli1995}
{\sc \au{Johnson, P.~R.} \& \au{Elimelech, M.}} \yr{1995}  \at{{Dynamics of
  Colloid Deposition in Porous Media: Blocking Based on Random Sequential
  Adsorption}}.  \jt{Langmuir}  \bvol{11},  \pg{801--812}.

\bibitem[Kang {\em et~al.\/}(2014)Kang, de~Anna, Nunes, Bijelic, Blunt \&
  Juanes]{kan-dea-nun-bij-blu-jua2014}
{\sc \au{Kang, P.~K.}, \au{de~Anna, P.}, \au{Nunes, J.~P.}, \au{Bijelic, B.},
  \au{Blunt, M.~J.} \& \au{Juanes, R.}} \yr{2014}  \at{Pore-scale intermittent
  velocity structure underpinning anomalous transport through 3-{D} porous
  media}.  \jt{Geophysical Research Letters}  \bvol{41},  \pg{6184--6190}.

\bibitem[Kang {\em et~al.\/}(2002)Kang, Zhang, Chen \& He]{kan-zha-che-he2002}
{\sc \au{Kang, Q.}, \au{Zhang, D.}, \au{Chen, S.} \& \au{He, X.}} \yr{2002}
  \at{{Lattice Boltzmann simulation of chemical dissolution in porous media}}.
  \jt{Physical Review E}  \bvol{65}~(036318).

\bibitem[Klages {\em et~al.\/}(2008)Klages, Radons \& Sokolov]{kla-rad-sok2008}
{\sc \au{Klages, R.}, \au{Radons, G.} \& \au{Sokolov, I.~M.}} \yr{2008} {\em
  Anomalous transport: foundations and applications\/}.  \publ{John Wiley \&
  Sons}.

\bibitem[af~Klinteberg {\em et~al.\/}(2019)af~Klinteberg, Askham \&
  Kropinski]{kli-ask-kro2019}
{\sc \au{af~Klinteberg, L.}, \au{Askham, T.} \& \au{Kropinski, M.~C.}}
  \yr{2019}  \at{{A Fast Integral Equation Method for the Two-Dimensional
  Navier-Stokes Equations}}.  \jt{arxiv}  \bvol{1908.07392}.

\bibitem[af~Klinteberg \& Tornberg(2018)]{kli-tor2018}
{\sc \au{af~Klinteberg, L.} \& \au{Tornberg, A.{-}K.}} \yr{2018}  \at{{Adaptive
  Quadrature by Expansion for Layer Potential Evaluation in Two Dimensions}}.
  \jt{SIAM Journal on Scientific Computing}  \bvol{40}~(3),  \pg{A1225--1249}.

\bibitem[Kl\"{o}ckner {\em et~al.\/}(2013)Kl\"{o}ckner, Barnett, Greengard \&
  O'Neil]{klo-bar-gre-one2013}
{\sc \au{Kl\"{o}ckner, A.}, \au{Barnett, A.}, \au{Greengard, L.} \& \au{O'Neil,
  M.}} \yr{2013}  \at{{Quadrature by expansion: A new method for the evaluation
  of layer potentials}}.  \jt{Journal of Computational Physics}  \bvol{252},
  \pg{332--349}.

\bibitem[Knudby \& Carrera(2005)]{knu-car2005}
{\sc \au{Knudby, C.} \& \au{Carrera, J.}} \yr{2005}  \at{On the relationship
  between indicators of geostatistical, flow and transport connectivity}.
  \jt{Advances in Water Resources}  \bvol{28},  \pg{405--421}.

\bibitem[Koch \& Brady(1988)]{koc-bra1988}
{\sc \au{Koch, D.~L.} \& \au{Brady, J.~F.}} \yr{1988}  \at{Anomalous diffusion
  in heterogeneous porous media}.  \jt{Physics of Fluids}  \bvol{31}~(5),
  \pg{965--973}.

\bibitem[Konikow \& Bredehoeft(1978)]{kon-bre1978}
{\sc \au{Konikow, L.~F.} \& \au{Bredehoeft, J.~D.}} \yr{1978} {\em Computer
  model of two-dimensional solute transport and dispersion in ground water\/},
  ,  \vol{vol.~7}.  \publ{US Government Printing Office}.

\bibitem[Koponen {\em et~al.\/}(1996)Koponen, Kataja \&
  Timonen]{kop-kat-tim1996}
{\sc \au{Koponen, A.}, \au{Kataja, M.} \& \au{Timonen, J.}} \yr{1996}
  \at{Tortuos flow in porous media}.  \jt{Physical Review E}  \bvol{54}~(1),
  \pg{406--410}.

\bibitem[Kutsovsky {\em et~al.\/}(1996)Kutsovsky, Scriven \&
  Davis]{kut-scr-dav-ham1995}
{\sc \au{Kutsovsky, Y.~E.}, \au{Scriven, L.~E.} \& \au{Davis, H.~T.}} \yr{1996}
   \at{{NMR imaging of velocity profiles and velocity distributions in bead
  packs}}.  \jt{Physics of Fluids}  \bvol{8}~(4),  \pg{863--871}.

\bibitem[Lachauss{\'e}e {\em et~al.\/}(2018)Lachauss{\'e}e, Bertho, Morize,
  Sauret \& Gondret]{lachaussee2018competitive}
{\sc \au{Lachauss{\'e}e, F.}, \au{Bertho, Y.}, \au{Morize, C.}, \au{Sauret, A.}
  \& \au{Gondret, P.}} \yr{2018}  \at{Competitive dynamics of two erosion
  patterns around a cylinder}.  \jt{Physical Review Fluids}  \bvol{3}~(1),
  \pg{012302}.

\bibitem[L{\'o}pez {\em et~al.\/}(2018)L{\'o}pez, Stickland \&
  Dempster]{lopez2018cfd}
{\sc \au{L{\'o}pez, A.}, \au{Stickland, M.~T.} \& \au{Dempster, W.~M.}}
  \yr{2018}  \at{{CFD study of fluid flow changes with erosion}}.  \jt{Computer
  Physics Communications}  \bvol{227},  \pg{27--41}.

\bibitem[Matyka {\em et~al.\/}(2008)Matyka, Khalili \& Koza]{mat-kha-koz2008}
{\sc \au{Matyka, M.}, \au{Khalili, A.} \& \au{Koza, Z.}} \yr{2008}
  \at{Tortuosity-porosity relation in porous media flow}.  \jt{Physical Review
  E}  \bvol{78}~(2),  \pg{026306}.

\bibitem[Miller {\em et~al.\/}(1998)Miller, Christakos, Imhoff, McBride \&
  Pedit]{mil-chr-imh-mcb-ped1998}
{\sc \au{Miller, C.~T.}, \au{Christakos, G.}, \au{Imhoff, P.~T.}, \au{McBride,
  J.~F.} \& \au{Pedit, J.~A.}} \yr{1998}  \at{Multiphase flow and transport
  modeling in heterogeneous porous media: challenges and approaches}.
  \jt{Advances in Water Resources}  \bvol{31}~(2),  \pg{77--120}.

\bibitem[Mitchell \& Spagnolie(2017)]{mit-spa2017}
{\sc \au{Mitchell, W.~H.} \& \au{Spagnolie, S.~E.}} \yr{2017}  \at{{A
  generalized traction integral equation for Stokes flow, with applications to
  near-wall particle mobility and viscous erosion}}.  \jt{Journal of
  Computational Physics}  \bvol{333},  \pg{462--482}.

\bibitem[Moore(2017)]{moo2017}
{\sc \au{Moore, M.~N.~J.}} \yr{2017}  \at{{Riemann-Hilbert Problems for the
  Shapes Formed by Bodies Dissolving, Melting, and Eroding in Fluid Flows}}.
  \jt{Communications on Pure and Applied Mathematics}  \bvol{70}~(9),
  \pg{1810--1831}.

\bibitem[Moore {\em et~al.\/}(2013)Moore, Ristroph, Childress, Zhang \&
  Shelley]{moore2013self}
{\sc \au{Moore, M.~N.~J.}, \au{Ristroph, L.}, \au{Childress, S.}, \au{Zhang,
  J.} \& \au{Shelley, M.J.}} \yr{2013}  \at{Self-similar evolution of a body
  eroding in a fluid flow}.  \jt{Physics of Fluids}  \bvol{25}~(11),
  \pg{116602}.

\bibitem[Morrow {\em et~al.\/}(2019)Morrow, King, Moroney \&
  McCue]{morrow2019moving}
{\sc \au{Morrow, L.~C.}, \au{King, J.~R.}, \au{Moroney, T.~J.} \& \au{McCue,
  S.~W.}} \yr{2019}  \at{Moving boundary problems for quasi-steady conduction
  limited melting}.  \jt{arxiv}  \bvol{1901.01247}.

\bibitem[Nilsen \& Storesletten(1990)]{nil-sto1990}
{\sc \au{Nilsen, T.} \& \au{Storesletten, L.}} \yr{1990}  \at{An analytical
  study on natural convection in isotropic and anisotropic porous channels}.
  \jt{Journal of Heat Transfer}  \bvol{112}~(2),  \pg{396--401}.

\bibitem[Parker \& Izumi(2000)]{par-izu2000}
{\sc \au{Parker, G.} \& \au{Izumi, N.}} \yr{2000}  \at{Purely erosional cyclic
  and solitary steps created by flow over a cohesive bed}.  \jt{Journal of
  Fluid Mechanics}  \bvol{419},  \pg{203--238}.

\bibitem[Power \& Miranda(1987)]{pow-mir1987}
{\sc \au{Power, H.} \& \au{Miranda, G.}} \yr{1987}  \at{Second kind integral
  equation formulation of stokes' flows past a particle of arbitrary shape}.
  \jt{SIAM Journal on Applied Mathematics}  \bvol{47}~(4),  \pg{689--698}.

\bibitem[Pozrikidis(1992)]{poz1992}
{\sc \au{Pozrikidis, C.}} \yr{1992} {\em Boundary Integral and Singularity
  Methods for Linearized Viscous Flow\/}.  \publ{New York, NY, USA: Cambridge
  University Press}.

\bibitem[Puyguiraud {\em et~al.\/}(2019)Puyguiraud, Gouze \&
  Dentz]{puy-gou-den2019}
{\sc \au{Puyguiraud, A.}, \au{Gouze, P.} \& \au{Dentz, M.}} \yr{2019}
  \at{Stochastic dynamics of lagrangian pore-scale velocities in
  three-dimensional porous media}.  \jt{Water Resources Research}
  \bvol{55}~(2),  \pg{1196--1217}.

\bibitem[Quaife \& Moore(2018)]{qua-moo2018}
{\sc \au{Quaife, B.} \& \au{Moore, M.~N.~J.}} \yr{2018}  \at{A
  boundary-integral framework to simulate viscous erosion of a porous medium}.
  \jt{Journal of Computational Physics}  \bvol{375},  \pg{1--21}.

\bibitem[Rees \& Storesletten(1995)]{ree-sto1995}
{\sc \au{Rees, D.~A.~S.} \& \au{Storesletten, L.}} \yr{1995}  \at{{The Effect
  of Anisotropic Permeability on Free Convective Boundary Layer Flow in Porous
  Media}}.  \jt{Transport in Porous Media}  \bvol{19},  \pg{79--92}.

\bibitem[Ristroph {\em et~al.\/}(2012)Ristroph, Moore, Childress, Shelley \&
  Zhang]{ris-moo-chi-she-zha2012}
{\sc \au{Ristroph, L.}, \au{Moore, M.~N.~J.}, \au{Childress, S.}, \au{Shelley,
  M.~J.} \& \au{Zhang, J.}} \yr{2012}  \at{Sculpting of an erodible body in
  flowing water}.  \jt{Proceedings of the National Academy of Sciences}
  \bvol{109}~(48),  \pg{19606--19609}.

\bibitem[Rycroft \& Bazant(2016)]{rycroft2016asymmetric}
{\sc \au{Rycroft, C.~H.} \& \au{Bazant, M.~Z.}} \yr{2016}  \at{Asymmetric
  collapse by dissolution or melting in a uniform flow}.  \jt{Proceedings of
  the Royal Society A: Mathematical, Physical and Engineering Sciences}
  \bvol{472},  \pg{20150531}.

\bibitem[Saffman(1959)]{saf1959}
{\sc \au{Saffman, P.~G.}} \yr{1959}  \at{A theory of dispersion in a porous
  medium}.  \jt{Journal of Fluid Mechanics}  \bvol{6}~(3),  \pg{321--349}.

\bibitem[Siena {\em et~al.\/}(2019)Siena, Ilievand, Prill, Riva \&
  Guadagnini]{sie-ili-pri-riv-gua2019}
{\sc \au{Siena, M.}, \au{Ilievand, O.}, \au{Prill, T.}, \au{Riva, M.} \&
  \au{Guadagnini, A.}} \yr{2019}  \at{{Identification of Channeling in
  Pore-Scale Flows}}.  \jt{Geophysical Research Letters}  \bvol{46}~(6),
  \pg{3270--3278}.

\bibitem[Tang {\em et~al.\/}(2015)Tang, Valocchi \& Werth]{tan-val-wer2015}
{\sc \au{Tang, Y.}, \au{Valocchi, A.~J.} \& \au{Werth, C.~J.}} \yr{2015}  \at{A
  hybrid pore-scale and continuum-scale model for solute diffusion, reaction,
  and biofilm development in porous media}.  \jt{Water Resources Research}
  \bvol{51},  \pg{1846--1859}.

\bibitem[Trefethen \& Weideman(2014)]{tre-wei2014}
{\sc \au{Trefethen, L.~N.} \& \au{Weideman, J.~A.~C.}} \yr{2014}  \at{{The
  Exponentially Convergent Trapezoidal Rule}}.  \jt{SIAM Review}
  \bvol{56}~(3),  \pg{385--458}.

\bibitem[Wan \& Fell(2004)]{wan-fel2004}
{\sc \au{Wan, C.~F.} \& \au{Fell, R.}} \yr{2004}  \at{{Investigation of Rate of
  Erosion of Soils in Embankment Dams}}.  \jt{Journal of Geotechnical and
  Geoenvironmental Engineering}  \bvol{130}~(4),  \pg{373--380}.

\bibitem[Western {\em et~al.\/}(2001)Western, Bl\"{o}schl \&
  Grayson]{wes-blo-gra2001}
{\sc \au{Western, A.~W.}, \au{Bl\"{o}schl, G.} \& \au{Grayson, R.~B.}}
  \yr{2001}  \at{Toward capturing hydrologically significant connectivity in
  spatial patterns}.  \jt{Water Resources Research}  \bvol{37}~(1),
  \pg{83--97}.

\bibitem[Wykes {\em et~al.\/}(2018)Wykes, Huang \& Ristroph]{wykes2018self}
{\sc \au{Wykes, M.~S.~D.}, \au{Huang, J.~M.} \& \au{Ristroph, G.~A.~Hajjar~L.}}
  \yr{2018}  \at{Self-sculpting of a dissolvable body due to gravitational
  convection}.  \jt{Physical Review Fluids}  \bvol{3}~(4),  \pg{043801}.

\end{thebibliography}

\end{document}